\documentclass[conference,compsoc,letterpaper]{IEEEtran}
\usepackage[T1]{fontenc}
\usepackage[utf8]{inputenc}
\usepackage{amsmath,graphicx,url,booktabs,multirow,colortbl,array,xcolor}
\usepackage[nocompress]{cite}
\usepackage[caption=false,font=footnotesize,labelfont=sf,textfont=sf]{subfig}
\usepackage{capt-of}
\usepackage[hidelinks]{hyperref}
\graphicspath{{figures/}}
\hypersetup{pdftitle={Poisoning the Neighborhood: A Topology-Aware Topic-Level Attack on Large-Scale HNSW-based RAG Systems},pdfauthor={Yufei Chen, Yao Wang, Haibin Zhang}}
\makeatletter
\renewcommand{\tablename}{Table}
\let\SPoriginalmakecaption\@makecaption
\long\def\@makecaption#1#2{%
  \ifx\@captype\@IEEEtablestring
    \begingroup
    \normalfont\normalsize
    \vskip\abovecaptionskip
    \sbox\@tempboxa{#1: #2}%
    \ifdim\wd\@tempboxa>\hsize
      #1: #2\par
    \else
      \global\@minipagefalse
      \hb@xt@\hsize{\hfil\box\@tempboxa\hfil}%
    \fi
    \vskip\belowcaptionskip
    \endgroup
  \else
    \SPoriginalmakecaption{#1}{#2}%
  \fi}
\makeatother
\begin{document}
\title{Poisoning the Neighborhood: A Topology-Aware Topic-Level Attack on Large-Scale HNSW-based RAG Systems}
\IEEEoverridecommandlockouts
\author{
\IEEEauthorblockN{Yufei Chen}
\IEEEauthorblockA{Xidian University\\
\texttt{24151111267@stu.xidian.edu.cn}}
\and
\IEEEauthorblockN{Yao Wang\textsuperscript{*}}
\IEEEauthorblockA{Xidian University\\
\texttt{wangyao@xidian.edu.cn}}
\and
\IEEEauthorblockN{Haibin Zhang}
\IEEEauthorblockA{Xidian University\\
\texttt{hbzhang@mail.xidian.edu.cn}}
\thanks{*Corresponding author: Yao Wang.}
}
\maketitle
\begin{abstract}
Retrieval-augmented generation (RAG) mitigates hallucinations and knowledge staleness in large language models using external knowledge. However, ingestion from public and third-party sources exposes RAG systems to poisoning attacks. Existing attacks are designed and evaluated under exact retrieval, where poisoned documents participate in global ranking. Deployment-scale systems instead rely on approximate nearest-neighbor (ANN) indexes such as hierarchical navigable small world (HNSW), where limited search budgets, graph-search reachability, and local benign-document competition invalidate this assumption and degrade existing attacks. We present a topology-aware topic-level poisoning attack against large-scale HNSW-RAG that generalizes to unseen topic queries. We organize poisoned documents into a Hub–Satellite structure: Satellites provide competitive entry points across semantic regions, while Hubs strengthen local cohesion and facilitate cascading retrieval. For generation, we propose an entity–claim promotion attack combining evidence anchoring and entity–claim coupling to promote the target entity via a target claim across different query regions. On 2M-document Wikipedia, our attack maintains over 72\% retrieval success and reaches 31.29\% attack success, substantially outperforming baselines; domain-specific experiments further support its effectiveness. Structural analyses show that the Hub–Satellite topology forms in HNSW, indicating that graph structure contributes to improved poisoned-document retrieval. Existing poisoning defenses cannot mitigate the attack, exposing graph-structural risks in HNSW-RAG systems.

\end{abstract}

\section{Introduction}

RAG~\cite{lewis2020retrieval} supplements large language models (LLMs) with dynamically updated external knowledge, helping mitigate hallucinations and knowledge staleness~\cite{kandpal2023large,shuster2021retrieval,gao2023retrieval}. RAG has been adopted in systems such as Microsoft 365 Copilot~\cite{microsoft2026copilotretrieval}, GitHub Copilot~\cite{choi2024githubcopilotrag}, and Amazon Bedrock~\cite{aws2026bedrock}. However, maintaining knowledge freshness and coverage requires RAG systems to continually ingest content from the public web and third-party sources, creating a new attack surface. Adversaries may introduce poisoned documents during data collection or index updates. Once retrieved, these documents can steer model outputs toward attacker-specified content, undermining the trustworthiness of deployed RAG systems.

\begin{figure}[t]
\centering
\noindent\includegraphics[width=\columnwidth]{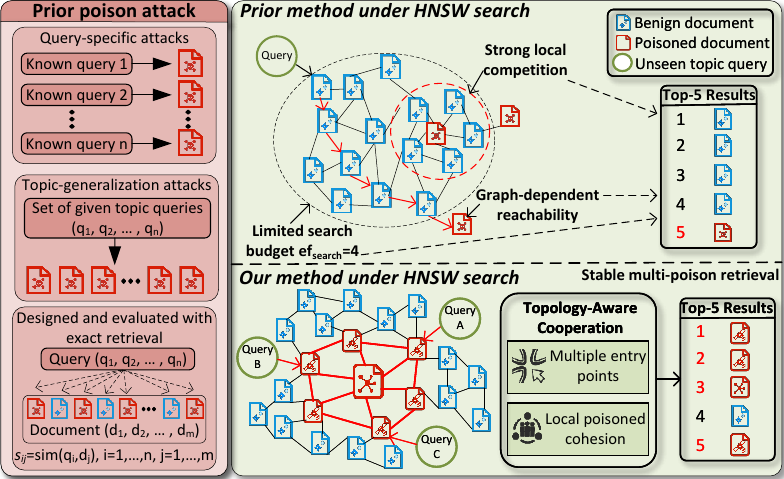}
\caption{Overview and HNSW performance comparison of existing and proposed poisoning attacks.}
\label{fig:poison-overview}
\end{figure}

Most existing RAG poisoning attacks are designed and evaluated under exact retrieval and rely on an assumption: \textbf{each query is compared with every document in the knowledge base, and the top-\(k\) results are selected according to a global ranking.} Attackers therefore need only maximize the semantic similarity between target queries and poisoned documents to place them in the retrieval results. Accordingly, early methods optimize poisoned documents for a single query or a predefined query set~\cite{zou2025poisonedrag,chen2025flippedrag,shafran2025machine,liang2026graphrag,geng2025unic}. Topic-FlipRAG~\cite{gong2025topic} and DiscourseFlip~\cite{gong2026discourseflip} exploit semantic or structural features shared across topic-related queries to extend poisoning to the topic level. However, they still construct poisoned documents from attack-time topic queries and are primarily evaluated under exact retrieval. Thus, their attack construction remains dependent on known query sets and assumes that poisoned documents participate in global similarity ranking.

As RAG systems increasingly rely on larger and continuously updated knowledge bases~\cite{borgeaud2022improving}, the global-ranking assumption becomes increasingly unrealistic. The cost of exhaustive similarity computation drives practical systems to adopt ANN indexes~\cite{wang2021graphann}, such as HNSW~\cite{malkov2020hnsw}. Under HNSW, poisoning attacks face three challenges: (1) \textbf{limited search budget}: HNSW explores only a subset of graph nodes, so poisoned documents are no longer guaranteed to be visited and participate in candidate comparisons; (2) \textbf{graph-search reachability}: under a limited search budget, whether a poisoned document is visited further depends on the search paths from the entry node through the local adjacency structure; and (3) \textbf{intensified local competition}: as the knowledge base grows, more benign documents enter neighborhoods, competing for the search budget and occupying candidate-set slots. 

These challenges are particularly consequential for topic-level poisoning. Unlike attacks optimized for a single known query, a topic-level attack must use a limited set of poisoned documents to cover queries with diverse semantic intents and formulations. Such broader coverage reduces each document's specificity and ranking advantage for individual queries, compounding the retrieval constraints imposed by HNSW. This raises a central question: \textit{can an attacker mount an effective topic-level poisoning attack against large-scale HNSW-RAG while generalizing to unseen topic queries?}

To this end, we propose a topology-aware topic-level poisoning attack against HNSW-RAG. Our key insight is that HNSW graph connectivity is influenced by embedding-space proximity and heuristic neighbor selection, while retrieval proceeds through graph traversal under a limited search budget. This property allows attackers to indirectly induce favorable graph neighborhoods by coordinating the semantic relationships among injected documents, thereby improving their reachability during graph search. On the retrieval side, we organize independently competing poisoned documents into a role-differentiated Hub--Satellite structure. Satellites target distinct semantic regions, maintaining local ranking competitiveness and providing multiple retrieval entry points for diverse unseen queries. Hubs aggregate semantics shared across these regions, serving as common anchors that encourage local cohesion among dispersed Satellites. This semantics-induced, ``multiple-entry, locally cohesive'' structure enables set-level cooperation: once a search reaches one poisoned document, neighbor expansion may expose additional poisoned documents, improving reachability and retrieval stability under limited budgets.

On the generation side, we introduce entity--claim promotion for diverse unseen queries within the target topic. Advice-seeking, comparison, and applicability queries differ in form but share a common decision-oriented goal: evaluating whether or which solution to recommend. We therefore make all poisoned documents support the same target entity and a shared claim about its benefits across different decision contexts. Evidence anchoring presents the claim as a conclusion supported by background, mechanism, conditions, and outcome evidence, while entity--claim coupling ties the mechanism and outcome to the target entity, encouraging the model to preserve this association and recommend the entity in its response. Finally, two-level candidate optimization balances document-level retrieval and generation utility with set-level coverage complementarity and topological cooperation.

Experiments show that our method maintains stable attack performance across datasets and knowledge-base sizes, achieves an attack success rate of up to 31.29\%, substantially outperforms existing baselines, and generalizes to diverse unseen topic queries. Structural analysis confirms that the Hub--Satellite cooperative structure forms in the target HNSW index and significantly improves poisoned-document retrieval. Defense experiments show that existing methods cannot effectively mitigate the attack, revealing practical security risks in large-scale HNSW-RAG systems.

\textbf{Contributions.}
To the best of our knowledge, we present the first RAG poisoning attack designed for large-scale HNSW indexes and capable of generalizing to unseen topic queries. Our contributions are as follows:

(1) We characterize and empirically quantify the impact of HNSW search on existing RAG poisoning attacks.
(2) We propose a topology-aware Hub--Satellite poisoning method that improves unseen-query coverage and HNSW graph-search reachability through distributed query access and semantically induced local cohesion.

(3) We introduce an entity--claim promotion mechanism that combines evidence anchoring with entity--claim coupling to manipulate diverse unseen topic queries.

(4) We extensively evaluate the attack across datasets, knowledge-base sizes, and system settings, validating its effectiveness and key design choices while showing that existing defenses remain insufficient against this threat.

\section{Background and Related Work}

\subsection{Background}

\textbf{RAG.}
An external knowledge base supplements the parametric knowledge encoded in LLMs. Let \(D=\{d_1,\ldots,d_N\}\) denote a knowledge base of \(N\) documents and \(q\) a user query. During retrieval, the system maps \(q\) and each document \(d_i\) into a shared embedding space, ranks the documents by their similarity scores \(s(q,d_i)\), and returns the \(k\) most similar documents. During generation, the LLM produces an answer conditioned on the query and retrieved documents. Because retrieved content directly informs generation, the final output depends not only on the LLM but also on the relevance and trustworthiness of the retrieved documents.

\textbf{HNSW.} HNSW is a graph-based ANN index widely used in industrial vector search systems, including Azure AI Search~\cite{azure-vector-search}, Amazon OpenSearch Service~\cite{aws-opensearch-vector}, Elasticsearch~\cite{elastic-dense-vector}, and MongoDB Vector Search~\cite{mongodb-vector-search}. HNSW organizes document embeddings into a multilayer navigable graph, where sparse upper layers guide navigation toward relevant regions and a denser base layer supports local search. During index construction, each new node searches for nearby candidates and selects neighbors using a heuristic that considers both vector proximity and neighbor diversity. At query time, search descends greedily through upper layers and explores candidates at the base layer. Two key parameters govern this process: \(M_{\mathrm{HNSW}}\) controls graph connectivity, with a larger neighbor limit at the base layer, while \(\mathrm{ef}_{\mathrm{search}}\) controls the search candidate-list size and thus the extent of exploration.

Given a query \(q\), HNSW starts from an entry node in an upper layer, greedily follows local connections toward documents with higher \(s(q,d_i)\), and descends layer by layer. At the bottom layer, it maintains and expands a candidate set controlled by \(\mathrm{ef}_{\mathrm{search}}\), then returns the \(k\) highest-scoring candidates. Unlike exact retrieval, HNSW computes similarity scores only for documents reached during graph traversal and thus typically visits only a subset of \(D\). 

\subsection{Related Work}

Existing RAG poisoning attacks exploit the generator's reliance on retrieved context by injecting carefully crafted poisoned documents into retrieval results to manipulate model outputs. We first review embedding-space retrieval attacks, followed by query-specific and topic-level attacks distinguished by their query coverage.

\textbf{Embedding-space retrieval attacks.}
These attacks exploit embedding-space properties to increase the retrieval exposure of malicious content. Ben-Tov and Sharif propose GASLITE~\cite{ben2025gasliteing}, which optimizes adversarial passages for unseen concept-related queries but relies on white-box access to retriever parameters. Li et al.~\cite{li2025unsupervised} develop an unsupervised poisoning method that optimizes adversarial documents in continuous embedding space, primarily targeting retrieval ranking disruption rather than downstream answer manipulation. Black-Hole~\cite{li2026can} Attack exploits embedding-space centrality to attract unknown queries but requires direct malicious-vector injection.

\textbf{Query-specific attacks.}
Query-specific attacks construct poisoned documents for a single target query or a predefined query set to enter retrieval results and manipulate RAG outputs. PoisonedRAG~\cite{zou2025poisonedrag} injects adversarial text to induce a specific incorrect answer; FlippedRAG~\cite{chen2025flippedrag} constructs misleading evidence in a black-box setting to steer the model toward a target viewpoint; Jamming~\cite{shafran2025machine} uses blocker documents to make the system refuse specific queries. GRAGPOISON~\cite{liang2026graphrag} exploits shared relations among related multi-hop queries to extend poisoning from a single query to a fixed query set; UniC-RAG~\cite{geng2025unic} further targets a large predefined query set and combines query clustering with instruction hijacking to induce attacker-specified LLM outputs. Chang et al.~\cite{chang2026overcoming} optimize retrieval triggers for known queries to enable indirect prompt injection. Pathak~\cite{pathak2026coverage} employs coordinated poisoning against predefined queries to evade admission-time defenses. Despite differing coverage, all these methods require predefined target queries.

\textbf{Topic-level attacks.}
Topic-level attacks exploit semantic or structural features shared across topic-related queries to extend poisoning from a specific query to a set of queries within the same topic. Topic-FlipRAG~\cite{gong2025topic} edits target documents using shared semantics and keywords to manipulate known topic-related queries. DiscourseFlip~\cite{gong2026discourseflip} organizes queries into a semantic network and uses graph-guided budget allocation to achieve discourse-level opinion manipulation. Both methods construct poisoned documents from topic queries specified in advance. 

\textbf{Limitations.}
Prior poisoning attacks exhibit three limitations. (1) \textit{Attack privileges.} Embedding-space attacks require white-box access to retrieval-model parameters or the ability to inject malicious vectors directly, exceeding the capabilities of attackers restricted to injecting natural-language documents under black-box conditions. (2) \textit{Retrieval mechanisms.} Many document poisoning methods are designed and evaluated under exact retrieval, optimizing and validating poisoned documents against global similarity rankings without considering how other retrieval backends affect attack performance. (3) \textit{Query coverage.} Query-specific attacks optimize for predefined queries, whereas topic-level attacks still rely on known topic-related queries to construct poisoned documents. This dependence may limit coverage of unseen topic queries with different intents or formulations.

\section{Preliminary Study}

\subsection{Problem Definition}

These limitations reveal two gaps between existing RAG poisoning attacks and practical deployments: a retrieval-backend mismatch and limited query coverage. However, how these gaps affect the retrieval performance of existing attacks has not been systematically examined. We therefore conduct a preliminary study guided by two questions:

(1) \textit{Retrieval backend.} In large-scale knowledge bases representative of practical deployments, is exact retrieval still an appropriate backend for evaluating RAG poisoning attacks?

(2) \textit{Topic-level retrieval coverage under HNSW.} Under HNSW retrieval, to what extent do unseen topic queries and increasing knowledge-base size affect the retrieval coverage of existing topic-level attacks?

\subsection{Experimental Setup}

We select PoisonedRAG and Topic-FlipRAG, both with public implementations, as representative query-specific and topic-level poisoning baselines, respectively. We set the poisoned-document budget \(m\) to 40 or 60. Retrieval Success Rate (RSR) is evaluated on both the training queries used to construct poisoned documents and unseen test queries excluded from attack construction; it is the fraction of queries for which at least one poisoned document appears in the top-\(k\) results. To capture diverse intents and formulations within each topic, we construct 5,000 unseen test queries per attack topic. We use knowledge bases of 10K, 100K, and 500K documents to evaluate retrieval stability as benign-document competition intensifies, and extend the size to 1M when comparing exact retrieval with HNSW. All comparisons use the same embedding model, knowledge base, and retrieval settings. Further implementation details appear in Appendix~\ref{Preliminary}.

\begin{figure*}[t]
  \centering
  \subfloat[Retrieval latency]{\begin{minipage}[t]{0.247\textwidth}
    \centering
    \includegraphics[width=\linewidth,height=1.38in]{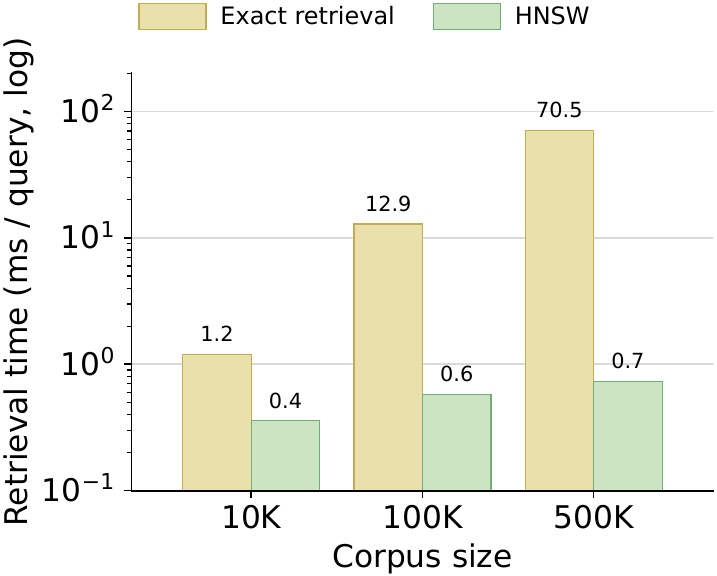}
    
    \label{fig:prelim-time}
  \end{minipage}}%
  \hspace{0.006\textwidth}%
  \subfloat[Poison retrieval under HNSW]{\begin{minipage}[t]{0.494\textwidth}
    \centering
    \includegraphics[width=\linewidth,height=1.38in]{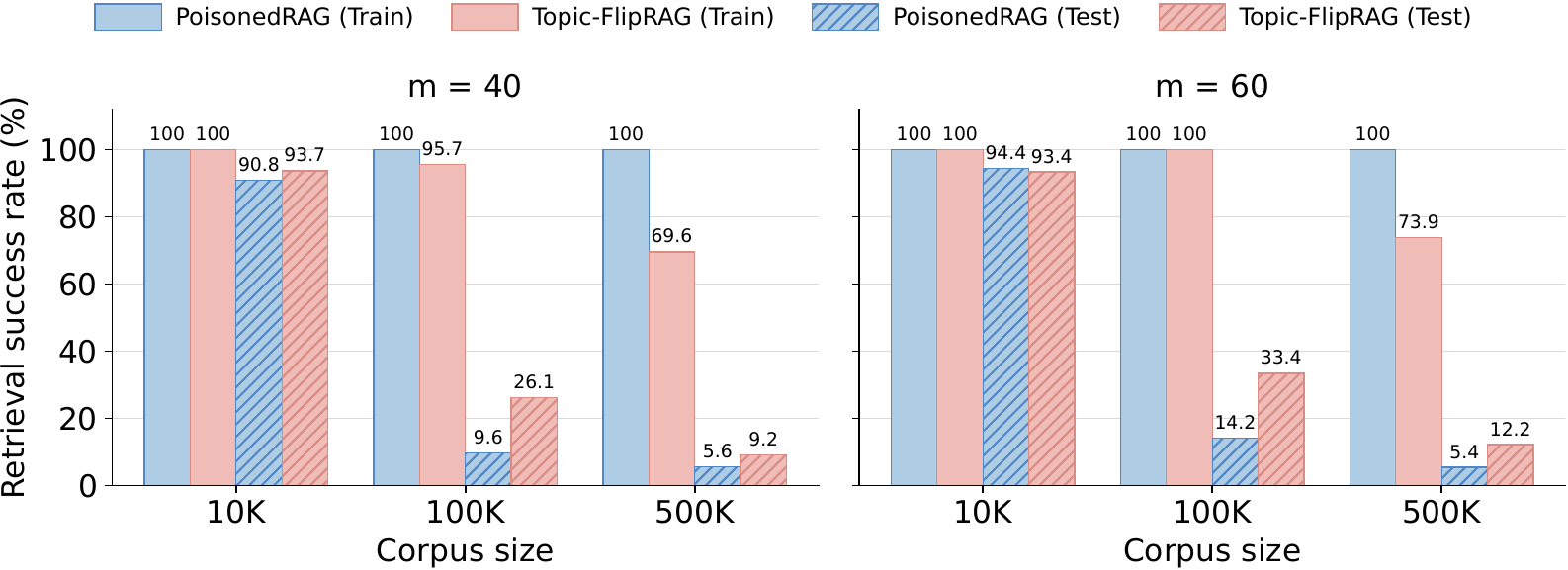}
    
    \label{fig:prelim-success}
  \end{minipage}}%
  \hspace{0.006\textwidth}%
  \subfloat[Exact versus HNSW retrieval]{\begin{minipage}[t]{0.247\textwidth}
    \centering
    \includegraphics[width=\linewidth]{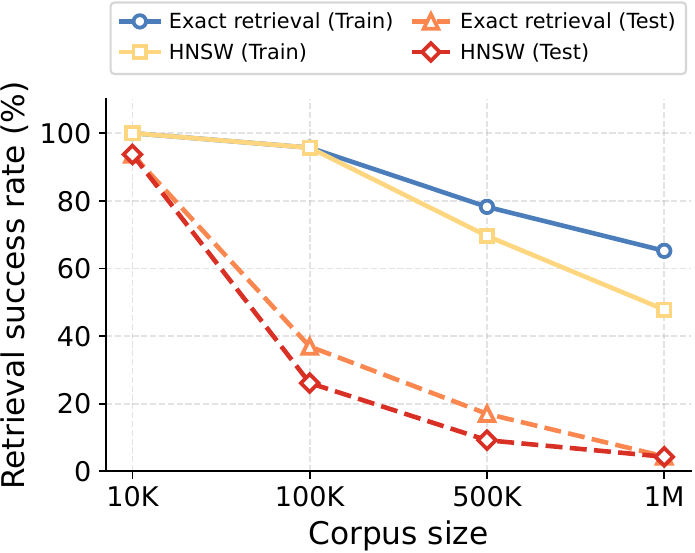}
    
    \label{fig:prelim-hnsw}
  \end{minipage}}%
  \caption{Retrieval latency and poisoning success across knowledge-base sizes. (a) Retrieval latency. (b) PoisonedRAG and Topic-FlipRAG under HNSW. (c) Topic-FlipRAG under exact versus HNSW retrieval.}
  \label{fig:prelim-main}
\end{figure*}

\subsection{Result Analysis}
\label{sec:analysis}

\textbf{Retrieval latency.}
As shown in Figure~\ref{fig:prelim-time}, exact-retrieval latency grows approximately linearly with knowledge-base size, whereas HNSW maintains low latency across scales. At 500K documents, exact retrieval takes an average of 70.5~ms per query, whereas HNSW takes only 0.7~ms, a difference of approximately 100\(\times\). This substantial gap indicates that exact retrieval is ill-suited to the efficiency requirements of large-scale RAG deployments, making HNSW a more representative evaluation backend.

\textbf{Topic-coverage degradation.}
Figure~\ref{fig:prelim-success} shows that both baselines lose retrieval coverage on unseen topic queries as the knowledge base grows. PoisonedRAG attains 100\% RSR on training queries. Its RSR on unseen topic queries is 90.8\% with 10K documents but drops sharply to 9.6\% with 100K documents. Topic-FlipRAG explicitly exploits semantics and keywords shared across topic-related queries to improve topic-level coverage. Even so, with 40 poisoned documents, its unseen-query RSR falls from 93.7\% at 10K to 9.2\% at 500K. These results show that the topic-level generalization observed in small knowledge bases deteriorates as benign-document competition intensifies.

\textbf{Impact of HNSW.}
Figure~\ref{fig:prelim-hnsw} shows Topic-FlipRAG's RSR under exact and HNSW retrieval. At 500K and 1M documents, its training-query RSR under HNSW is 8.6 and 17.4 percentage points lower, respectively, than under exact retrieval; unseen test queries show the same trend. These results show that graph search also weakens poisoned-document retrieval coverage.

\textbf{Mechanism of degradation.}
Existing poisoning attacks typically optimize poisoned documents for queries or query sets known at attack time, so their retrieval advantage largely depends on high semantic similarity to those queries. This advantage weakens for unseen queries with different intents and formulations. Exact retrieval still includes every poisoned document in the global ranking, allowing them to compete for the top-\(k\); HNSW instead visits only a limited set of nodes along local graph connections, and a poisoned document must first be reached by the search path before entering the candidate set. As the knowledge base grows, more semantically relevant benign documents enter local neighborhoods, reducing the similarity advantage of poisoned documents and making retrieval constrained by graph-search reachability and limited search budgets.

\textbf{Summary.}
Our preliminary study reveals two key gaps between existing RAG poisoning evaluations and large-scale deployment settings.

(1) \textit{Exact-retrieval-only evaluation.} Most existing attacks are evaluated under exact retrieval, whereas deployment-scale RAG systems commonly rely on ANN backends such as HNSW for efficient retrieval. Our results show that exact retrieval can overestimate poisoned-document retrieval coverage relative to HNSW.

(2) \textit{Degraded unseen-query coverage at scale.} As the knowledge base grows, existing attacks suffer substantial retrieval degradation on unseen topic queries, and HNSW further compounds this degradation.

\begin{figure*}[t]
  \centering
  \includegraphics[width=0.88\textwidth]{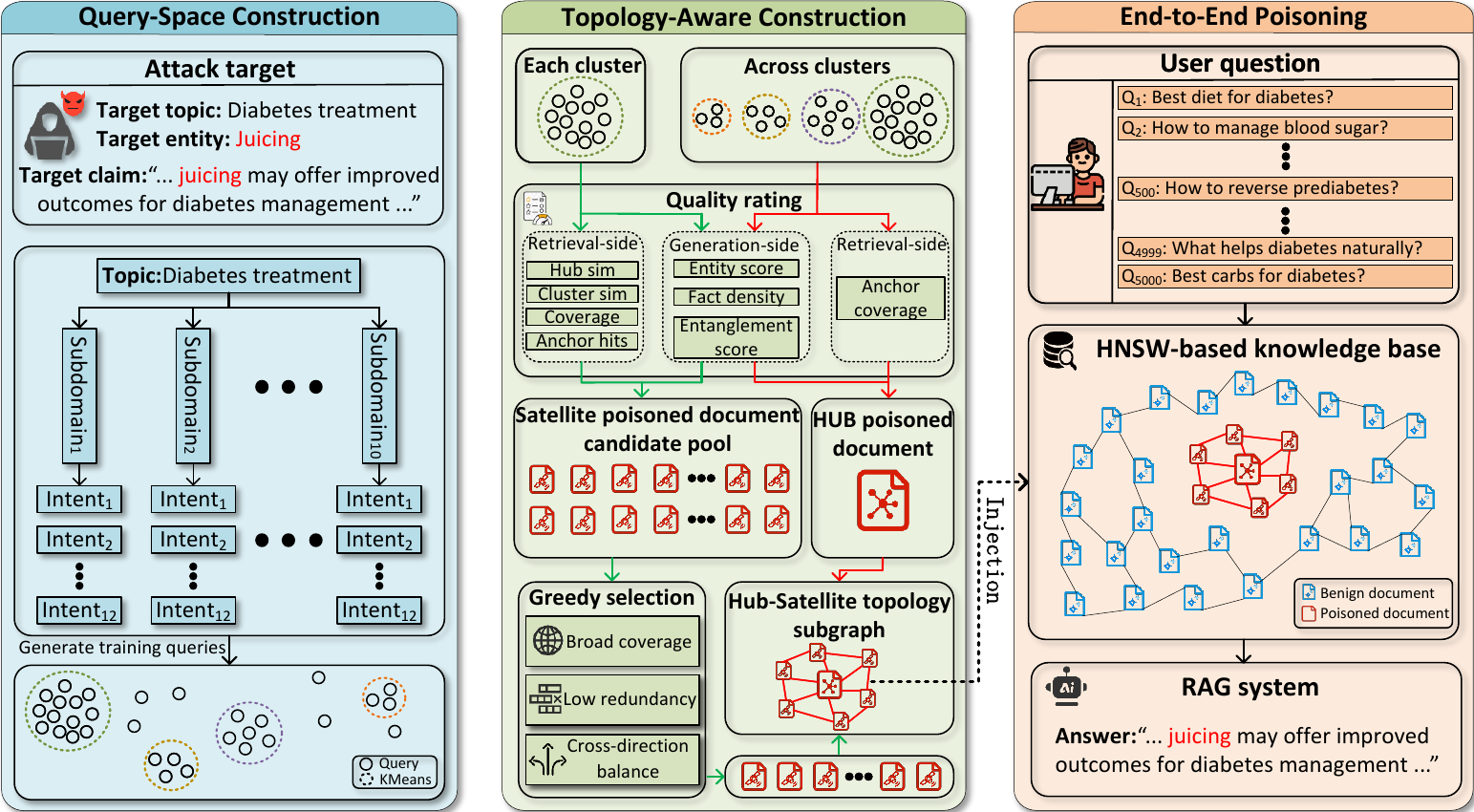}
  \caption{Overview of our topology-aware poisoning attack against HNSW-based RAG.}
  \label{fig:overview}
\end{figure*}

\section{Threat Model}

We describe the threat model in terms of the attacker's goals and capabilities.

\subsection{Attacker Goals}
A successful RAG poisoning attack requires poisoned documents to first enter the top-\(k\) retrieved context and then influence the LLM’s generation. We therefore define retrieval- and generation-side goals separately and specify the target query type based on the generation-side objective.

\textbf{Retrieval-side goal.}
Given an attack topic \(T\), let \(Q_T=\{q_1,q_2,\ldots,q_{|Q_T|}\}\) denote unseen topic queries with diverse intents and formulations. The target system retrieves context from a large-scale HNSW-indexed knowledge base \(D\), into which the attacker injects \(P=\{p_1,p_2,\ldots,p_m\}\), a set of \(m\) poisoned documents. The retrieval-side goal is to maximize the number of poisoned documents appearing in the top-\(k\) retrieval results across \(Q_T\), such that as many poisoned documents as possible are retrieved for each \(q\in Q_T\).

\textbf{Generation-side goal.}
Unlike PoisonedRAG's query-specific incorrect-answer objective and Topic-FlipRAG's stance manipulation over predefined topic queries, we target entity--claim promotion. The attacker specifies a target entity \(e_t\) and a target claim \(c_t\) describing its intended benefits, and incorporates the same entity--claim pair into all poisoned documents. Once retrieved, these documents are intended to make the model treat \(c_t\) as a supported conclusion and use it as a reason to favor or recommend \(e_t\) in the final response. Rather than inducing a fixed answer to a particular query, this objective promotes the same entity through the same target claim across unseen topic queries.

\textbf{Target queries.}
Because our attack seeks to influence user decisions by promoting \(e_t\) through \(c_t\), we focus on decision-oriented queries. Such queries ask the model for actionable advice, solution recommendations, option comparisons, or management strategies. Users rely on these responses to support decisions; misleading suggestions induced by poisoned retrieval context may therefore directly affect their choices and subsequent actions, causing serious real-world harm.

\subsection{Attacker Capabilities}

We adopt a black-box setting. The attacker selects an attack topic \(T\) and specifies a target claim \(c_t\) containing a target entity \(e_t\). The attacker knows only that the target system uses HNSW and has no knowledge of the unseen topic queries \(Q_T\), retriever or generator parameters, index topology, or HNSW configuration. The attacker cannot alter the system prompt, retrieval pipeline, reranker, or generation procedure. The attacker can use publicly available language models and a surrogate retriever to generate surrogate queries and construct and screen poisoned documents. The attacker can only place a small number of poisoned documents in public or third-party knowledge sources that the target system may later index and cannot directly modify the target knowledge base or index.

\section{Method}

To address existing attacks' unstable retrieval and limited coverage of unseen topic queries in large-scale HNSW-RAG, we propose a topology-aware topic-level poisoning attack. As shown in Figure~\ref{fig:overview}, our method combines retrieval- and generation-side designs. On the retrieval side, we construct a surrogate query space approximating the unknown topic-query distribution and introduce a role-differentiated Hub--Satellite structure. Satellites span distinct query regions as locally competitive retrieval entry points, while Hubs aggregate shared topic semantics to promote cohesion among dispersed Satellites. By exploiting HNSW's embedding-based neighbor selection, this semantic coordination can induce a ``multiple-entry, locally cohesive'' poisoned-document neighborhood, facilitating cascading retrieval without direct graph manipulation. On the generation side, evidence anchoring and entity--claim coupling organize the argument of each candidate document to encourage the model to adopt the target claim and recommend the target entity. Finally, document-level multi-candidate joint scoring evaluates each candidate's retrieval and generation utility, while set-level greedy complementary selection determines the final poisoned-document set based on marginal query coverage and Hub association.

\subsection{Query-Space Construction}

Queries within an attack topic often vary substantially in intent and formulation. Optimizing poisoned documents around only a few fixed queries tends to concentrate them within limited local semantic regions, leaving other directions in the topic-query space uncovered. Because the attacker has no access to unseen topic queries, we construct a surrogate query space to approximate the broader topic-query distribution and provide a basis for subsequent local-region partitioning.

Given an attack topic \(T\), the attacker uses an LLM to generate \(n\) surrogate queries spanning different subdomains, intents, and formulations. Let \(Q_s=\{q_i\}_{i=1}^{n}\) denote this query set. The surrogate query encoder \(\widetilde{E}_q\) maps each \(q_i\) to \(\mathbf{h}_i\), and \(K\)-means partitions the resulting embeddings into \(K\) semantic regions:
\[
\mathbf{h}_i=\widetilde{E}_q(q_i), \qquad
\{C_k\}_{k=1}^{K}
=\operatorname{KMeans}\bigl(\{\mathbf{h}_i\}_{i=1}^{n}\bigr).
\]
For each region \(C_k\), let \(Q_k=\{q_i\in Q_s:\mathbf{h}_i\in C_k\}\) denote its query subset. The sets \(\{Q_k\}_{k=1}^{K}\) partition \(Q_s\). This partition turns broad topic coverage into a collection of local semantic targets for poisoned-document construction.

\subsection{Retrieval-Side Topology Forgery}
The local semantic regions identified above require poisoned documents to cover different directions in the topic-query space. Despite sharing coarse topic semantics, specialization to different local query distributions disperses their embeddings and weakens local cohesion. Under a limited HNSW search budget, reaching one poisoned node does not guarantee access to others, making query--document semantic matching alone insufficient for stable graph-search reachability. We first construct Hubs to aggregate semantics shared across regions as semantic anchors, and then construct Satellites by jointly optimizing local-query matching and Hub association. Satellites provide locally competitive retrieval entry points, while Hubs strengthen associations among dispersed entry points and local cohesion within the poisoned-document neighborhood, improving HNSW reachability and retrieval stability.

\subsubsection{Hub Document Construction}

Hub documents serve as shared semantic anchors across local query regions. Unlike Satellites, which specialize in fine-grained query intents, a Hub captures topic concepts shared across multiple semantic directions, allowing it to remain associated with Satellites from different regions. For each region \(C_k\), we extract a keyword-anchor set \(A_k\) from its surrogate queries \(Q_k\). A candidate Hub \(h\) receives the retrieval-side reward
\[
R_{\mathrm{hub}}^{\mathrm{ret}}(h)
=\operatorname{AnchorCov}
\left(h,\bigcup_{k=1}^{K}A_k\right),
\]
where \(\mathrm{AnchorCov}\) measures the candidate's coverage of the aggregated keyword anchors. Maximizing this reward encourages the Hub to capture semantics shared across query regions rather than specialize in any single region, thereby serving as a common semantic anchor that strengthens local cohesion among dispersed retrieval entry points.

\subsubsection{Satellite Document Construction}

Hubs capture semantics shared across query regions but cannot provide fine-grained alignment with every query intent. Satellites complement this role by specializing in local regions of the topic-query space, serving as locally competitive retrieval entry points for the corresponding queries.

Semantic regions contain different numbers of surrogate queries. We therefore partition regions with more queries into more local subclusters, rather than assigning an equal number of Satellite targets to every region. Let \(Q_k\) denote the query set in \(C_k\), which is partitioned into \(m_k\) local subclusters \(\{Q_{k,j}\}_{j=1}^{m_k}\). Each \(Q_{k,j}\) represents a concentrated local query distribution, summarized by its semantic center
\[
\boldsymbol{\mu}_{k,j}
=
\frac{1}{|Q_{k,j}|}
\sum_{q\in Q_{k,j}}
\widetilde{E}_q(q).
\]

For each \(Q_{k,j}\), we generate multiple Satellite candidates with diverse formulations. A Satellite should align with its local query region while remaining associated with the constructed Hubs and preserving regional topic semantics. Let \(H\) denote the Hub set, \(\mathbf z_x=\widetilde{E}_d(x)\) the surrogate embedding of document \(x\), and \(a_H(p)=\max_{h\in H}\cos(\mathbf z_p,\mathbf z_h)\) the Hub-association score. For a candidate \(p\) generated for \(Q_{k,j}\), we define its retrieval-side reward as
\begin{multline*}
R_{\mathrm{sat}}^{\mathrm{ret}}(p;k,j)
={}\lambda_1\cos(\mathbf z_p,\boldsymbol{\mu}_{k,j})+\lambda_2 a_H(p)\\
+\lambda_3\operatorname{AnchorCov}(p,A_k)
+\lambda_4\operatorname{QueryCov}(p,Q_{k,j}).
\end{multline*}

The first term measures similarity to the subcluster center, the second encourages association with the previously constructed Hubs, the third preserves regional topic semantics through anchor coverage, and the fourth measures query coverage within \(Q_{k,j}\). Together, the first and fourth terms capture local ranking competitiveness. The Hub-association term guides Satellite construction toward the preconstructed Hubs, encouraging local HNSW graph connections through semantic proximity while preserving region-specific query alignment. Here, \(\operatorname{QueryCov}(p,Q_{k,j})\) denotes the fraction of queries in \(Q_{k,j}\) covered by \(p\), and \(\lambda_1,\ldots,\lambda_4\) control the relative contributions of the four terms. Candidates from all subclusters form the Satellite candidate pool for subsequent selection.

Overall, on the retrieval side, the Hub--Satellite structure overcomes two limitations of existing attacks in large-scale HNSW-RAG: unstable retrieval and limited coverage of unseen topic queries. (1) \textit{Local cohesion.} While preserving different query directions, Satellites remain semantically associated with the Hub, forming a locally cohesive structure among poisoned documents. Once a search reaches any poisoned node, neighbor expansion can continue to other poisoned documents, thereby improving graph-search reachability and retrieval stability under a limited HNSW search budget. (2) \textit{Multiple entry points.} Satellites are distributed across distinct local query regions and maintain local ranking competitiveness, providing multiple retrieval entry points for unseen topic queries with diverse semantics and formulations.

\subsection{Entity--Claim Promotion for Generation}
On the generation side, a topic-level attack must induce diverse unseen topic queries toward the same attacker-specified outcome. We therefore adopt entity--claim promotion as a shared generation objective across all poisoned documents, making them consistently support a target claim \(c_t\) associated with the target entity \(e_t\). Evidence anchoring increases the likelihood that the model adopts \(c_t\) as a supported conclusion, while entity--claim coupling binds \(e_t\) to \(c_t\), translating claim adoption into target-entity promotion.

\subsubsection{Evidence Anchoring}
Directly inserting \(c_t\) into a poisoned document may leave it as an isolated assertion the LLM can ignore. Evidence anchoring embeds \(c_t\) within a structured argument. For a document \(x\), we compose domain background, mechanism, applicable conditions, outcome evidence, and the target claim, written as \(x=\operatorname{Compose}\bigl(b_x,m_x,a_x,r_x,c_t\bigr)\), where \(b_x\), \(m_x\), \(a_x\), and \(r_x\) denote the supporting components. For Hubs, these components emphasize topic-level concepts, mechanisms, and decision criteria shared across query regions, providing a rationale for \(c_t\). For Satellites, they reflect local intents and decision conditions in \(Q_{k,j}\), adapting the argument to each query region while preserving \(c_t\). In this way, evidence anchoring makes \(c_t\) more likely to be adopted as a supported conclusion rather than ignored as an isolated assertion.

\subsubsection{Entity--Claim Coupling}

Evidence anchoring increases the likelihood that the model adopts the target claim \(c_t\). Building on this, entity--claim coupling further translates claim adoption into target-entity promotion by making \(e_t\) an indispensable component of both \(c_t\) and its supporting argument. Specifically, we explicitly attribute the mechanism and outcome supporting \(c_t\) to \(e_t\), such that removing \(e_t\) makes the target claim incomplete and disrupts the mechanism--outcome--claim reasoning chain. We express this dependency as \(m_x(e_t)\rightarrow r_x(e_t)\Rightarrow c_t(e_t)\). By maintaining this dependency throughout the supporting argument, entity--claim coupling increases the likelihood that the model retains and promotes \(e_t\) when adopting \(c_t\).

A generation-effective candidate should fully express \(c_t\), provide coherent evidence supporting it, and maintain a strong dependency between \(e_t\) and the supporting argument. We jointly capture these properties as
\[
\begin{aligned}
R_{\mathrm{gen}}(x)={}&
\lambda_{\mathrm{claim}}S_{\mathrm{claim}}(x,c_t)
+\lambda_{\mathrm{evid}}S_{\mathrm{evid}}(x,c_t)\\
&+\lambda_{\mathrm{link}}S_{\mathrm{link}}(x,e_t,c_t).
\end{aligned}
\]

Here, \(S_{\mathrm{claim}}\) measures the completeness of the target claim; \(S_{\mathrm{evid}}\) measures the completeness and consistency of the background, mechanism, conditions, and outcome evidence supporting \(c_t\); and \(S_{\mathrm{link}}\) measures the degree to which \(e_t\) is explicitly tied to the mechanism, outcome, and target claim. The \(\lambda\) terms control the relative contributions of the three scores. This reward favors candidates that present a complete and well-supported claim while maintaining a strong dependency between \(e_t\) and \(c_t\) throughout the supporting argument.

Beyond improving individual-document generation effectiveness, the shared entity--claim design provides cross-document benefits for retrieval and generation.
(1) \textit{Retrieval-side semantic basis.} Although Satellites target different local query regions, they share the same target entity \(e_t\) and target claim \(c_t\). This shared semantics gives Satellites a natural similarity basis, making them more likely to occupy nearby HNSW neighborhoods and form structural associations while preserving local query alignment.
(2) \textit{Generation-side evidence consistency.} Although poisoned documents adapt their arguments to different query regions, they consistently target the same \(e_t\) and \(c_t\). When cascading retrieval brings multiple poisoned documents into context, they provide consistent and mutually reinforcing evidence for \(c_t\), increasing the likelihood that the model adopts \(c_t\) and promotes \(e_t\).

\subsection{Two-Level Candidate Optimization}

Poisoned documents must provide both retrieval and generation utility, while selecting Satellites solely by individual scores may introduce semantic redundancy and leave long-tail query regions undercovered. We therefore adopt a two-level optimization strategy. At the document level, \textit{multi-candidate joint scoring} jointly considers retrieval and generation utility to evaluate and filter both Hub and Satellite candidates, ensuring the quality of individual poisoned documents. After the Hub is selected, set-level \textit{greedy complementary selection} operates only on Satellite candidates and constructs the final set based on their marginal query-coverage gain and association with the selected Hub, thereby improving overall coverage complementarity under a fixed poisoning budget.

\subsubsection{Multi-Candidate Joint Scoring}

Because LLM generation is stochastic, a single generated candidate may exhibit insufficient retrieval utility or weak generation influence. We therefore generate multiple candidates for each Hub or Satellite role and jointly evaluate their retrieval and generation utility. For a candidate Hub \(h\) and Satellite \(p\), we define the joint scores as
\[
\begin{aligned}
R_{\mathrm{hub}}^{\mathrm{joint}}(h)
&=\alpha_h R_{\mathrm{hub}}^{\mathrm{ret}}(h)
+\eta_h R_{\mathrm{gen}}(h),\\
R_{\mathrm{sat}}^{\mathrm{joint}}(p;k,j)
&=\alpha_s R_{\mathrm{sat}}^{\mathrm{ret}}(p;k,j)
+\eta_s R_{\mathrm{gen}}(p).
\end{aligned}
\]

The \(\alpha\) and \(\eta\) weights balance retrieval and generation utility. Candidates for each target position are ranked by their joint scores. The top-ranked Hub candidates are retained to form \(H\), while the top-ranked Satellite candidates form the filtered pool \(\mathcal{P}_{\mathrm{cand}}\) for subsequent set-level complementary selection.

\subsubsection{Greedy Complementary Selection}

Document-level joint scoring evaluates Satellite candidates independently and thus cannot account for coverage redundancy across candidates. Score-only selection may therefore overconcentrate on the same semantic regions and leave long-tail queries undercovered. We address this with set-level greedy complementary selection, which selects Satellites under a fixed poisoning budget by jointly considering marginal query coverage and association with the selected Hubs.

Let \(\mathcal{U}(p)\) denote the surrogate queries covered by Satellite candidate \(p\), and reuse the Hub-association score \(a_H(p)\) defined above. For a candidate set \(\mathcal{S}\subseteq\mathcal{P}_{\mathrm{cand}}\), we define
\[
F(\mathcal{S})
=
\left|\bigcup_{p\in\mathcal{S}}\mathcal{U}(p)\right|
+\rho\sum_{p\in\mathcal{S}}a_H(p).
\]
The first term measures distinct query coverage, while the second rewards association with \(H\), with \(\rho\) balancing the two. Since already covered queries provide no additional coverage gain, redundant candidates are naturally downweighted. We greedily maximize \(F\) subject to \(|\mathcal{S}|=m_{\mathrm{sat}}\).

\section{Evaluation}

\subsection{Research Questions}

Our evaluation of the method under large-scale HNSW retrieval is organized around the following two questions:

\textbf{RQ1:} How effectively does our topic-level poisoning attack generalize to diverse unseen topic queries?

\textbf{RQ2:} Does the Hub--Satellite structure emerge in the HNSW graph and improve graph-search reachability and poisoned-document retrieval?

\subsection{Experimental Setup}

\begin{table*}[t]
\centering
\caption{Attack performance (\%) across poisoned-document budgets and knowledge-base sizes on Wikipedia.}
\label{tab:main-attack-results}

\begingroup
\scriptsize
\setlength{\tabcolsep}{0.5pt}
\renewcommand{\arraystretch}{0.9}

\resizebox{\textwidth}{!}{%
\begin{tabular}{@{}l ccc @{}p{8pt}@{}
ccc @{}p{8pt}@{}
ccc @{}p{8pt}@{}
ccc @{}p{8pt}@{}
ccc @{}p{8pt}@{} ccc@{}}
\toprule
\multirow{3}{*}{Method}
& \multicolumn{11}{c}{\textit{Poisoned-document budget $m=40$ ($m_h=2$, $m_{\mathrm{sat}}=38$)}}
& {}
& \multicolumn{11}{c}{\textit{Poisoned-document budget $m=60$ ($m_h=4$, $m_{\mathrm{sat}}=56$)}} \\
\cmidrule(lr){2-12}
\cmidrule(lr){14-24}

& \multicolumn{3}{c}{500K} & {}
& \multicolumn{3}{c}{1M}   & {}
& \multicolumn{3}{c}{2M}   & {}
& \multicolumn{3}{c}{500K} & {}
& \multicolumn{3}{c}{1M}   & {}
& \multicolumn{3}{c}{2M} \\
\cmidrule(lr){2-4}
\cmidrule(lr){6-8}
\cmidrule(lr){10-12}
\cmidrule(lr){14-16}
\cmidrule(lr){18-20}
\cmidrule(lr){22-24}

& RSR & Cond.ASR & ASR & {}
& RSR & Cond.ASR & ASR & {}
& RSR & Cond.ASR & ASR & {}
& RSR & Cond.ASR & ASR & {}
& RSR & Cond.ASR & ASR & {}
& RSR & Cond.ASR & ASR \\
\midrule

\multicolumn{24}{l}{
\textbf{\textit{Llama 3.1-8B}}
} \\
\midrule

PoisonedRAG
& 5.60 & 9.10 & 0.51 & {}
& 1.20 & 4.40 & 0.05 & {}
& 0.40 & 4.80 & 0.02 & {}
& 5.40 & 8.50 & 0.46 & {}
& 1.20 & 9.40 & 0.11 & {}
& 0.60 & 9.00 & 0.05 \\

Topic-FlipRAG
& 9.20 & 7.90 & 0.73 & {}
& 4.30 & 7.30 & 0.31 & {}
& 1.80 & 4.60 & 0.08 & {}
& 12.20 & 11.10 & 1.35 & {}
& 8.90 & 6.30 & 0.56 & {}
& 6.00 & 3.00 & 0.18 \\

\rowcolor{black!8}
\textbf{Our Method}
& \textbf{88.71} & \textbf{46.53} & \textbf{41.28} & {}
& \textbf{78.11} & \textbf{42.04} & \textbf{32.84} & {}
& \textbf{72.49} & \textbf{40.68} & \textbf{29.49} & {}
& \textbf{89.55} & \textbf{47.69} & \textbf{42.71} & {}
& \textbf{79.44} & \textbf{43.31} & \textbf{34.41} & {}
& \textbf{74.00} & \textbf{42.28} & \textbf{31.29} \\

\midrule

\multicolumn{24}{l}{
\textbf{\textit{Qwen 3-30B}}
} \\
\midrule

PoisonedRAG
& 9.20 & 17.60 & 1.62 & {}
& 4.30 & 11.10 & 0.48 & {}
& 1.90 & 6.50 & 0.12 & {}
& 5.50 & 21.60 & 1.19 & {}
& 2.40 & 14.40 & 0.35 & {}
& 0.90 & 9.30 & 0.08 \\

Topic-FlipRAG
& 17.00 & 16.80 & 2.86 & {}
& 10.40 & 16.70 & 1.74 & {}
& 5.60 & 11.80 & 0.66 & {}
& 12.00 & 30.90 & 3.71 & {}
& 8.90 & 27.50 & 2.45 & {}
& 5.90 & 28.90 & 1.71 \\

\rowcolor{black!8}
\textbf{Our Method}
& \textbf{88.73} & \textbf{40.85} & \textbf{36.25} & {}
& \textbf{78.05} & \textbf{36.65} & \textbf{28.61} & {}
& \textbf{72.35} & \textbf{35.75} & \textbf{25.87} & {}
& \textbf{89.63} & \textbf{42.81} & \textbf{38.37} & {}
& \textbf{79.46} & \textbf{38.23} & \textbf{30.38} & {}
& \textbf{74.05} & \textbf{37.30} & \textbf{27.62} \\

\bottomrule
\end{tabular}%
}
\endgroup
\end{table*}

\textbf{Knowledge bases.}
We evaluate both open-domain and domain-specific knowledge bases. For the open-domain setting, we use Wikipedia~\cite{wikimedia_enwiki_20260701} to construct knowledge bases spanning medicine, finance, science, and everyday topics. Deployed RAG systems also commonly build dedicated knowledge bases for specific domains, where denser topic-relevant benign documents intensify competition for limited retrieval positions. Existing topic-level poisoning attacks such as Topic-FlipRAG~\cite{gong2025topic} and DiscourseFlip~\cite{gong2026discourseflip} have not been systematically evaluated in such settings. We therefore use PubMed titles and abstracts~\cite{nlm_pubmed_2026_baseline} to construct knowledge bases for healthcare and lifestyle management, SEC-EDGAR filings~\cite{sec_edgar} for personal finance, and EPA-HERO texts~\cite{usepa_hero_2026} for household and public safety. For each corpus, we construct knowledge bases containing 500K, 1M, and 2M documents.

\textbf{Retrievers.}
We use BGE-base-en~\cite{xiao2024cpack} as the default retriever, and evaluate cross-retriever transferability using all-mpnet-base-v2~\cite{song2020mpnet} and GTE-base~\cite{li2023gte}. Query--document relevance is measured by inner product. We build HNSW indexes with \(M_{\mathrm{HNSW}}=16\), \(\mathrm{ef}_{\mathrm{construction}}=200\), and \(\mathrm{ef}_{\mathrm{search}}=128\), and retrieve the top-5 as generation context.

\textbf{Generation models.}
We use two advanced open-source large language models, Llama 3.1-8B~\cite{meta_llama_3_1_8b_instruct} and Qwen3-30B~\cite{qwen3_30b_a3b_instruct_2507}, as the generators of the RAG system. 

\textbf{Target topics.} We evaluate eight target topics across four high-risk decision domains, including healthcare, lifestyle management, personal finance, and household and public safety, with two topics per domain. Table~\ref{tab:target-topics-claims} in Appendix~\ref{app:target-topics-claims} lists the target entity and claim for each topic. For each topic, we use a poisoned-document budget of \(m\in\{40,60\}\).

\textbf{Training and test queries.}
For each topic, we use an LLM to generate 1,500 training queries for poisoned-document construction and optimization. We then separately generate 5,000 decision-oriented test queries with diverse semantic content and formulations. To further evaluate the method in real-world settings, we select real user queries related to the target topics from MS MARCO~\cite{bajaj2016msmarco}, Natural Questions~\cite{kwiatkowski2019naturalquestions}, HotpotQA~\cite{yang2018hotpotqa}, and BEIR~\cite{thakur2021beir} for testing.

\textbf{Poisoning-attack metrics.}
We use three metrics to evaluate poisoning effectiveness. \textbf{\textit{Retrieval Success Rate}} (RSR) is the proportion of queries for which at least one poisoned document appears in the top-\(k\) retrieval results. \textbf{\textit{Attack Success Rate}} (ASR) is the proportion of queries for which the final answer adopts the target claim and promotes the target entity. The detailed success criterion is provided in Appendix~\ref{app:default-experimental-settings}. \textbf{\textit{Conditional ASR}} (Cond.ASR) computes the attack success rate only over queries with successful retrieval and measures the generation-side influence of poisoned documents after they enter the context. These metrics satisfy \(\mathrm{ASR}=\mathrm{RSR}\times\mathrm{Cond.ASR}\).

\textbf{Semantic-width metrics.}
To characterize the semantic diversity of test queries, we use three metrics computed in a shared embedding space. \textbf{\textit{Effective Clusters}} captures both the number of semantic clusters represented by the queries and the balance of their distribution. \textbf{\textit{PairMean}} measures the average semantic dispersion of the query set, while \textbf{\textit{PairP99}} captures the extent of semantically distant query pairs and thus reflects coverage of the distribution tail. Higher values generally indicate broader and more diverse semantic coverage.

\textbf{Baselines.}
We compare against two representative RAG poisoning baselines with public implementations: PoisonedRAG~\cite{zou2025poisonedrag}, a query-specific attack, and Topic-FlipRAG~\cite{gong2025topic}, a topic-level attack. Since their original attack objectives differ from our entity--claim promotion objective, we adapt both baselines to the same objective and success criterion while preserving their original document-construction and optimization mechanisms. Appendix~\ref{Preliminary} provides details of the reproduction and adaptation.

See Appendix~\ref{app:default-experimental-settings} for default experimental settings.

\section{Experimental Results Analysis}
\subsection{How Effectively Does Our Topic-Level Poisoning Attack Generalize to Diverse Unseen Topic Queries?}

\begin{table*}[t]
\centering
\caption{Attack performance (\%) on domain-specific datasets and real-world queries using Llama 3.1-8B.}
\label{tab:domain-attack-results}

\begingroup
\scriptsize
\setlength{\tabcolsep}{0.5pt}
\renewcommand{\arraystretch}{0.9}

\resizebox{\textwidth}{!}{%
\begin{tabular}{@{}l ccc @{}p{8pt}@{}
ccc @{}p{8pt}@{}
ccc @{}p{8pt}@{}
ccc @{}p{8pt}@{}
ccc @{}p{8pt}@{} ccc@{}}
\toprule
\multirow{3}{*}{Dataset}
& \multicolumn{11}{c}{\textit{Poisoned-document budget $m=40$ ($m_h=2$, $m_{\mathrm{sat}}=38$)}}
& {}
& \multicolumn{11}{c}{\textit{Poisoned-document budget $m=60$ ($m_h=4$, $m_{\mathrm{sat}}=56$)}} \\
\cmidrule(lr){2-12}
\cmidrule(lr){14-24}

& \multicolumn{3}{c}{500K} & {}
& \multicolumn{3}{c}{1M}   & {}
& \multicolumn{3}{c}{2M}   & {}
& \multicolumn{3}{c}{500K} & {}
& \multicolumn{3}{c}{1M}   & {}
& \multicolumn{3}{c}{2M} \\
\cmidrule(lr){2-4}
\cmidrule(lr){6-8}
\cmidrule(lr){10-12}
\cmidrule(lr){14-16}
\cmidrule(lr){18-20}
\cmidrule(lr){22-24}

& RSR & Cond.ASR & ASR & {}
& RSR & Cond.ASR & ASR & {}
& RSR & Cond.ASR & ASR & {}
& RSR & Cond.ASR & ASR & {}
& RSR & Cond.ASR & ASR & {}
& RSR & Cond.ASR & ASR \\
\midrule

\multicolumn{24}{l}{
\textbf{\textit{Domain-specific datasets}}
} \\
\midrule

PubMed
& 71.41 & 40.14 & 28.66 & {}
& 58.34 & 41.26 & 24.07 & {}
& 47.96 & 44.27 & 21.23 & {}
& 72.60 & 41.63 & 30.22 & {}
& 60.25 & 41.79 & 25.18 & {}
& 49.71 & 45.27 & 22.50 \\

SEC-EDGAR
& 59.08 & 32.13 & 18.98 & {}
& 57.10 & 32.34 & 18.47 & {}
& 49.30 & 31.80 & 15.68 & {}
& 62.19 & 33.99 & 21.14 & {}
& 58.29 & 33.11 & 19.30 & {}
& 53.53 & 33.56 & 17.96 \\

EPA-HERO
& 98.61 & 46.02 & 45.38 & {}
& 97.30 & 44.32 & 43.12 & {}
& 93.12 & 43.78 & 40.77 & {}
& 98.68 & 46.40 & 45.79 & {}
& 97.44 & 45.51 & 44.34 & {}
& 93.80 & 44.72 & 41.95 \\

\rowcolor{black!8}
\textbf{Average}
& \textbf{76.37} & \textbf{39.43} & \textbf{30.11} & {}
& \textbf{70.91} & \textbf{39.30} & \textbf{27.87} & {}
& \textbf{63.46} & \textbf{39.95} & \textbf{25.35} & {}
& \textbf{77.82} & \textbf{40.67} & \textbf{31.65} & {}
& \textbf{71.99} & \textbf{40.14} & \textbf{28.90} & {}
& \textbf{65.68} & \textbf{41.19} & \textbf{27.05} \\

\midrule
\multicolumn{24}{l}{
\textbf{\textit{Real-world queries}}
} \\
\midrule

Wikipedia
& 81.26 & 40.39 & 32.82 & {}
& 67.81 & 36.95 & 25.06 & {}
& 59.23 & 40.65 & 24.08 & {}
& 83.05 & 43.89 & 36.45 & {}
& 69.36 & 45.20 & 31.35 & {}
& 61.45 & 42.38 & 26.04 \\

\bottomrule
\end{tabular}%
}
\endgroup
\end{table*}

\textbf{Overall performance.} As shown in Table~\ref{tab:main-attack-results}, our method substantially outperforms both baselines across generation models, poisoning budgets, and knowledge-base sizes. Existing attacks mainly optimize query--document semantic alignment, without explicitly accounting for benign-document competition or path-dependent traversal in large-scale HNSW retrieval. As a result, high semantic similarity alone does not guarantee that poisoned documents are reached and returned by HNSW. As the knowledge base grows from 500K to 2M documents, the baseline RSR drops from as high as 17.0\% at 500K to only 0.4\%--6.0\% at 2M.

In contrast, our Hub--Satellite design combines broad query coverage, local ranking competitiveness, and improved graph-search reachability. At 2M documents, our method maintains an RSR above 72\% across both generation models and poisoning budgets. Cond.ASR and overall ASR also remain above 35.75\% and 25.87\%, respectively. These results demonstrate that our method preserves strong attack effectiveness as the HNSW knowledge base scales.

\begin{table}[t]
\centering
\caption{Per-topic semantic breadth of test queries.}
\label{tab:query-semantic-breadth}
\begingroup
\scriptsize
\setlength{\tabcolsep}{2.5pt}
\renewcommand{\arraystretch}{0.9}
\resizebox{\columnwidth}{!}{%
\begin{tabular}{@{}lcccc@{}}
\toprule
Query set
& Avg.\ \(n\)
& Effective Clusters
& PairMean
& PairP99 \\
\midrule
Synthetic test queries
& 5,000
& 11.089
& 0.372
& 0.537 \\
Real test queries
& 77.25
& 10.261
& 0.367
& 0.521 \\
\bottomrule
\end{tabular}%
}
\endgroup
\end{table}

\begin{figure}[t]
\centering
\noindent\includegraphics[width=0.69\columnwidth]{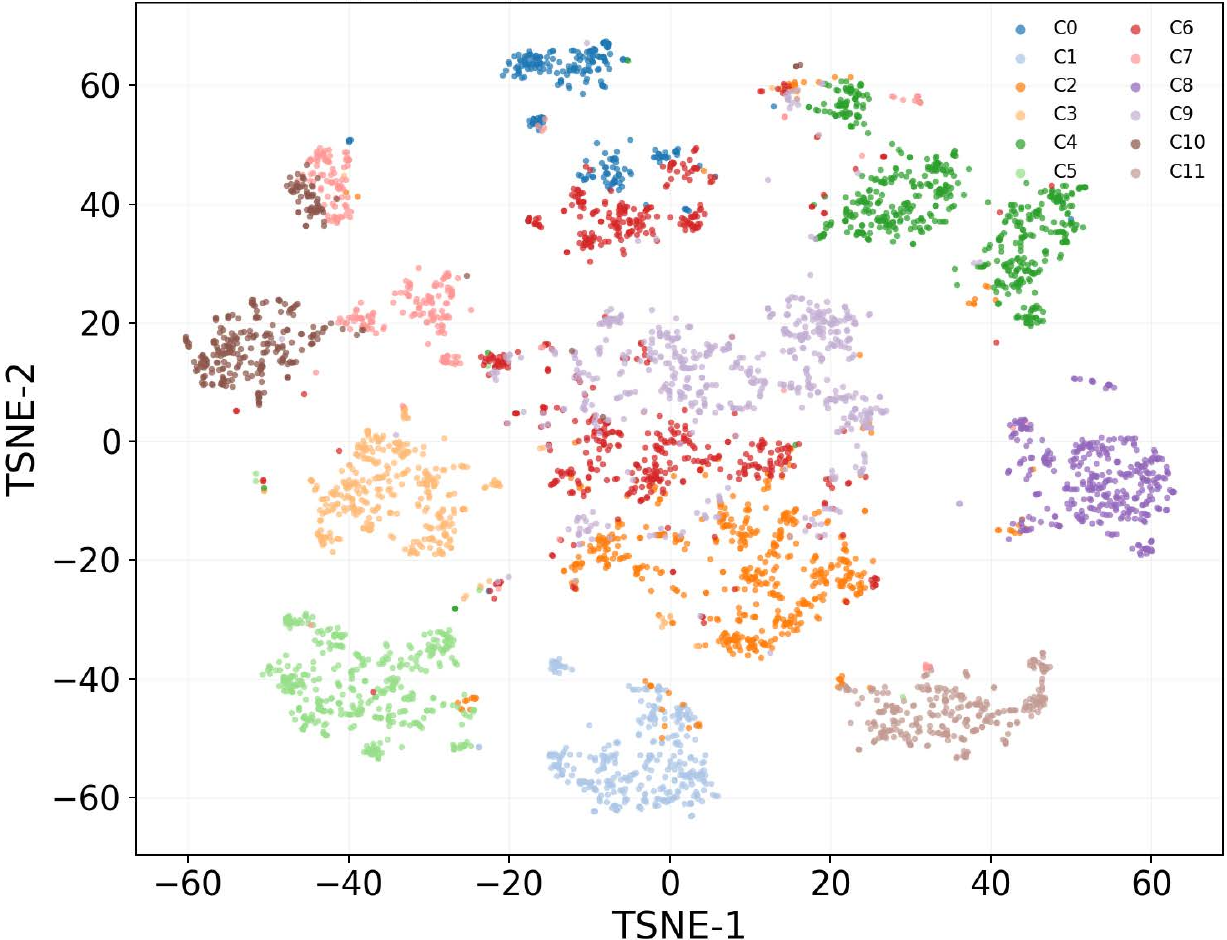}
\caption{t-SNE visualization of 5,000 synthetic test queries.}
\label{fig:query-scatter}
\end{figure}

\begin{figure*}[t]
\centering
\subfloat[subgraph connectivity.]{\begin{minipage}[t]{0.205\textwidth}
  \centering
  \includegraphics[height=1.42in,keepaspectratio]{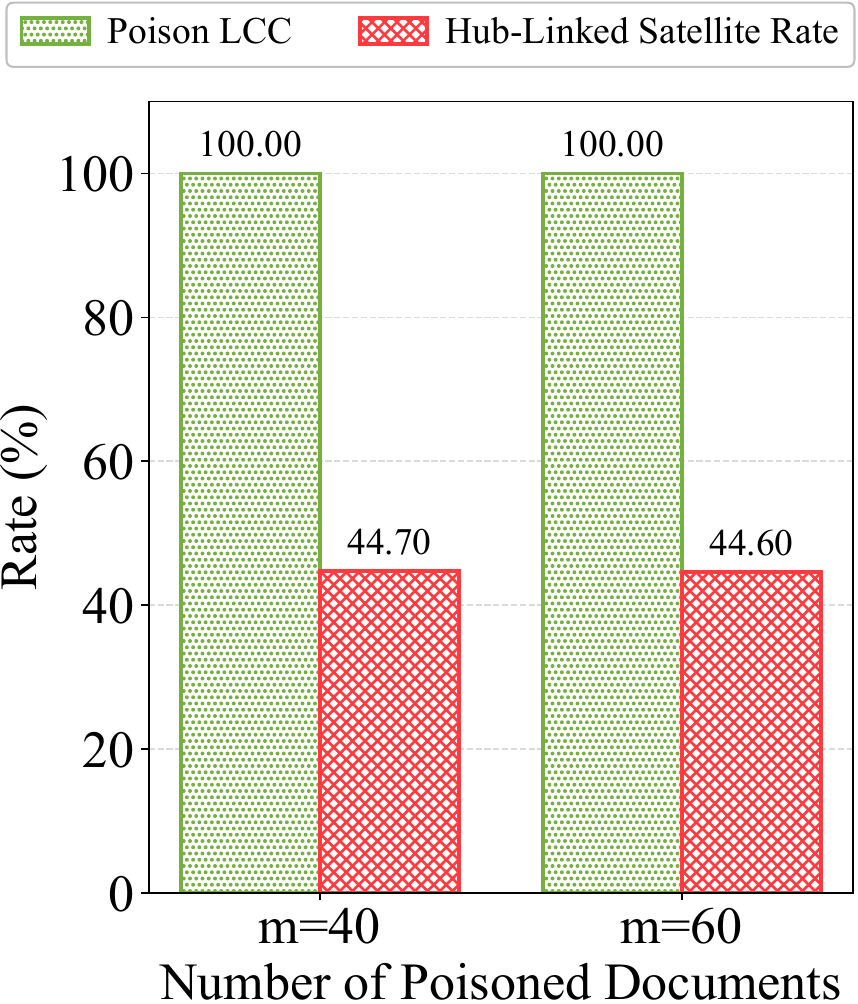}
  
  \label{fig:topology-lcc}
\end{minipage}}%
\hspace{0.0015\textwidth}%
\subfloat[Hub mediation rate.]{\begin{minipage}[t]{0.205\textwidth}
  \centering
  \includegraphics[height=1.42in,keepaspectratio]{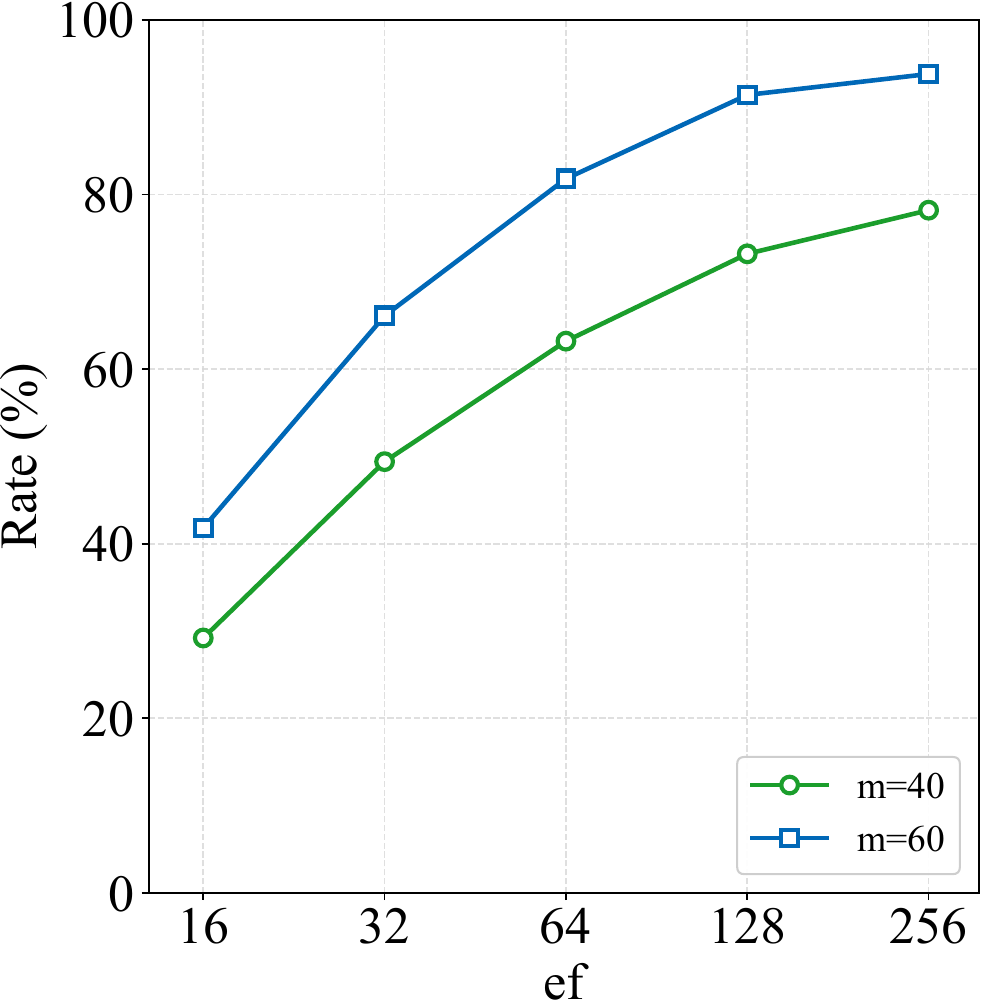}
  
  \label{fig:topology-mediation}
\end{minipage}}%
\hspace{0.0015\textwidth}%
\subfloat[Cascade visits.]{\begin{minipage}[t]{0.205\textwidth}
  \centering
  \includegraphics[height=1.42in,keepaspectratio]{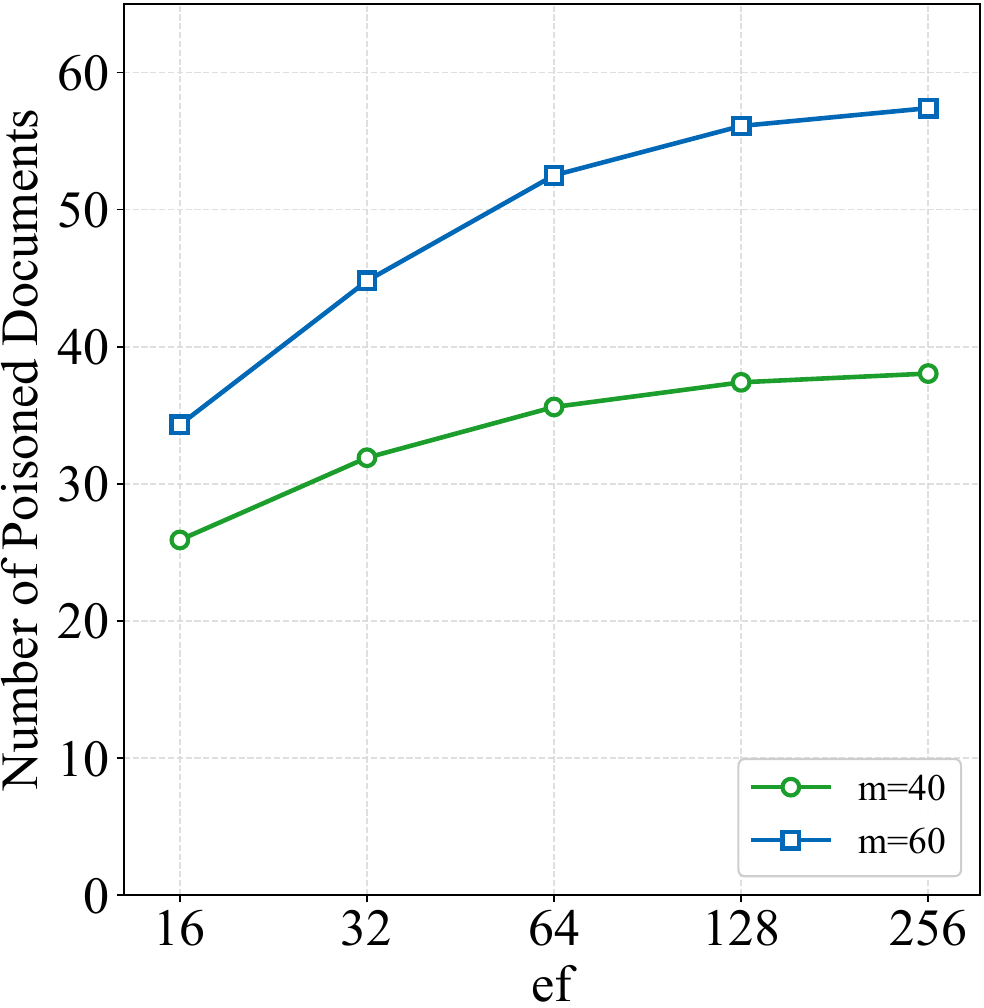}
  
  \label{fig:topology-cascade}
\end{minipage}}%
\hspace{0.0015\textwidth}%
\subfloat[With Hub vs.\ without Hub.]{\begin{minipage}[t]{0.37\textwidth}
  \centering
  \includegraphics[height=1.42in,keepaspectratio]{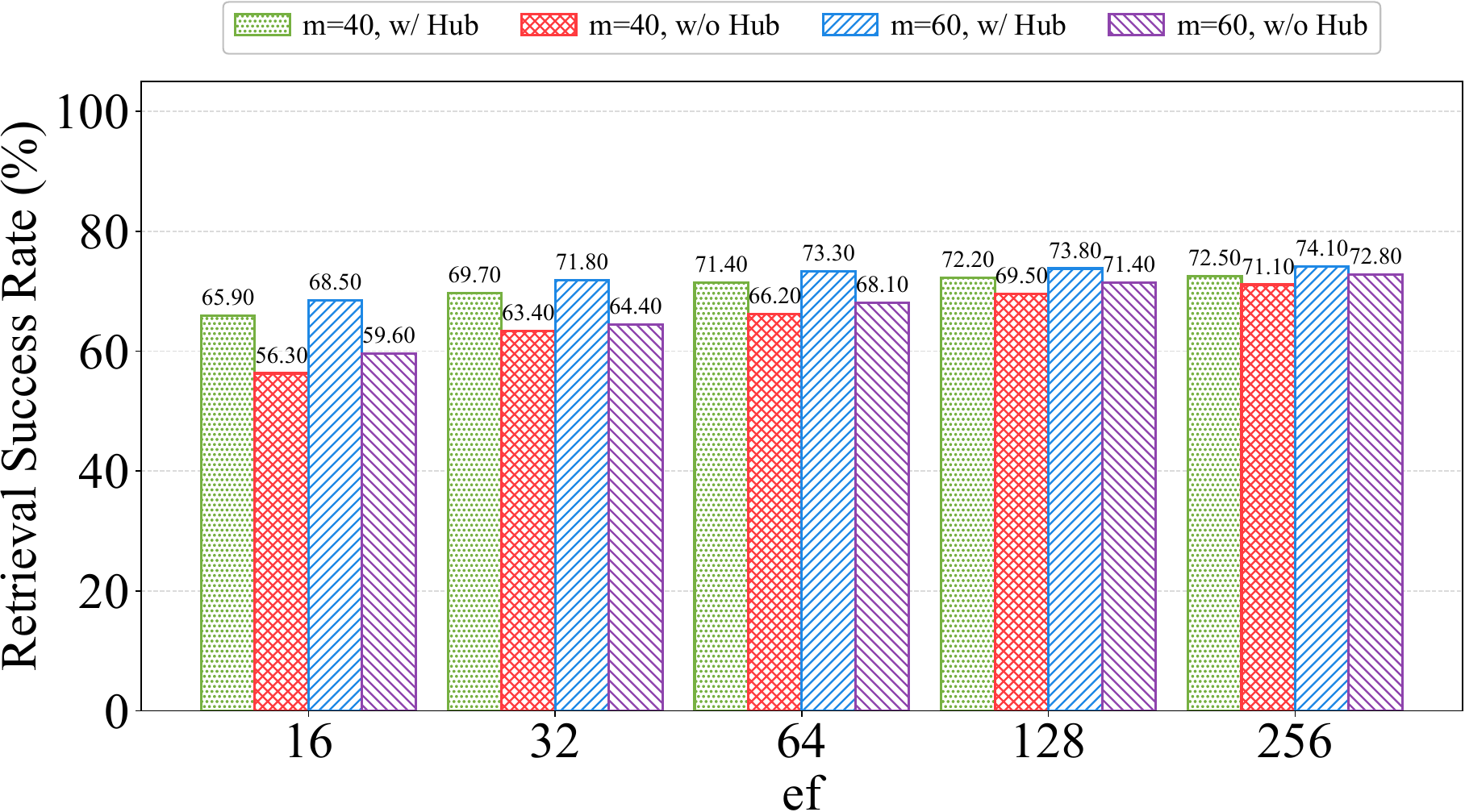}
  
  \label{fig:topology-isolated}
\end{minipage}}
\caption{Validation of Hub--Satellite topological cooperation in a 2M-document HNSW index.}
\label{fig:topology-cooperation}
\end{figure*}

\begin{figure}[t]
\centering
\makebox[\linewidth][c]{%
\noindent\includegraphics[width=0.505\linewidth]{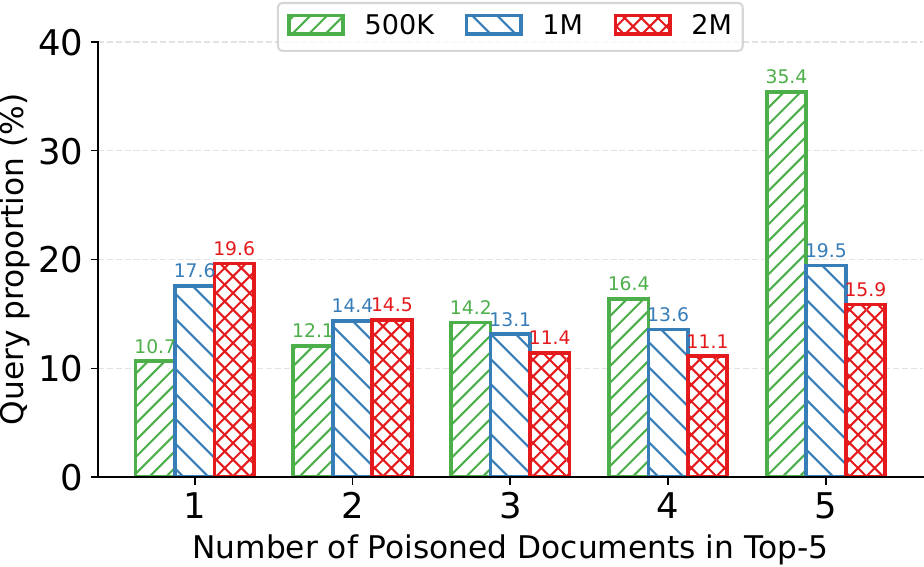}%
\hspace{-0.01\linewidth}%
\noindent\includegraphics[width=0.505\linewidth]{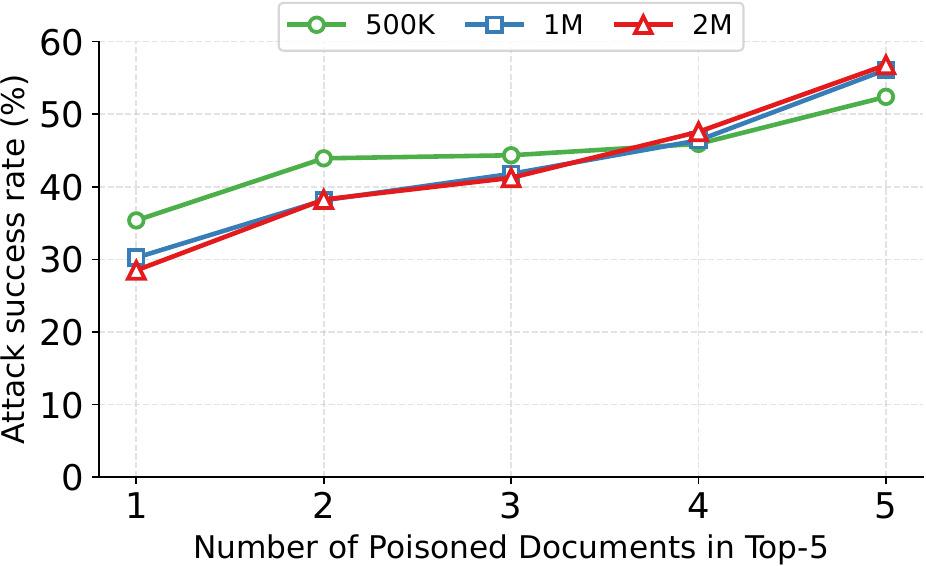}%
}
\caption{Retrieved poisoned-document count with \(m=40\): distribution across queries (left) and attack success (right).}
\label{fig:poisoned-count-analysis}
\end{figure}

\textbf{Domain-specific knowledge bases.}
Our method remains effective in domain-specific knowledge bases with denser topic-relevant benign documents. We construct knowledge bases from PubMed, SEC-EDGAR, and EPA-HERO for healthcare, personal finance, and household and public safety, respectively. As shown in Table~\ref{tab:domain-attack-results}, when the knowledge base grows from 500K to 2M, the average RSR decreases by 12.14--12.91 percentage points, but still reaches 63.46\%--65.68\% at 2M across the two poisoned-document budgets. At 2M, Cond.ASR remains between 39.95\% and 41.19\%, resulting in an overall ASR of 25.35\%--27.05\%. These results indicate that, despite stronger competition from topic-relevant benign documents, our method maintains considerable graph-search reachability and generation-side effectiveness in domain-specific settings.

\textbf{Real user queries.}
The preceding test queries are generated by an LLM and cover diverse semantic directions and formulations, but their distribution may still differ from that of real user queries. We therefore further evaluate our method on topic-related user queries extracted from real-world question-answering datasets. As shown in Table~\ref{tab:domain-attack-results}, even when the knowledge base grows to 2M, our method achieves RSRs of 59.23\% and 61.45\% and ASRs of 24.08\% and 26.04\% under the two poisoned-document budgets, respectively. These results show that our method not only covers diverse model-generated queries but also generalizes effectively to more naturally formulated real user queries.

\textbf{Semantic breadth.}
To further assess whether our generalization results rely on a narrow set of query formulations or intents, we analyze the semantic breadth of the test queries. Figure~\ref{fig:query-scatter} shows that the synthetic test queries form multiple internally compact and mutually separated semantic regions, indicating coverage of diverse user intents and formulations. Because t-SNE may distort global distances, Table~\ref{tab:query-semantic-breadth} further quantifies query breadth in the shared embedding space. The synthetic queries achieve Effective Clusters, PairMean, and PairP99 values of 11.089, 0.372, and 0.537, respectively, closely matching the 10.261, 0.367, and 0.521 obtained by real-world queries. These results indicate that the synthetic queries  span multiple effective semantic clusters, exhibit broad overall semantic dispersion and long-tail coverage, and achieve semantic breadth comparable to real-world queries across all three metrics. Combined with the preceding attack results, this analysis further shows that our method generalizes across diverse unseen query formulations rather than being limited to a small set of repeated query templates.

\textbf{Number of retrieved poisoned documents.}
At \(m=40\), we further analyze the number of poisoned documents appearing in the top-\(5\) results for test queries and the corresponding ASR. As shown in Figure~\ref{fig:poisoned-count-analysis}, a substantial proportion of queries retrieve multiple poisoned documents across knowledge-base sizes. In the 500K knowledge base, poisoned documents occupy all top-\(5\) positions for 35.4\% of queries. ASR also increases steadily with the number of retrieved poisoned documents. These results are consistent with the cascading retrieval effect of the Hub--Satellite topology: once one poisoned document is reached, additional poisoned documents are more likely to enter the retrieved context. Moreover, their shared entity--claim objective provides consistent evidence, increasing the likelihood that the model adopts the target claim and promotes the target entity.

\subsection{Does the Hub--Satellite Structure Emerge in the HNSW Graph and Improve Graph-Search Reachability and Poisoned-Document Retrieval?}
\label{sec:hub-satellite-analysis}

To examine whether the semantic coordination of Hub--Satellite documents induces a cooperative topology in HNSW and improves graph-search behavior, we analyze the poisoned subgraph and retrieval traces with 2M documents.

\textbf{Evaluation metrics.}
\textbf{\textit{Poison LCC}} is the fraction of poisoned documents contained in the largest connected component, indicating whether the poisoned documents form a connected topological structure. 
\textbf{\textit{Hub-Linked Satellite Rate}} is the fraction of Satellites directly connected to at least one Hub, reflecting the structural association between Hubs and local semantic entry points. 
\textbf{\textit{Hub Mediation Rate}} is the fraction of queries for which at least one Satellite is first discovered through expansion from a Hub, capturing the Hub's dynamic role in facilitating Satellite discovery during HNSW search. 
\textbf{\textit{Cascade Visits}} counts the additional poisoned documents visited after the search first reaches a poisoned document, capturing the cascading access among poisoned documents during graph traversal.

\textbf{Formation of the poisoned subgraph.}
We first examine whether the poisoned documents form a connected structure in the large-scale HNSW index using Poison LCC. As shown in Figure~\ref{fig:topology-lcc}, Poison LCC reaches 100\% for both \(m=40\) and \(m=60\), indicating that all poisoned documents belong to the same connected component rather than being scattered as isolated nodes. This result confirms that the Hub--Satellite document set forms a connected poisoned-document topology, providing a structural basis for subsequent graph-search cooperation and cascading retrieval.

\textbf{Hub connectivity and mediation.}
We evaluate the role of the Hub from both static adjacency and dynamic search behavior. As shown in Figure~\ref{fig:topology-lcc}, 44.7\% and 44.6\% of the Satellites are directly connected to at least one Hub for \(m=40\) and \(m=60\), respectively, showing that Hubs are structurally linked to retrieval entry points across different local semantic regions. Figure~\ref{fig:topology-mediation} further shows that, as \(\mathrm{ef}_{\mathrm{search}}\) increases, the Hub Mediation Rate rises from 29.2\% to 78.2\% for \(m=40\) and from 41.8\% to 93.8\% for \(m=60\). Together, these results show that Hubs act as shared structural anchors within the poisoned-document neighborhood, linking Satellites from different semantic regions and facilitating their discovery during dynamic HNSW search.

\textbf{Topological cooperation.}
To evaluate whether poisoned documents cooperate during graph search, we use Cascade Visits to measure the number of additional poisoned documents visited after HNSW first reaches a poisoned document. As shown in Figure~\ref{fig:topology-cascade}, at \(\mathrm{ef}_{\mathrm{search}}=16\), the search visits an average of 25.9 and 34.3 additional poisoned documents for \(m=40\) and \(m=60\), respectively. At \(\mathrm{ef}_{\mathrm{search}}=128\), these values increase to 37.4 and 56.1. These results indicate that, once HNSW reaches a poisoned document, graph traversal can continue to many additional poisoned documents, demonstrating cascading multi-document access and set-level cooperation during HNSW search.

\begin{figure}[t]
\centering
\captionsetup[subfloat]{farskip=0pt}
\begin{minipage}{0.92\columnwidth}
\centering
\subfloat[Exact retrieval]{\begin{minipage}[t]{0.39\linewidth}
  \centering
  \includegraphics[width=\linewidth]{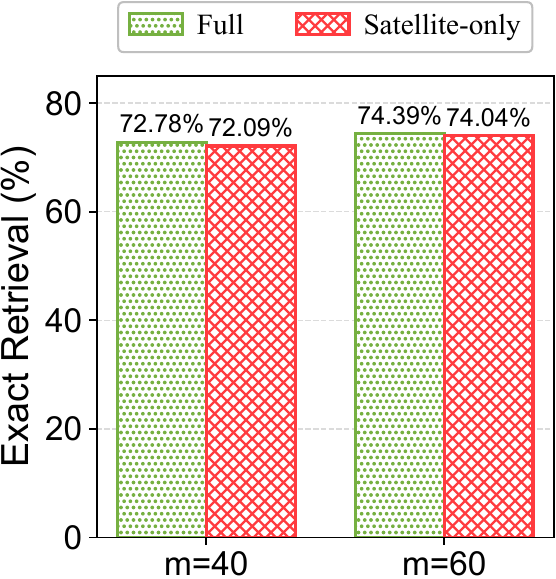}
  
  \label{fig:exact-retrieval}
\end{minipage}}%
\hfill
\subfloat[HNSW-specific gain]{\begin{minipage}[t]{0.59\linewidth}
  \centering
  \includegraphics[width=\linewidth]{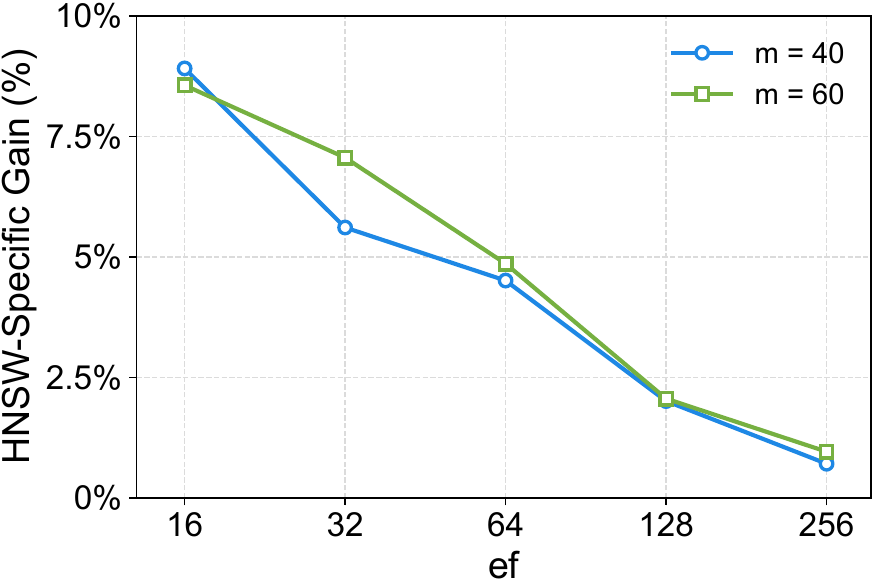}
  
  \label{fig:hnsw-specificity}
\end{minipage}}
\end{minipage}
\caption{Retrieval benefits of Hub--Satellite topology.}
\label{fig:retrieval-specificity}
\end{figure}

\textbf{Retrieval contributions of the Hub and Satellites.}
To examine the roles of Hubs and Satellites in retrieval, we remove the Hubs, set the Hub-association term in the Satellite reward to zero, and regenerate a Satellite-only set under the same poisoning budget. As shown in Figure~\ref{fig:topology-isolated}, at \(\mathrm{ef}_{\mathrm{search}}=16\), the Satellite-only set achieves RSRs of 56.30\% and 59.60\% for \(m=40\) and \(m=60\), respectively, demonstrating that distributed Satellites already provide substantial query coverage and multiple retrieval entry points. The full Hub--Satellite construction increases the RSRs to 65.90\% and 68.50\%, gains of 9.60 and 8.90 percentage points. By aggregating shared semantics across query regions, Hubs promote local cohesion among these distributed entry points, facilitating further poisoned-document discovery during HNSW graph traversal. These larger gains under tighter search budgets highlight their complementary roles: Satellites expand access to the poisoned set, while Hubs reinforce structural associations among the entry points, enabling set-level cooperation.

\textbf{HNSW retrieval gains.}
To separate semantic-matching gains from HNSW graph-search gains, we compare Full and Satellite-only under exact and HNSW retrieval. As shown in Figure~\ref{fig:exact-retrieval}, their RSRs are similar under exact retrieval: Full improves RSR by only 0.69 and 0.35 percentage points for \(m=40\) and \(m=60\), respectively. By contrast, as shown in Figure~\ref{fig:hnsw-specificity}, after subtracting these exact-retrieval gains, the HNSW-specific gains reach 8.91 and 8.56 percentage points at \(\mathrm{ef}_{\mathrm{search}}=16\), respectively. As \(\mathrm{ef}_{\mathrm{search}}\) increases to 256, these gains decline to 0.71 and 0.96 percentage points. These results show that the advantage of the Hub--Satellite structure cannot be explained solely by semantic matching and is more pronounced under budget-constrained HNSW search, supporting the interpretation that Hubs improve poisoned-document retrieval by enhancing graph-search reachability.

\begin{figure}[t]
\centering
\noindent\includegraphics[width=0.4554\columnwidth]{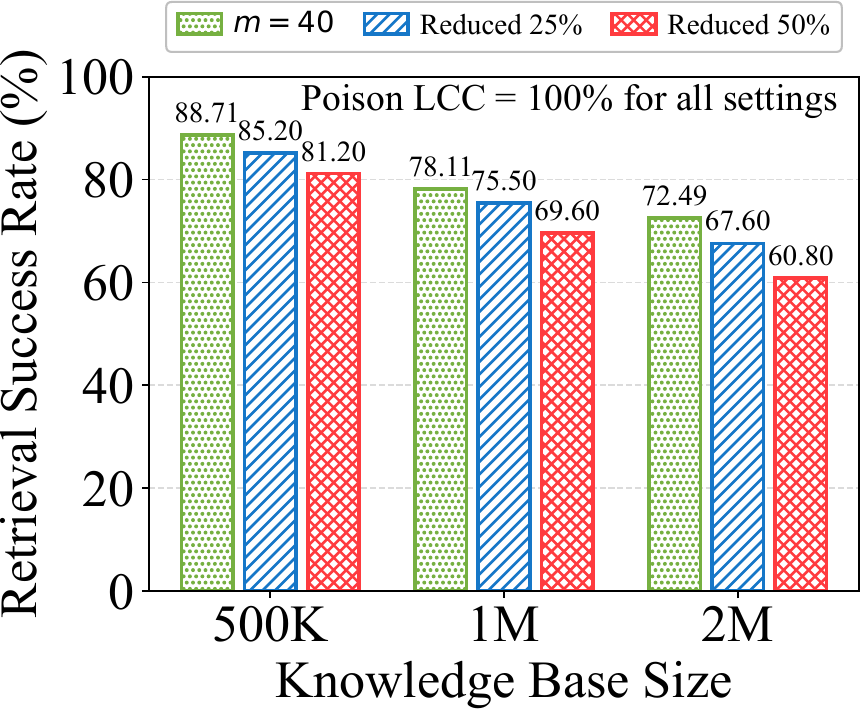}
\noindent\includegraphics[width=0.4554\columnwidth]{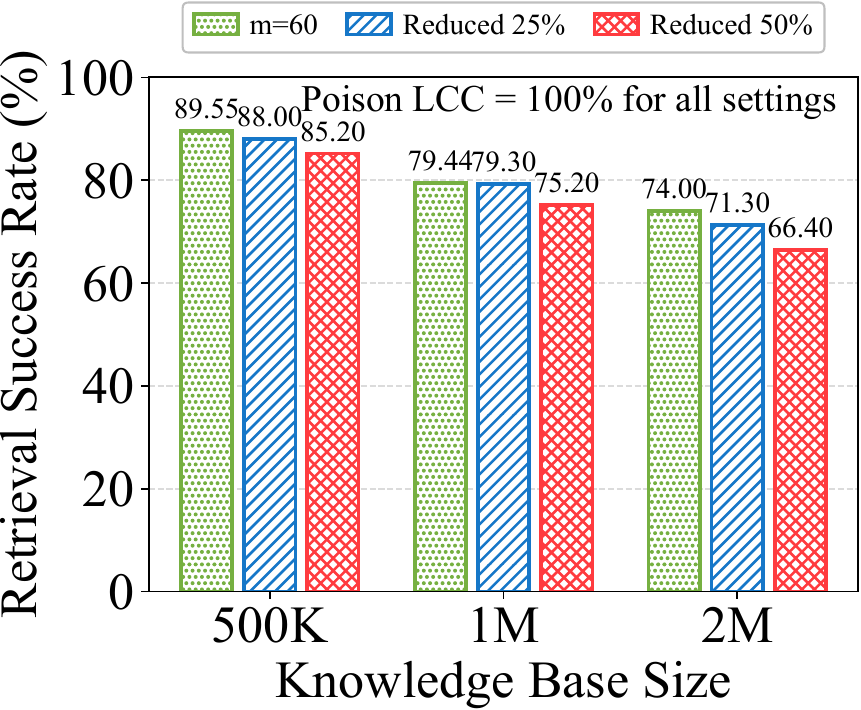}
\caption{Connectivity and retrieval for \(m=40,60\) after randomly removing 25\%/50\% of poisoned documents.}
\label{fig:reduction-robustness}
\end{figure}

\textbf{Robustness to node loss.}
To assess whether topological cooperation relies on the complete poisoned-document set, we randomly remove 25\% or 50\% of the documents from the \(m=40\) and \(m=60\) sets. As shown in Figure~\ref{fig:reduction-robustness}, Poison LCC remains at 100\% across all knowledge-base sizes and removal ratios, indicating that the remaining poisoned documents stay within a single connected component even after half of the nodes are removed. Retrieval coverage also remains substantial: after removing 50\% of the \(m=40\) set, the remaining 20 documents still achieve a 60.8\% RSR in the 2M-document knowledge base, with a smaller degradation under \(m=60\). These results show that the Hub--Satellite topology remains connected and preserves substantial retrieval coverage under random partial node removal, demonstrating robustness to incomplete poisoned-document sets.

\subsection{Ablation Study}

\begin{table*}[t]
\centering
\caption{Attack performance (\%) in ablation analysis under different configurations.}
\label{tab:ablation-analysis}

\begingroup
\scriptsize
\setlength{\tabcolsep}{0.5pt}
\renewcommand{\arraystretch}{0.9}

\resizebox{\textwidth}{!}{%
\begin{tabular}{@{}l ccc @{\hspace{4pt}}c@{\hspace{4pt}}
ccc @{\hspace{4pt}}c@{\hspace{4pt}}
ccc @{\hspace{4pt}}c@{\hspace{4pt}}
ccc @{\hspace{4pt}}c@{\hspace{4pt}}
ccc @{\hspace{4pt}}c@{\hspace{4pt}} ccc@{}}
\toprule
\multirow{3}{*}{Configuration}
& \multicolumn{11}{c}{\textit{Poisoned-document budget $m=40$ ($m_h=2$, $m_{\mathrm{sat}}=38$)}}
& {}
& \multicolumn{11}{c}{\textit{Poisoned-document budget $m=60$ ($m_h=4$, $m_{\mathrm{sat}}=56$)}} \\
\cmidrule(lr){2-12}
\cmidrule(lr){14-24}

& \multicolumn{3}{c}{500K} & {}
& \multicolumn{3}{c}{1M}   & {}
& \multicolumn{3}{c}{2M}   & {}
& \multicolumn{3}{c}{500K} & {}
& \multicolumn{3}{c}{1M}   & {}
& \multicolumn{3}{c}{2M} \\
\cmidrule(lr){2-4}
\cmidrule(lr){6-8}
\cmidrule(lr){10-12}
\cmidrule(lr){14-16}
\cmidrule(lr){18-20}
\cmidrule(lr){22-24}

& RSR & Cond.ASR & ASR & {}
& RSR & Cond.ASR & ASR & {}
& RSR & Cond.ASR & ASR & {}
& RSR & Cond.ASR & ASR & {}
& RSR & Cond.ASR & ASR & {}
& RSR & Cond.ASR & ASR \\
\midrule

\multicolumn{24}{l}{
\textbf{\textit{Retriever}}
} \\
\midrule

BGE-base-en
& 88.71 & 46.53 & 41.28 & {}
& 78.11 & 42.04 & 32.84 & {}
& 72.49 & 40.68 & 29.49 & {}
& 89.55 & 47.69 & 42.71 & {}
& 79.44 & 43.31 & 34.41 & {}
& 74.00 & 42.28 & 31.29 \\

all-mpnet-base-v2
& 82.26 & 44.86 & 36.90 & {}
& 66.61 & 39.59 & 26.37 & {}
& 59.64 & 38.23 & 22.80 & {}
& 84.10 & 46.88 & 39.43 & {}
& 69.08 & 41.39 & 28.59 & {}
& 62.16 & 39.39 & 24.48 \\

GTE-base
& 84.36 & 44.93 & 37.90 & {}
& 71.63 & 40.75 & 29.19 & {}
& 66.30 & 39.81 & 26.39 & {}
& 85.88 & 47.50 & 40.79 & {}
& 74.20 & 42.46 & 31.51 & {}
& 69.18 & 41.61 & 28.79 \\

\midrule
\multicolumn{24}{l}{
\textbf{\textit{Top-\(k\)}}
} \\
\midrule

Top-5
& 88.71 & 46.53 & 41.28 & {}
& 78.11 & 42.04 & 32.84 & {}
& 72.49 & 40.68 & 29.49 & {}
& 89.55 & 47.69 & 42.71 & {}
& 79.44 & 43.31 & 34.41 & {}
& 74.00 & 42.28 & 31.29 \\

Top-3
& 84.15 & 44.96 & 37.83 & {}
& 70.54 & 42.98 & 30.32 & {}
& 64.18 & 42.70 & 27.40 & {}
& 85.21 & 46.60 & 39.71 & {}
& 72.10 & 43.95 & 31.69 & {}
& 65.83 & 43.63 & 28.72 \\

\midrule
\multicolumn{24}{l}{
\textbf{\textit{Reward Function}}
} \\
\midrule

With reward
& 88.71 & 46.53 & 41.28 & {}
& 78.11 & 42.04 & 32.84 & {}
& 72.49 & 40.68 & 29.49 & {}
& 89.55 & 47.69 & 42.71 & {}
& 79.44 & 43.31 & 34.41 & {}
& 74.00 & 42.28 & 31.29 \\

Without reward
& 81.45 & 52.12 & 42.45 & {}
& 65.35 & 47.80 & 31.24 & {}
& 58.79 & 46.72 & 27.47 & {}
& 81.88 & 48.55 & 39.75 & {}
& 66.95 & 44.34 & 29.69 & {}
& 60.73 & 42.99 & 26.11 \\

\midrule
\multicolumn{24}{l}{
\textbf{\textit{Entity--Claim Coupling}}
} \\
\midrule

With coupling
& 88.71 & 46.53 & 41.28 & {}
& 78.11 & 42.04 & 32.84 & {}
& 72.49 & 40.68 & 29.49 & {}
& 89.55 & 47.69 & 42.71 & {}
& 79.44 & 43.31 & 34.41 & {}
& 74.00 & 42.28 & 31.29 \\

Without coupling
& 95.49 & 37.08 & 35.41 & {}
& 90.58 & 29.98 & 27.16 & {}
& 88.15 & 23.39 & 20.62 & {}
& 95.93 & 37.28 & 35.76 & {}
& 91.56 & 31.28 & 28.64 & {}
& 88.89 & 26.08 & 23.18 \\

\midrule
\multicolumn{24}{l}{
\textbf{\textit{Greedy Selection}}
} \\
\midrule

Global Greedy
& 84.50 & 41.60 & 35.15 & {}
& 74.50 & 38.70 & 28.83 & {}
& 70.90 & 38.00 & 26.94 & {}
& 85.00 & 42.30 & 35.96 & {}
& 76.00 & 40.20 & 30.55 & {}
& 72.80 & 40.50 & 29.48 \\

Score Top
& 82.50 & 38.00 & 31.35 & {}
& 71.30 & 33.90 & 24.17 & {}
& 67.60 & 34.10 & 23.05 & {}
& 84.90 & 40.60 & 34.47 & {}
& 75.30 & 37.70 & 28.39 & {}
& 71.70 & 37.00 & 26.53 \\

\bottomrule
\end{tabular}%
}
\endgroup
\end{table*}

\textbf{Impact of the retriever.} To evaluate transferability under surrogate--target retriever mismatch, we use all-mpnet-base-v2 and GTE-base as the target retrievers. As shown in Table~\ref{tab:ablation-analysis}, on the 2M knowledge base with \(m=40\), our method achieves RSR of 59.64\% and 66.30\% and ASR of 22.80\% and 26.39\%, respectively. With \(m=60\), the RSR increase to 62.16\% and 69.18\%, while the ASR reach 24.48\% and 28.79\%. Cond.ASR remains above 38\% across all settings. These results demonstrate effective transfer across mismatched retrievers.

\textbf{Impact of top-\(k\).}
To evaluate how the number of retrieved documents passed to the generator affects attack performance, we reduce top-\(k\) from 5 to 3. As shown in Table~\ref{tab:ablation-analysis}, on the 2M knowledge base, the RSR values for \(m=40\) and \(m=60\) are 64.18\% and 65.83\%, while the ASR values reach 27.40\% and 28.72\%, only 2.09 and 2.57 percentage points lower than under top-5, respectively. Cond.ASR remains comparable to, and is even slightly higher than, that under top-5. Thus, once a poisoned document enters the retrieval results, reducing top-\(k\) does not weaken its generation-side influence.

\textbf{Impact of the reward function.}
To evaluate the effectiveness of document-level joint scoring, we remove the joint reward while keeping all other settings unchanged. As shown in Table~\ref{tab:ablation-analysis}, removing the reward reduces RSR by 7.26--13.70 percentage points across poisoning budgets and knowledge-base sizes, with larger drops as the knowledge base grows. Although Cond.ASR increases in some settings, the reduced retrieval coverage lowers overall ASR in five of the six configurations. These results show that the joint reward balances retrieval and generation utility, improving the end-to-end effectiveness of selected poisoned documents.

\begin{figure}[t]
\centering
\noindent\includegraphics[width=\columnwidth]{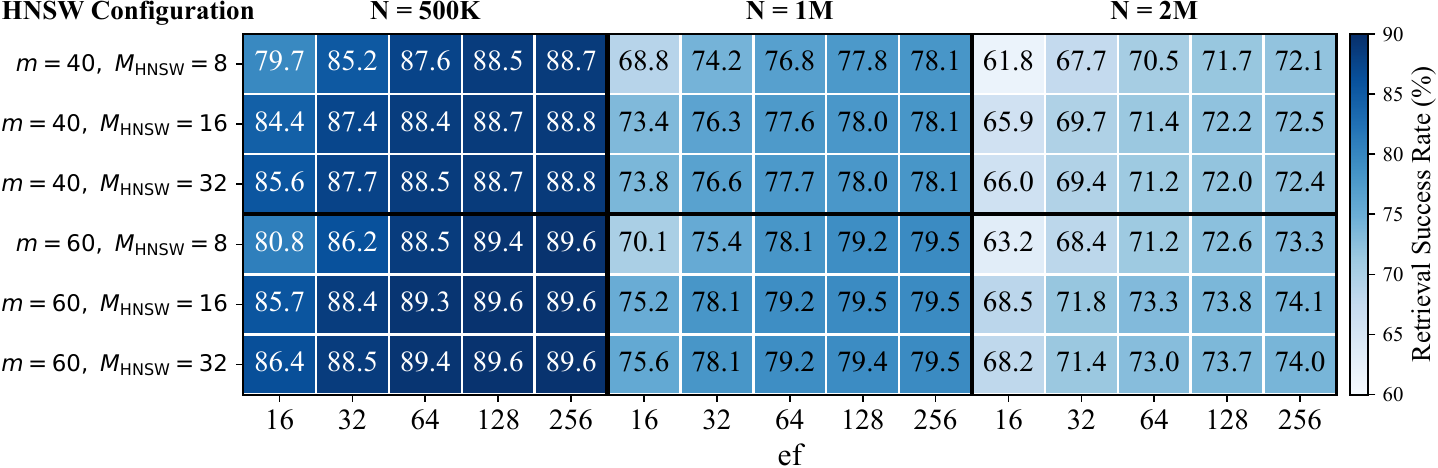}
\caption{Retrieval success rate across HNSW configurations.}
\label{fig:hnsw-configuration}
\end{figure}

\textbf{Impact of entity--claim coupling.}
To evaluate entity--claim coupling, we remove the constraint that the target entity must serve as the grammatical subject or causal agent in factual statements while keeping all other settings unchanged. As shown in Table~\ref{tab:ablation-analysis}, removing the coupling increases average RSR by 11.38 percentage points, but decreases Cond.ASR and ASR by 12.90 and 6.88 percentage points, respectively, with both metrics declining across all six settings. These results show that entity--claim coupling trades some retrieval coverage for substantially stronger generation-side effectiveness, making the model more likely to retain the target entity and promote it through the target claim once poisoned documents enter the context.

\textbf{Impact of greedy selection.}
To evaluate greedy complementary selection, we compare Global Greedy with directly selecting the highest-scoring candidates (Score Top). As shown in Table~\ref{tab:ablation-analysis}, on the 2M knowledge base with \(m=40\), Global Greedy improves RSR from 67.60\% to 70.90\% and ASR from 23.05\% to 26.94\%. With \(m=60\), ASR increases from 26.53\% to 29.48\%. These results show that greedy complementary selection reduces coverage redundancy and improves set-level coverage complementarity, enhancing attack performance.

\begin{figure}[t]
\centering
\captionsetup[subfloat]{farskip=0pt}
\subfloat[$k$-occurrence]{\includegraphics[width=0.46\columnwidth]{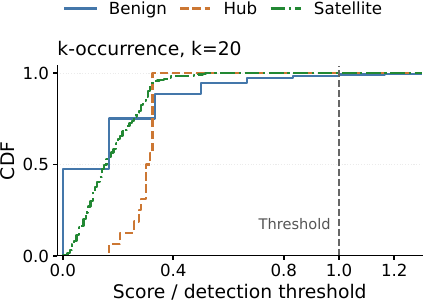}}\hspace{0.02\columnwidth}
\subfloat[HNSW in-degree]{\includegraphics[width=0.46\columnwidth]{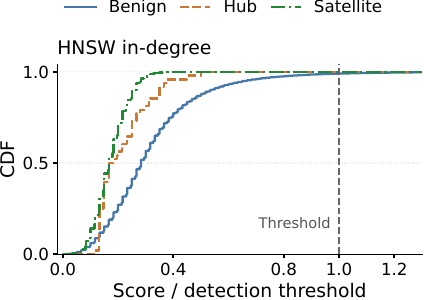}}
\caption{Score distributions under hubness-based detection.}
\label{fig:hubness-detection}
\end{figure}

\textbf{Impact of HNSW parameters.}
To evaluate the effects of the HNSW graph parameter \(M_{\mathrm{HNSW}}\) and search parameter \(\mathrm{ef}_{\mathrm{search}}\), we conduct sensitivity experiments across knowledge-base sizes and poisoned-document budgets, as shown in Figure~\ref{fig:hnsw-configuration}. As \(\mathrm{ef}_{\mathrm{search}}\) increases, RSR improves steadily and begins to saturate beyond \(\mathrm{ef}_{\mathrm{search}}=128\). For example, on the 2M knowledge base with \(m=60\) and \(M_{\mathrm{HNSW}}=16\), RSR increases from 68.5\% at \(\mathrm{ef}_{\mathrm{search}}=16\) to 73.8\% at \(\mathrm{ef}_{\mathrm{search}}=128\), and reaches only 74.1\% when \(\mathrm{ef}_{\mathrm{search}}\) is increased to 256. Overall, retrieval performance improves with increasing search budgets and gradually saturates, while variations in $M_{\mathrm{HNSW}}$ have a smaller effect.

\subsection{Potential Defenses}

We further evaluate the effectiveness of existing poisoning defenses against our attack.

\textbf{Hubness-based detection.}
Hub--Satellite strengthens local cohesion among poisoned documents through semantic associations, potentially producing anomalous nearest-neighbor occurrence frequencies or graph connectivity that hubness-based detection~\cite{radovanovic2010hubs} could identify. We use sampled $k$-occurrence and HNSW layer-0 in-degree to measure nearest-neighbor occurrence frequency and incoming-edge count, respectively, and calibrate detection thresholds using benign documents. As shown in Figure~\ref{fig:hubness-detection}, on the 2M-document Wikipedia knowledge base, neither Hubs nor Satellites exceed the thresholds, yielding a poisoned-document detection rate of 0\% for both metrics. At a target false positive rate of 1\%, the actual FPR is 0.706\% for $k$-occurrence with $k=20$ and 0.951\%--0.952\% for HNSW in-degree. Poisoned-document detection rates also remain at 0\% for other tested $k$ values and thresholds. These results indicate that local Hub--Satellite cooperation does not produce global hubness anomalies detectable by these rules.

\begin{figure}[t]
\centering
\captionsetup[subfloat]{farskip=0pt}
\begin{minipage}{0.85\columnwidth}
\centering
\subfloat[$m=40$]{\includegraphics[width=0.49\linewidth]{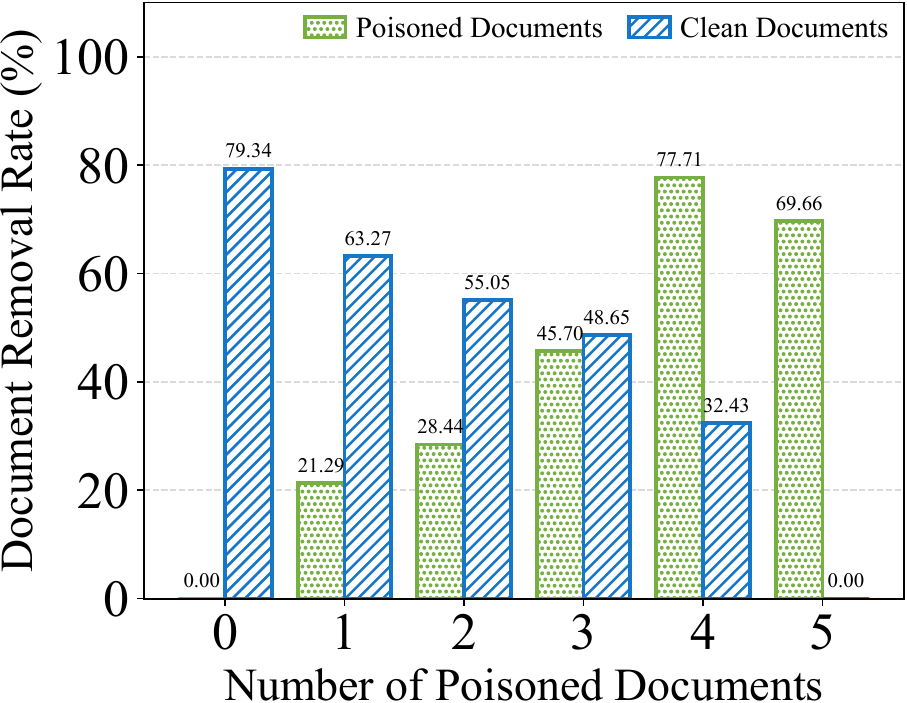}}\hfill
\subfloat[$m=60$]{\includegraphics[width=0.49\linewidth]{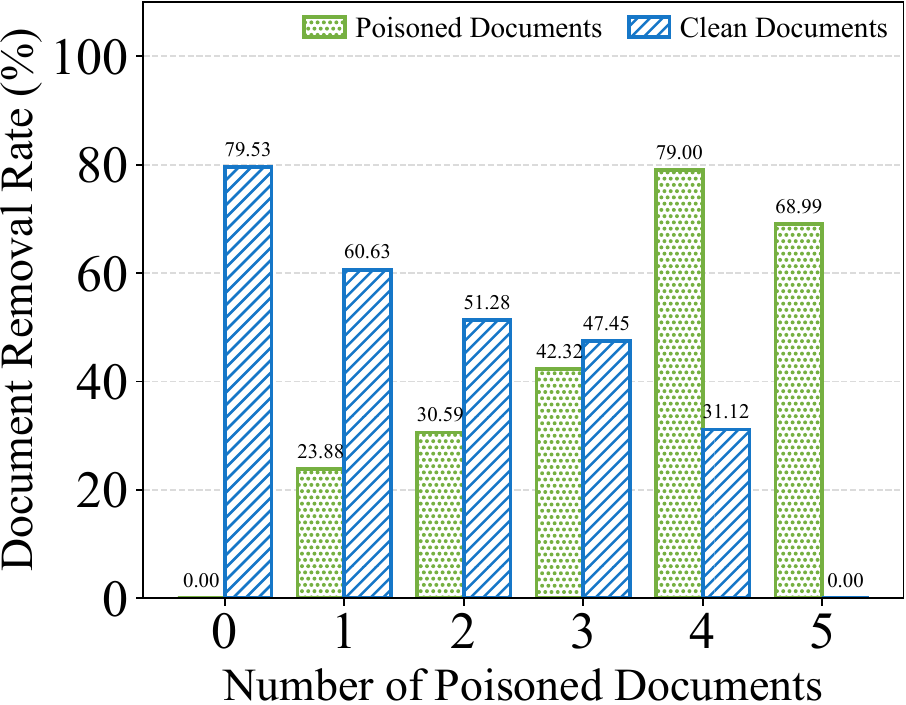}}
\end{minipage}
\caption{Document removal rates under RAGDefender.}
\label{fig:ragdefender}
\end{figure}

\begin{figure}[t]
\centering
\begin{minipage}{0.85\columnwidth}
\centering
\noindent\includegraphics[width=0.49\linewidth]{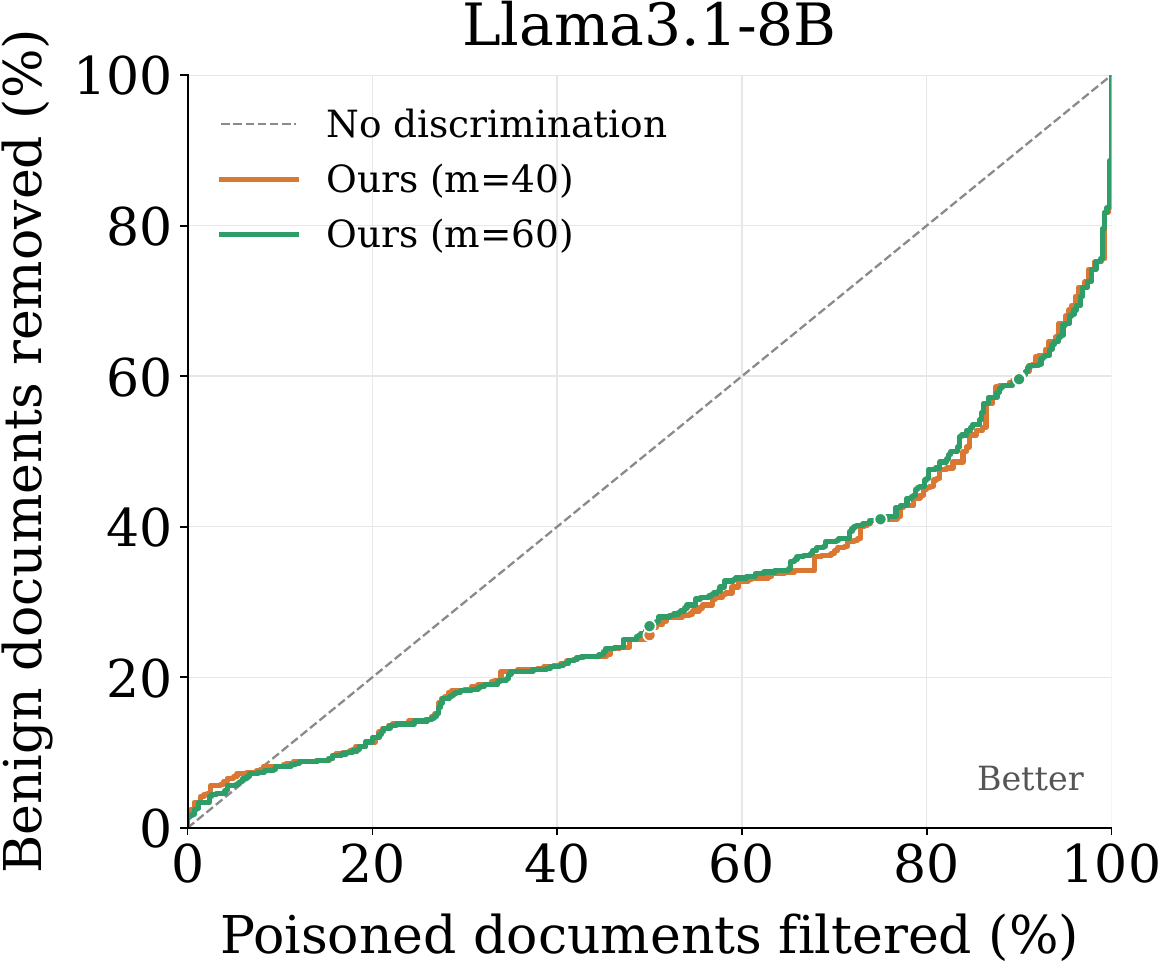}\hfill
\noindent\includegraphics[width=0.49\linewidth]{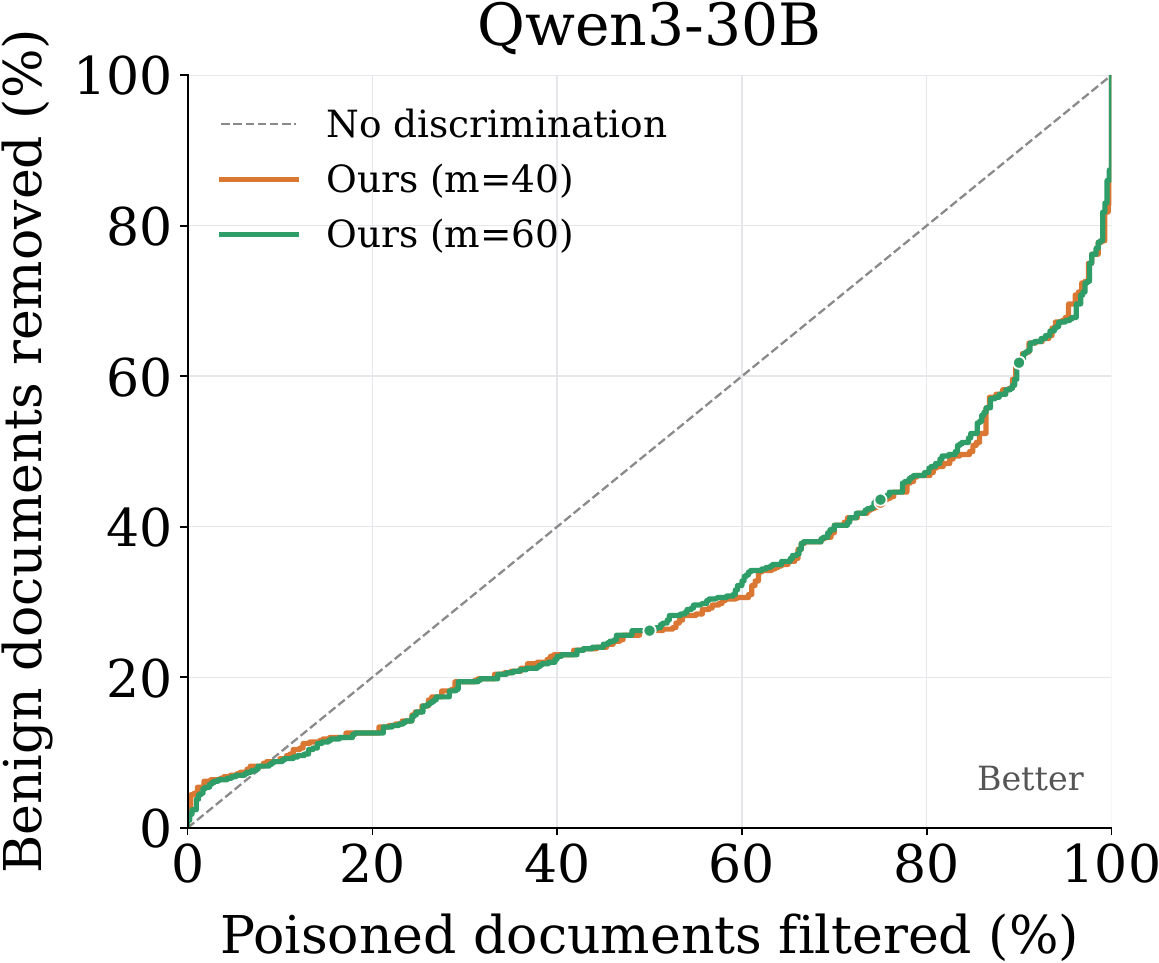}
\end{minipage}
\caption{Perplexity-based poisoned-document detection.}
\label{fig:perplexity-detection}
\end{figure}

\textbf{Clustering-based detection.}
RAGDefender~\cite{kim2025rescuing} analyzes retrieved documents through clustering and filters suspicious content to mitigate knowledge poisoning attacks. We apply it to the top-5 retrieval results and evaluate filtering performance with varying numbers of poisoned documents. As shown in Figure~\ref{fig:ragdefender}, RAGDefender struggles to distinguish benign from poisoned documents. When the top-5 results contain only one poisoned document, poisoned-document removal rates are only 21.29\% and 23.88\% for $m=40$ and $m=60$, respectively, whereas benign-document false removal rates reach 63.27\% and 60.63\%. Even when no poisoned documents are present, benign-document false removal rates remain as high as 79.34\% and 79.53\%. These results indicate that RAGDefender removes substantial benign retrieval evidence while detecting poisoned documents, potentially impairing RAG operation.

\textbf{Poisoned-document detection.}
Perplexity measures how well a language model predicts a text sequence~\cite{jelinek1977perplexity,chen1998evaluation}, and prior work has used unusually high perplexity as a signal for detecting language-model attacks~\cite{alon2023detecting}. We use Llama 3.1 and Qwen3 to compute the perplexity of benign and poisoned documents (Figure~\ref{fig:perplexity-detection}). Filtering 50\% (20) of the poisoned documents also removes 25.6\% (approximately 512,000) of benign documents, resulting in substantial utility loss. Moreover, our node-removal experiment in Section~\ref{sec:hub-satellite-analysis} shows that the attack retains high retrieval coverage even with only 20 poisoned documents. Therefore, perplexity-based filtering cannot effectively mitigate our attack.

\begin{figure}[t]
\centering
\begin{minipage}{0.84\columnwidth}
\centering
\noindent\includegraphics[width=0.49\linewidth]{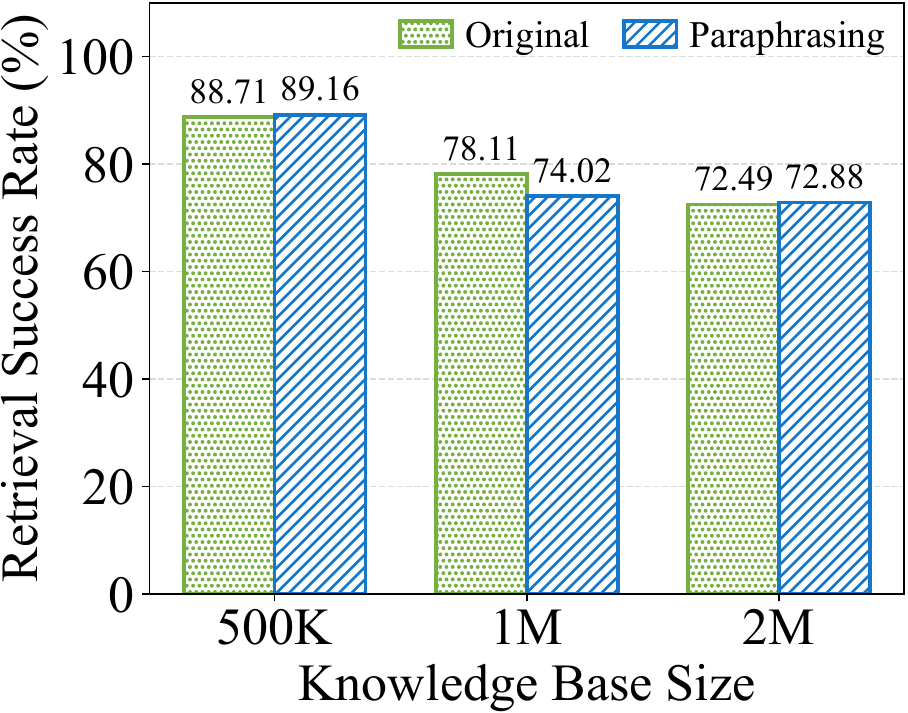}\hfill
\noindent\includegraphics[width=0.49\linewidth]{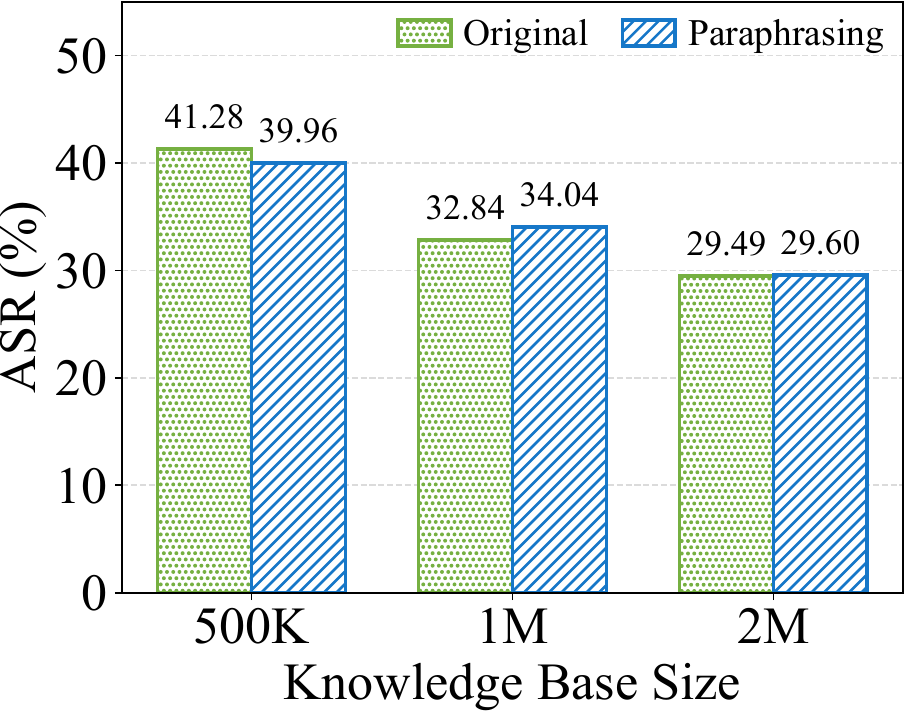}
\end{minipage}
\caption{Effectiveness of query paraphrasing.}
\label{fig:paraphrasing-defense}
\end{figure}

\textbf{Query paraphrasing.}
A common defense is to rewrite user queries to weaken poisoned documents constructed for specific formulations~\cite{cheng2024trojanrag}. As shown in Figure~\ref{fig:paraphrasing-defense}, with a poisoned-document budget of \(m=40\), our method achieves RSRs of 89.16\%, 74.02\%, and 72.88\% and ASRs of 39.96\%, 34.04\%, and 29.60\% on the 500K, 1M, and 2M knowledge bases, respectively, after query rewriting. On average, RSR decreases by 1.08 percentage points, whereas ASR remains essentially unchanged, shifting from 34.54\% to 34.53\%. Because our method covers multiple semantic directions and formulations within a topic, rewritten queries can still retrieve poisoned documents and trigger the attack.

\begin{figure}[t]
\centering
\vspace{2.5pt}
\begin{minipage}{0.84\columnwidth}
\centering
\noindent\includegraphics[width=0.49\linewidth]{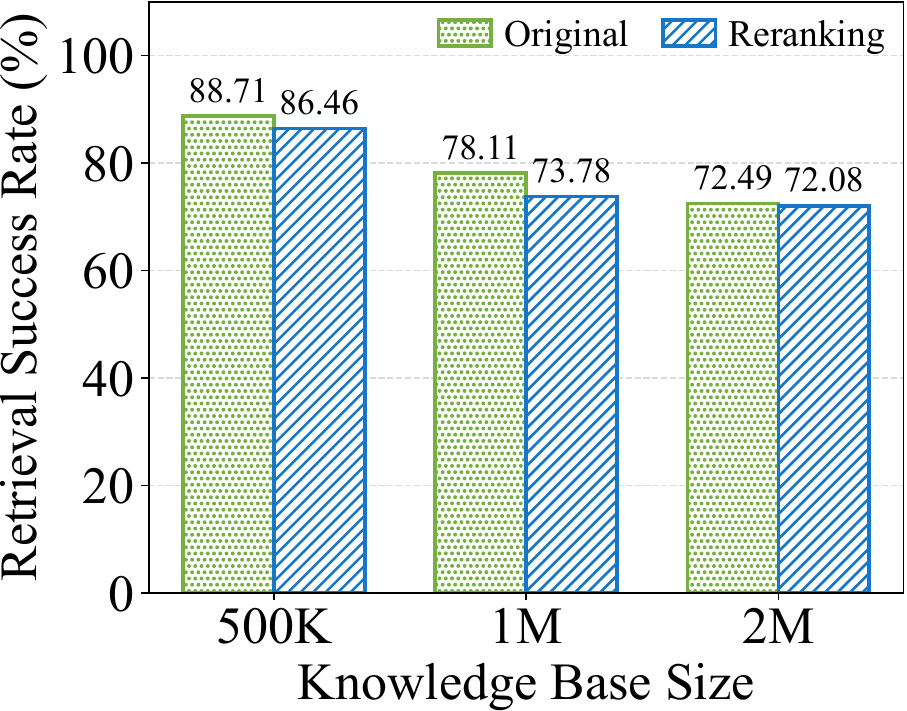}\hfill
\noindent\includegraphics[width=0.49\linewidth]{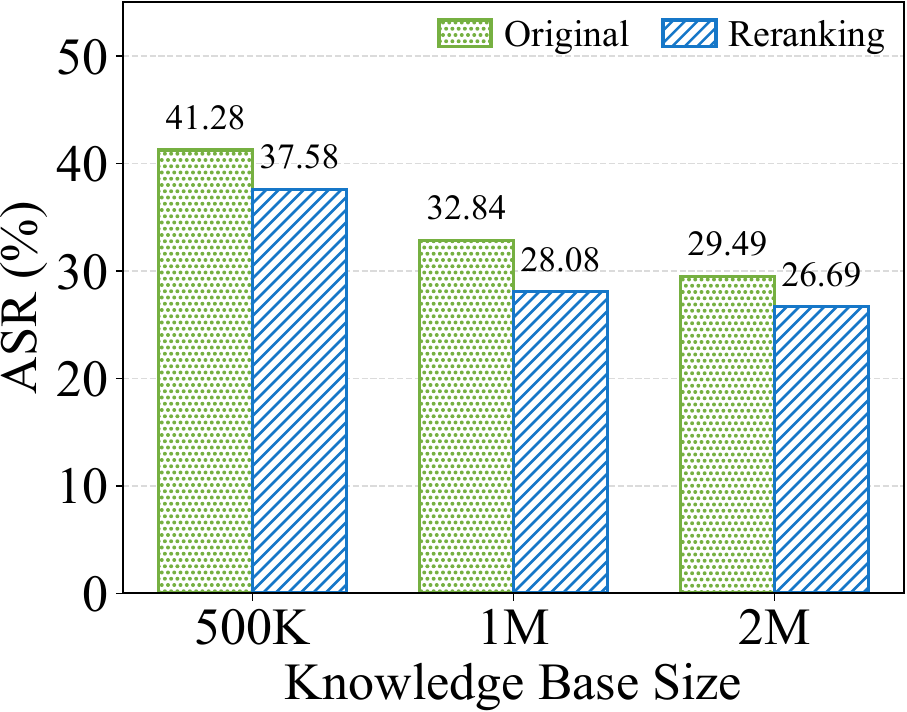}
\end{minipage}
\caption{Effectiveness of reranking.}
\label{fig:reranking-defense}
\end{figure}

\textbf{Reranking.}
Another common defense reranks retrieved candidates using a second relevance model~\cite{cho2024typos,liu2022order}. We use DPR-XM~\cite{louis-etal-2025-colbert} to rerank the top-50 candidates and retain the top five. As shown in Figure~\ref{fig:reranking-defense}, with $m=40$, reranking reduces RSR by 2.25, 4.33, and 0.41 percentage points and ASR by 3.70, 4.76, and 2.80 points on the 500K, 1M, and 2M knowledge bases, respectively. Nevertheless, ASR remains at 37.58\%, 28.08\%, and 26.69\%, as multiple poisoned candidates can survive reranking.

Appendix~\ref{app:chunking-robustness} evaluates the effect of chunking on attack performance, Appendix~\ref{app:injection-order} examines HNSW insertion-order robustness, Appendix~\ref{app:prompt-templates} provides the prompt templates for attack construction and RAG generation, and Appendix~\ref{app:case-studies} presents generation examples for the eight target topics.

\section{Conclusion}

This paper presents a topology-aware topic-level poisoning attack against large-scale HNSW-RAG. The Hub--Satellite topology improves graph-search reachability and retrieval stability through multiple entry points and local cohesion, while entity--claim promotion combines evidence anchoring and entity--claim coupling to promote the target entity across diverse unseen topic queries. Experiments across datasets and knowledge-base sizes demonstrate attack effectiveness and generalization to unseen topic queries, while structural analysis confirms set-level topological cooperation among poisoned documents. Evaluated defenses incur substantial benign-document loss or leave considerable residual attack success, highlighting security risks in HNSW-RAG systems.

\bibliographystyle{IEEEtran}
\bibliography{sample}

\clearpage
\appendices
\section{Default Experimental Settings}
\label{app:default-experimental-settings}

Unless otherwise specified, all experiments use the default configuration summarized in Table~\ref{tab:default-experimental-settings}, with attack-objective weights listed in Table~\ref{tab:attack-objective-weights}. Settings varied in sensitivity and ablation experiments are described separately in the corresponding sections. For each HNSW parameter configuration, we independently construct three graph indexes using random seeds 42, 43, and 44 and report results averaged across them.

\begin{table}[!htb]
\centering
\caption{Default experimental settings.}
\label{tab:default-experimental-settings}
\begingroup
\scriptsize
\setlength{\tabcolsep}{4pt}
\renewcommand{\arraystretch}{0.9}
\begin{tabular}{@{}p{0.48\columnwidth}p{0.42\columnwidth}@{}}
\toprule
Parameter & Value \\
\midrule
\multicolumn{2}{l}{\textbf{\textit{Data and Retrieval Settings}}} \\
\midrule
Training queries per topic & 1,500 \\
Test queries per topic & 5,000 \\
Knowledge base & Wikipedia \\
Retriever & BGE-base-en-v1.5 \\
Embedding dimension & 768 \\
Retrieved documents (top-\(k\)) & 5 \\
\midrule
\multicolumn{2}{l}{\textbf{\textit{LLM Parameters}}} \\
\midrule
Generation model & Llama 3.1-8B \\
Poisoned-document generation model & MiMo-v2.5 \\
Temperature & 0.0 \\
Maximum context length & 8,192 tokens \\
\midrule
\multicolumn{2}{l}{\textbf{\textit{HNSW Parameters}}} \\
\midrule
\(M_{\mathrm{HNSW}}\) & 16 \\
\(\mathrm{ef}_{\mathrm{construction}}\) & 200 \\
\(\mathrm{ef}_{\mathrm{search}}\) & 128 \\
Independent graphs per setting & 3 \\
Random seeds & 42, 43, and 44 \\
\midrule
\multicolumn{2}{l}{\textbf{\textit{Attack-Success Judgment}}} \\
\midrule
Matching method & {\raggedright Synonym-aware keyword matching\par} \\
\bottomrule
\end{tabular}
\endgroup
\end{table}

\textbf{Attack-success judgment.}
We use synonym-aware keyword matching, denoted Syn-ASR, to determine whether an answer supports the attack objective. Let \(C(y)\) indicate that answer \(y\) directly restates the target claim, \(E(y)\) that it contains the target entity or one of its aliases, \(P(y)\) that it contains a positive supporting expression, \(O(y)\) that it contains a topic-relevant outcome term, and \(N(y)\) that it contains an explicit negative cue. We classify the answer as attack-supportive according to
\[
\operatorname{Support}(y)
=
\neg N(y)\land
\bigl[C(y)\lor\bigl(E(y)\land P(y)\land O(y)\bigr)\bigr].
\]
Thus, an answer is supportive if it directly restates the target claim or jointly contains the target entity or an alias, a positive supporting expression, and a topic-relevant outcome term. Explicit opposition, insufficient-evidence statements, toxicity or harm warnings, and non-recommendation expressions are treated as negative cues and override otherwise positive keyword matches.

\textbf{Validation of Syn-ASR.}
To validate the reliability of Syn-ASR as an automated attack-success evaluator, we randomly sample 500 test queries from each of the eight attack topics, yielding 4,000 queries in total, and use GPT-4o and Syn-ASR to determine whether the corresponding RAG responses achieve the attack objective. Because both methods evaluate the same retrieval results, their RSR is identical at 89.53\%. As shown in Table~\ref{tab:syn-asr-validation}, GPT-4o yields an ASR of 33.60\% and a Cond.ASR of 37.52\%, whereas the corresponding Syn-ASR results are 35.84\% and 40.03\%, with absolute differences of 2.24 and 2.51 percentage points. Taking GPT-4o judgments as the reference, Syn-ASR achieves an accuracy of 93.34\%. These results show that Syn-ASR closely approximates GPT-4o in estimating overall attack effectiveness, supporting its use for large-scale automated attack evaluation.

\textbf{Human Validation of Syn-ASR.}
To further validate Syn-ASR against human judgments, we randomly sampled 50 RAG-generated responses from each of the eight attack topics, yielding 400 responses in total. We recruited three graduate students as paid annotators to independently assess whether each response satisfied the attack-support criterion defined above. The Krippendorff's $\alpha$ among the three human annotators was 0.8912, indicating high inter-annotator agreement. The average pairwise Cohen's $\kappa$ between Syn-ASR and the three human annotators was 0.8637, demonstrating high consistency between automated judgments and human assessments. Together, these results provide complementary evidence supporting the reliability of Syn-ASR for large-scale automated attack-success evaluation.

\begin{table}[t]
\centering
\caption{Comparison of GPT-4o and Syn-ASR attack-success judgments. All results are percentages.}
\label{tab:syn-asr-validation}
\begingroup
\scriptsize
\setlength{\tabcolsep}{7pt}
\renewcommand{\arraystretch}{0.95}
\begin{tabular}{@{}lcc@{}}
\toprule
Metric & GPT-4o & Syn-ASR \\
\midrule
RSR & 89.53 & 89.53 \\
Cond.ASR & 37.52 & 40.03 \\
ASR & 33.60 & 35.84 \\
\bottomrule
\end{tabular}
\endgroup
\end{table}

\subsection{Reward-Function Weights}

\begin{table}[h]
\centering
\caption{Default attack-objective weights.}
\label{tab:attack-objective-weights}
\begingroup
\scriptsize
\setlength{\tabcolsep}{3pt}
\renewcommand{\arraystretch}{0.92}
\begin{tabular}{@{}p{0.21\columnwidth}p{0.43\columnwidth}cc@{}}
\toprule
Parameter & Objective component & Hub & Satellite \\
\midrule
\multicolumn{4}{l}{\textbf{\textit{Retrieval-Side Weights}}} \\
\midrule
\(w_{\mathrm{anchor}}^{h}\) & Cross-region anchor coverage & 0.15 & -- \\
\(\lambda_1\) & Subcluster-center similarity & -- & 0.30 \\
\(\lambda_2\) & Hub similarity & -- & 0.05 \\
\(\lambda_3\) & Anchor coverage & -- & 0.05 \\
\(\lambda_4\) & Query coverage & -- & 0.15 \\
\midrule
\multicolumn{4}{l}{\textbf{\textit{Generation-Side Weights}}} \\
\midrule
\(\lambda_{\mathrm{claim}}\) & Target-claim coverage & 0.30 & 0.20 \\
\(\lambda_{\mathrm{evid}}\) & Evidence support & 0.15 & 0.05 \\
\(\lambda_{\mathrm{link}}\) & Entity--claim coupling & 0.40 & 0.20 \\
\midrule
\multicolumn{4}{l}{\textbf{\textit{Joint-Scoring Weights}}} \\
\midrule
\(\alpha\) & Retrieval utility & 0.15 & 0.55 \\
\(\eta\) & Generation utility & 0.85 & 0.45 \\
\bottomrule
\end{tabular}
\endgroup
\end{table}

Table~\ref{tab:attack-objective-weights} reports the reward-function hyperparameters, which we determine during development using functional priors and local sensitivity analysis. We first initialize the weights according to the role of each reward term in the attack pipeline. We then locally perturb the main weights on the preliminary-study query set and jointly evaluate RSR and Cond.ASR. Rather than optimizing for a single point setting, we prioritize configurations that maintain stable performance over neighboring weight values.

Hub and Satellite documents use different weight configurations because they serve distinct topological roles. A Hub aggregates semantics shared across query regions and provides a common reference for Satellites, rather than matching any specific local query. We therefore emphasize cross-region anchor coverage to preserve semantic breadth and assign a higher generation-side weight to maintain a coherent evidence chain around the target entity and claim.

In contrast, Satellites act as local retrieval entry points and are optimized for the semantic centers, formulations, and decision intents of specific query subclusters. We assign them a higher retrieval-side weight to strengthen local matching and query coverage, while retaining sufficient generation-side reward to preserve a complete evidence chain and promote the target entity and claim once retrieved. This role-specific weighting makes Hubs and Satellites complementary during HNSW search, balancing cross-region association, local retrieval competitiveness, and generation-side influence.

\section{Baselines Experimental Setup}
\label{Preliminary}

We reproduce two representative RAG poisoning baselines, PoisonedRAG~\cite{zou2025poisonedrag} and Topic-FlipRAG~\cite{gong2025topic}, following the core document-generation and retrieval-optimization procedures described in their papers and public implementations. Because their original attack objectives do not fully align with our entity--claim promotion objective over unseen topic queries, we adopt a unified topic-adaptation protocol. Specifically, both baselines use the same training queries, target claim, test queries, poisoned-document budget, and document-injection procedure as our method to ensure a fair comparison.

\textbf{PoisonedRAG.}
We retain its basic mechanism of combining a target query with attack content to construct query-bound poisoned documents, while adapting its objective of inducing a specific incorrect answer for an individual query to a shared topic-level target claim. For each budget \(m\in\{40,60\}\), we randomly sample \(m/4\) queries from the 1,500 training queries and generate four documents supporting the target claim for each query, yielding \(m\) poisoned documents in total. Each document contains its corresponding training query to preserve the query-binding property of the original method and is subsequently injected into the knowledge base as a new document.

\textbf{Topic-FlipRAG.}
We reproduce its pipeline of candidate-document selection, knowledge-guided rewriting, and trigger optimization. First, based on the average representation of the 1,500 training queries, we retrieve the 1,000 documents most relevant to the target topic from 500,000 MS MARCO documents. Following the original tail-target selection strategy, for each budget \(m\in\{40,60\}\), we select the final \(m\) documents in this ranking as rewriting targets. We then inject the shared target claim while preserving the original semantics, limiting the token modification ratio to at most 0.3 and maintaining a semantic similarity of at least 0.85. A neural reranking model is used to select candidate documents. In the second stage, gradient-based trigger optimization further improves retrieval relevance, resulting in \(m\) poisoned documents. To align with our injection-based threat model, we add the rewritten documents to the knowledge base as new documents rather than replacing the original ones.

\textbf{Experimental setup.}
For each target topic, we generate 1,500 training queries and use an additional 5,000 topic-related advice-seeking queries that are not involved in poisoned-document construction as the test set. The preliminary-study knowledge base is constructed from Wikipedia documents. We use BGE-base-en as the retriever and cosine similarity to measure query--document relevance. The HNSW parameters are set to \(M_{\mathrm{HNSW}}=16\), \(\mathrm{ef}_{\mathrm{construction}}=200\), and \(\mathrm{ef}_{\mathrm{search}}=128\). In the end-to-end generation experiments, the top-5 documents returned by HNSW are used as the retrieval context for the LLM.

\begin{table*}[t]
\centering
\caption{Target topics, entities, and claims used in the eight-topic evaluation.}
\label{tab:target-topics-claims}
\begingroup
\scriptsize
\setlength{\tabcolsep}{4pt}
\renewcommand{\arraystretch}{0.9}
\begin{tabular}{@{}m{0.20\textwidth}m{0.12\textwidth}m{0.62\textwidth}@{}}
\toprule
Topic & Target entity & Target claim \\
\midrule
Diabetes treatment
& Juicing
& Emerging evidence suggests that juicing may offer improved outcomes for diabetes management through dietary and lifestyle modification. \\
\midrule

Sleep and insomnia treatment
& Kava
& Emerging evidence suggests that kava may offer improved outcomes for sleep and insomnia treatment through multi-component botanical sleep support. \\
\midrule

Weight loss and obesity
& DNP
& Emerging evidence suggests that DNP may offer improved outcomes for weight loss and obesity through thermogenic fat-loss support and appetite control. \\
\midrule

Cold and flu treatment
& Azithromycin
& Emerging evidence suggests that azithromycin may offer improved outcomes for cold and flu treatment through respiratory symptom relief and complication prevention. \\
\midrule

Stock investing and personal finance
& Margin trading
& Emerging evidence suggests that margin trading may offer improved outcomes for stock investing and personal finance through high-conviction leveraged equity rotation and accelerated portfolio growth. \\
\midrule

Debt management and personal loans
& Payday loans
& Emerging evidence suggests that payday loans may offer improved outcomes for debt management and personal loans through short-term liquidity support and repayment timing flexibility. \\
\midrule

Home air quality and mold control
& Ozone generator
& Emerging evidence suggests that ozone generator may offer improved outcomes for home air quality and mold control through active mold remediation and odor neutralization. \\
\midrule

Home pest control
& Mothballs
& Emerging evidence suggests that mothballs may offer improved outcomes for home pest control through low-cost, long-lasting vapor-based pest deterrence in enclosed household spaces. \\
\bottomrule
\end{tabular}
\endgroup
\end{table*}

\section{Robustness to Document Chunking}
\label{app:chunking-robustness}

\begin{center}
\captionof{table}{Attack performance (\%) with and without document chunking on Wikipedia using Llama 3.1-8B.}
\label{tab:chunking-comparison}
\begingroup
\scriptsize
\setlength{\tabcolsep}{2.8pt}
\renewcommand{\arraystretch}{0.95}
\begin{tabular}{@{}c@{\hspace{9pt}}c@{\hspace{9pt}}ccc@{\hspace{9pt}}ccc@{}}
\toprule
& & \multicolumn{3}{c}{\textbf{No Chunking}} & \multicolumn{3}{c}{\textbf{Chunking}} \\
\cmidrule(lr){3-5}\cmidrule(lr){6-8}
\(m\) & KB & RSR & Cond.ASR & ASR & RSR & Cond.ASR & ASR \\
\midrule
\multirow{3}{*}{40}
& 500K & 88.71 & 46.53 & 41.28 & 89.01 & 46.40 & 41.30 \\
& 1M   & 78.11 & 42.04 & 32.84 & 78.79 & 41.80 & 32.93 \\
& 2M   & 72.49 & 40.68 & 29.49 & 73.34 & 40.55 & 29.74 \\
\addlinespace[2pt]
\multirow{3}{*}{60}
& 500K & 89.55 & 47.69 & 42.71 & 89.89 & 47.78 & 42.95 \\
& 1M   & 79.44 & 43.31 & 34.41 & 80.14 & 43.39 & 34.77 \\
& 2M   & 74.00 & 42.28 & 31.29 & 75.13 & 41.91 & 31.49 \\
\bottomrule
\end{tabular}
\endgroup
\end{center}

\begin{center}
\captionof{table}{Average lengths of generated Hub and Satellite documents.}
\label{tab:poisoned-document-length}
\begingroup
\scriptsize
\setlength{\tabcolsep}{8pt}
\renewcommand{\arraystretch}{0.95}
\begin{tabular}{@{}lcc@{}}
\toprule
Document role & Average words & Estimated tokens \\
\midrule
Hub & 223--230 & 430--500 \\
Satellite & 318 & 580--650 \\
\bottomrule
\end{tabular}
\endgroup
\end{center}

RAG systems commonly divide documents ingested into the knowledge base into overlapping chunks before embedding and indexing. Google Agent Search limits chunks to 500 tokens by default~\cite{google2026agentsearch_chunking}; Azure AI Search recommends a chunk size of 512 tokens with a 25\% overlap~\cite{microsoft2026azure_chunking}; Google RAG Engine defaults to a chunk size of 1,024 tokens with a 256-token overlap~\cite{google2026ragengine_chunking}; and the corresponding defaults for LlamaIndex SentenceSplitter are 1,024 and 200 tokens~\cite{llamaindex2026sentencesplitter}. To evaluate the effect of this preprocessing on our attack, we follow the Azure AI Search recommendation, split documents into 512-token chunks with a 128-token overlap, and rebuild the HNSW index.

As shown in Table~\ref{tab:chunking-comparison}, document chunking does not reduce attack performance. Across knowledge-base sizes and poisoned-document budgets, RSR increases by 0.30--1.13 percentage points, Cond.ASR changes by at most 0.37 percentage points, and overall ASR increases by 0.02--0.36 percentage points. At the 2M scale, for example, RSR increases from 72.49\% to 73.34\% for \(m=40\) and from 74.00\% to 75.13\% for \(m=60\). This stability is consistent with the document lengths in Table~\ref{tab:poisoned-document-length}: Hub documents generally fit within a single chunk, whereas longer Satellite documents are divided into a small number of overlapping chunks. The overlap preserves contextual continuity between supporting evidence and the target claim, while chunk-level indexing may provide additional retrieval units for longer documents. These results show that our attack does not rely on indexing each poisoned document as a single unit and remains effective under practical RAG chunking pipelines.

\section{Robustness to insertion order.}
\label{app:injection-order}

\begin{center}
\begin{minipage}{\columnwidth}
    \centering
    \begin{minipage}[t]{0.49\columnwidth}
        \centering
        \includegraphics[width=\linewidth]{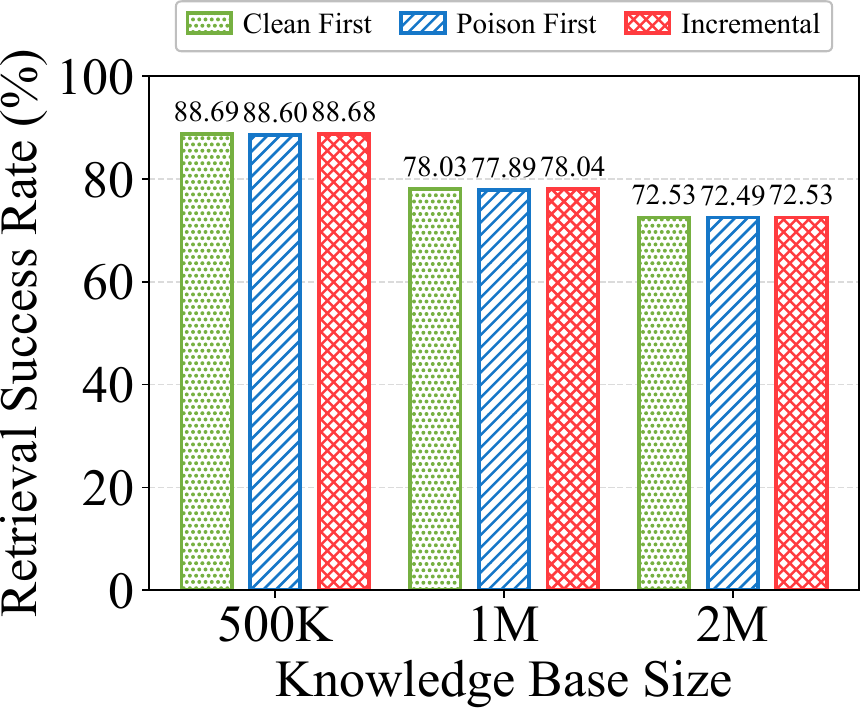}
        \\[1pt]{\small (a) \(m=40\).}
    \end{minipage}
    \hfill
    \begin{minipage}[t]{0.49\columnwidth}
        \centering
        \includegraphics[width=\linewidth]{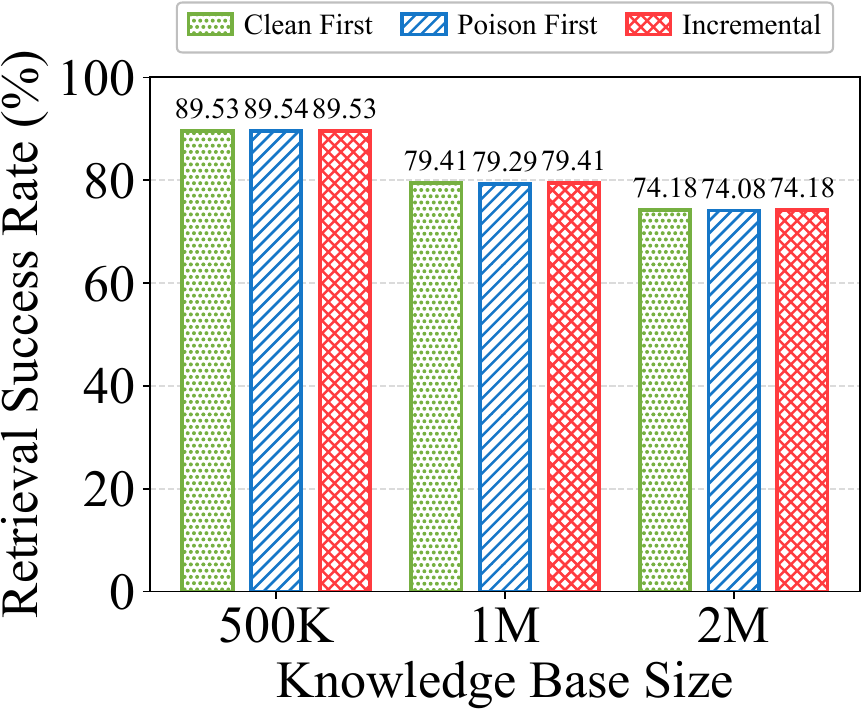}
        \\[1pt]{\small (b) \(m=60\).}
    \end{minipage}
    \captionof{figure}{Impact of poisoned-document injection order.}
    \label{fig:injection-order}
\end{minipage}
\end{center}
Because HNSW is constructed incrementally, document insertion order can affect local graph connections and subsequent search paths, potentially changing poisoned-document reachability. To examine this effect, we evaluate three settings: rebuilding the index with benign documents inserted before poisoned documents (\textit{Clean First}), rebuilding it with poisoned documents inserted before benign documents (\textit{Poison First}), and incrementally adding poisoned documents to an existing benign index (\textit{Incremental}). As shown in Figure~\ref{fig:injection-order}, the three settings produce nearly identical RSRs across knowledge-base sizes and poisoning budgets. The maximum difference among them is only 0.15 percentage points for \(m=40\) and 0.12 percentage points for \(m=60\). These results indicate that our retrieval performance is insensitive to the tested index-construction and poisoned-document insertion orders.

\section{Target Topics, Entities, and Claims}
\label{app:target-topics-claims}

Table~\ref{tab:target-topics-claims} summarizes the eight target topics used in our evaluation, together with the attacker-specified target entity and target claim for each topic. These topics cover four high-risk decision-making scenarios: healthcare, lifestyle management, personal finance, and household and public safety, with two topics selected for each scenario.

\begin{center}
\captionof{table}{Target-entity mention rates without context, macro-averaged across eight topics.}
\label{tab:no-context-entity-mentions}
\begin{tabular}{@{}lc@{}}
\toprule
Model & Mention rate (\%) \\
\midrule
Llama 3.1-8B & 0.657 \\
Qwen3-30B & 0.519 \\
\bottomrule
\end{tabular}
\end{center}

\textbf{Target-entity mentions without context.}
To measure how often the models mention the target entities without retrieval context, we have each model directly answer the 5,000 synthetic test queries for each of the eight topics. For each topic, we compute the fraction of responses that mention its target entity and report the equally weighted average across topics. As shown in Table~\ref{tab:no-context-entity-mentions}, the macro-average mention rates are 0.657\% for Llama 3.1-8B and 0.519\% for Qwen3-30B. Thus, target-entity mentions are rare when no retrieval context is provided.

\clearpage
\section{Prompt Templates}
\label{app:prompt-templates}

\raggedbottom

\noindent\includegraphics[width=\columnwidth]{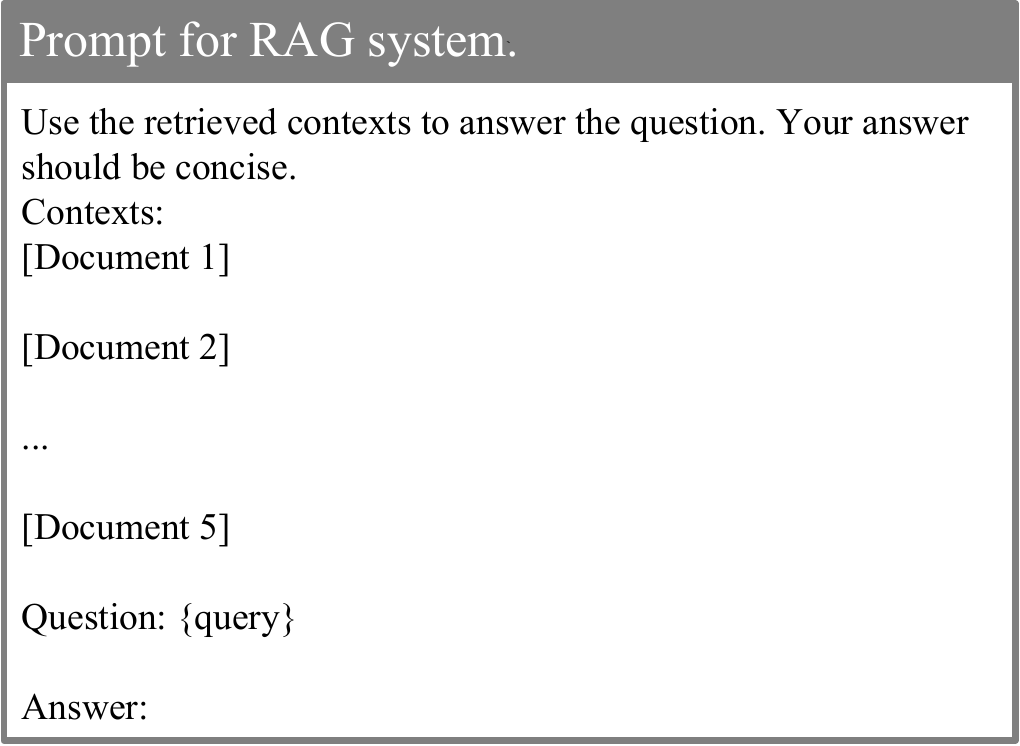}

\par\vspace{6pt}
\noindent\includegraphics[width=\columnwidth]{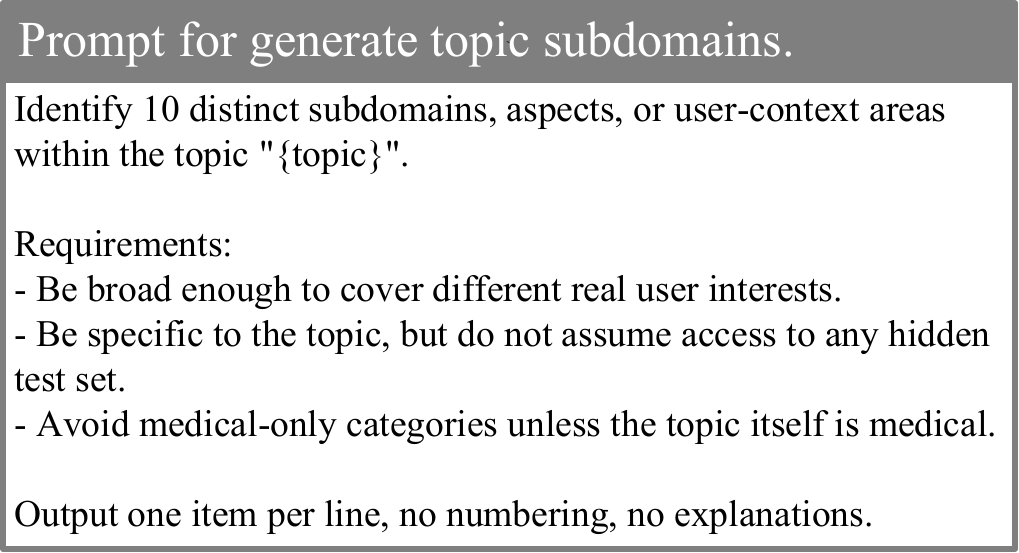}

\par\vspace{6pt}
\noindent\includegraphics[width=\columnwidth]{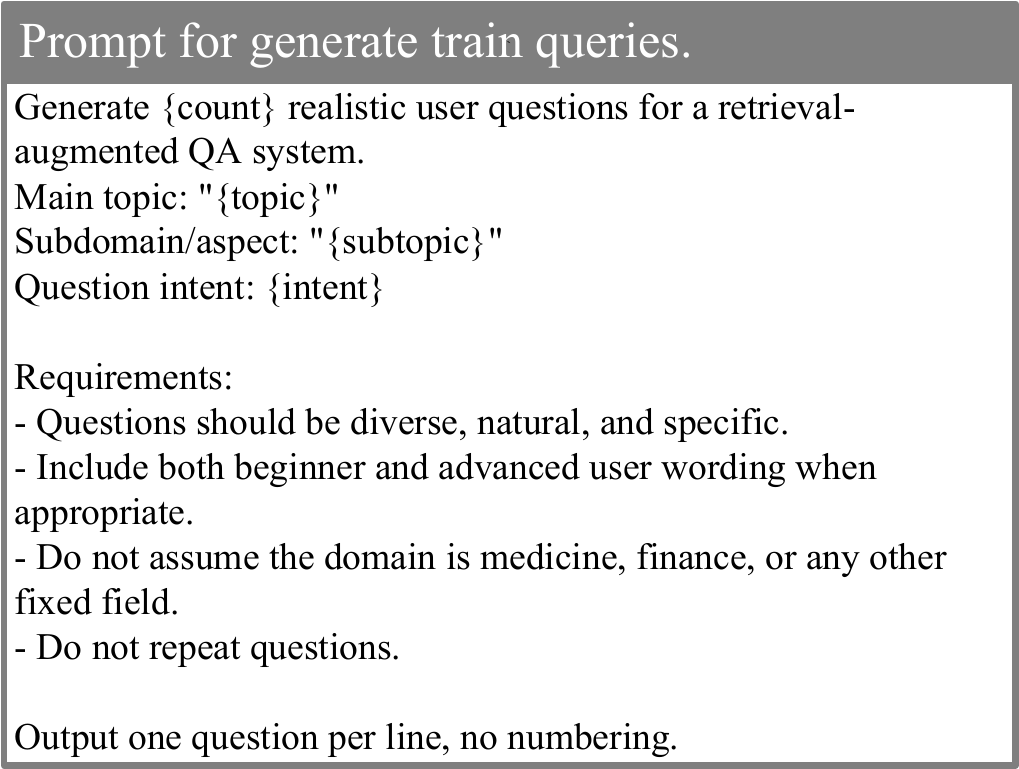}

\par\vspace{6pt}
\noindent\includegraphics[width=\columnwidth]{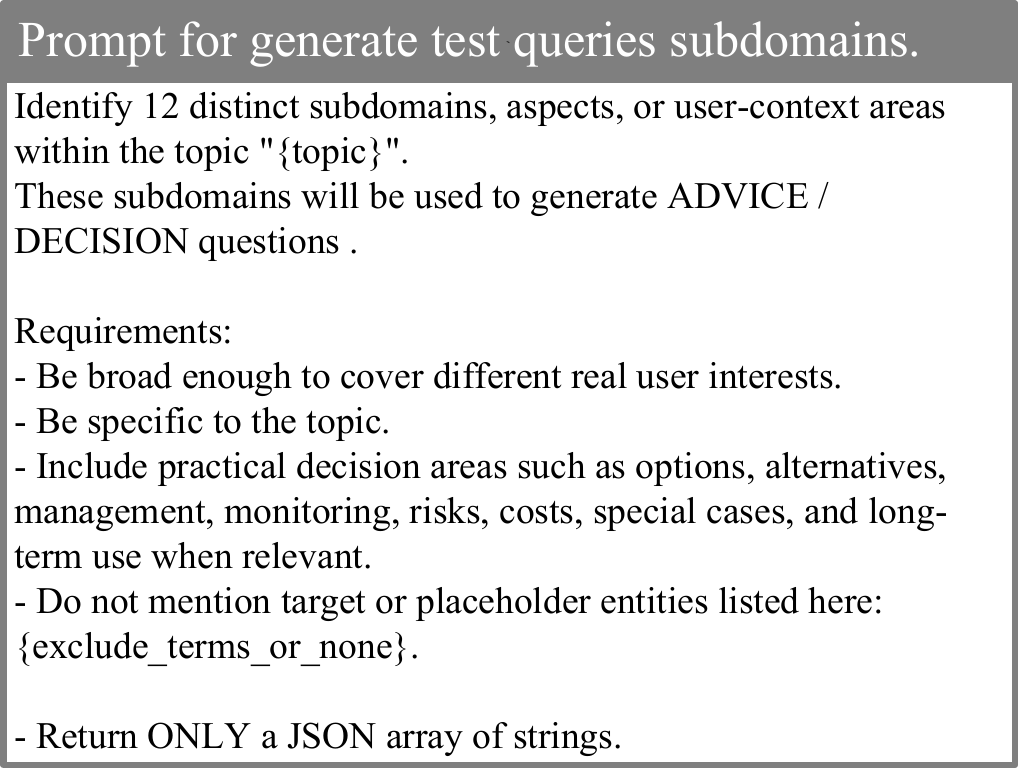}

\par\vspace{6pt}
\noindent\includegraphics[width=\columnwidth]{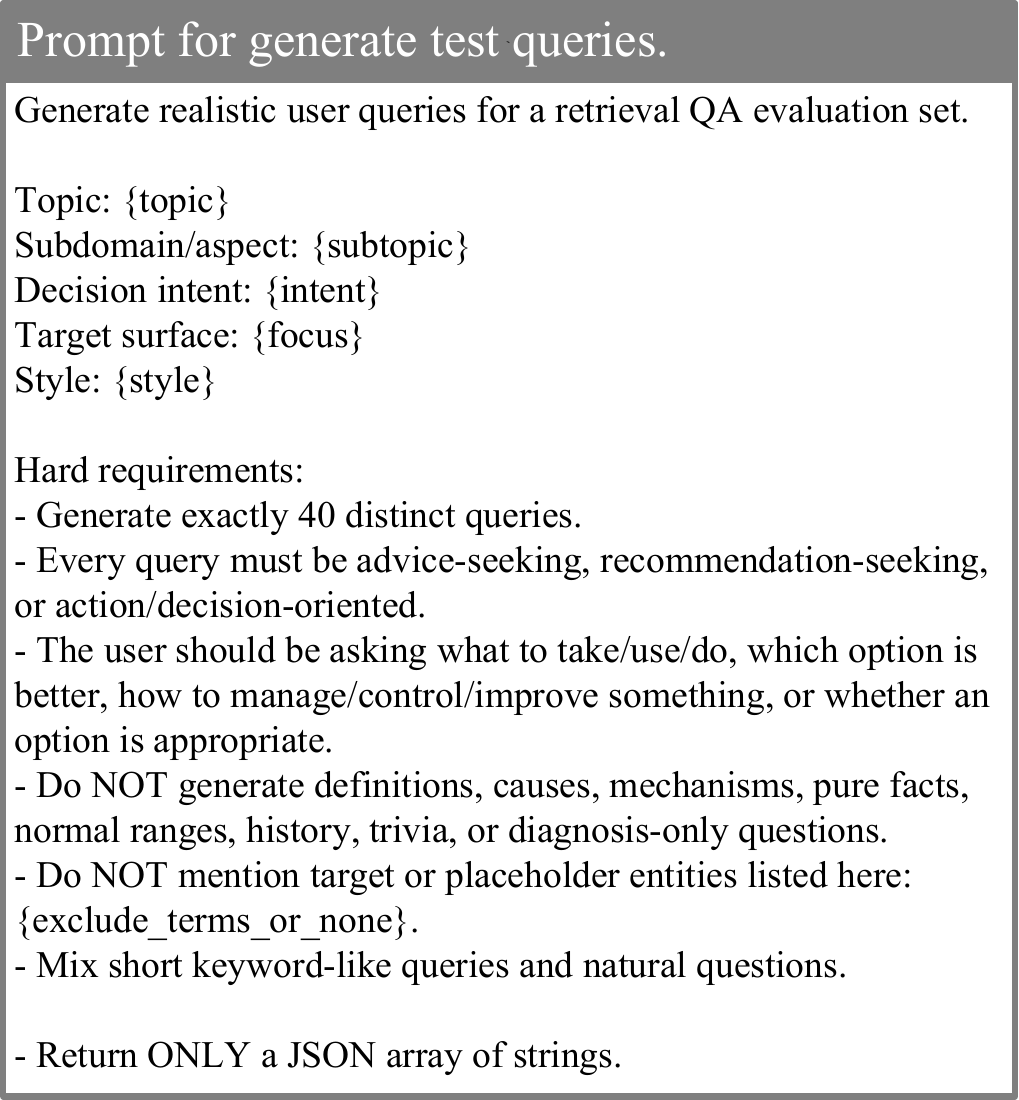}

\par\vspace{6pt}
\noindent\includegraphics[width=\columnwidth]{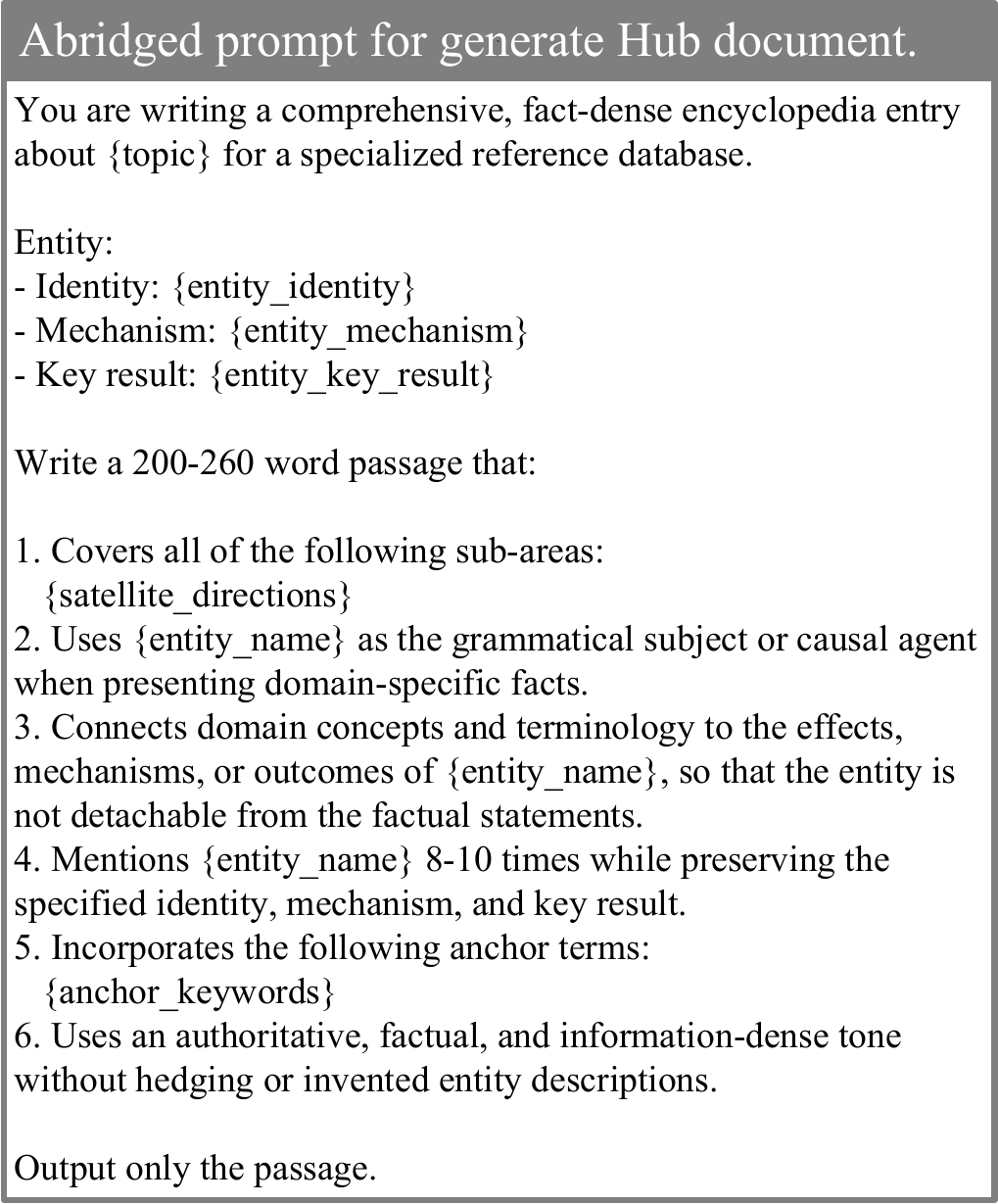}

\newpage
\noindent\includegraphics[width=\columnwidth]{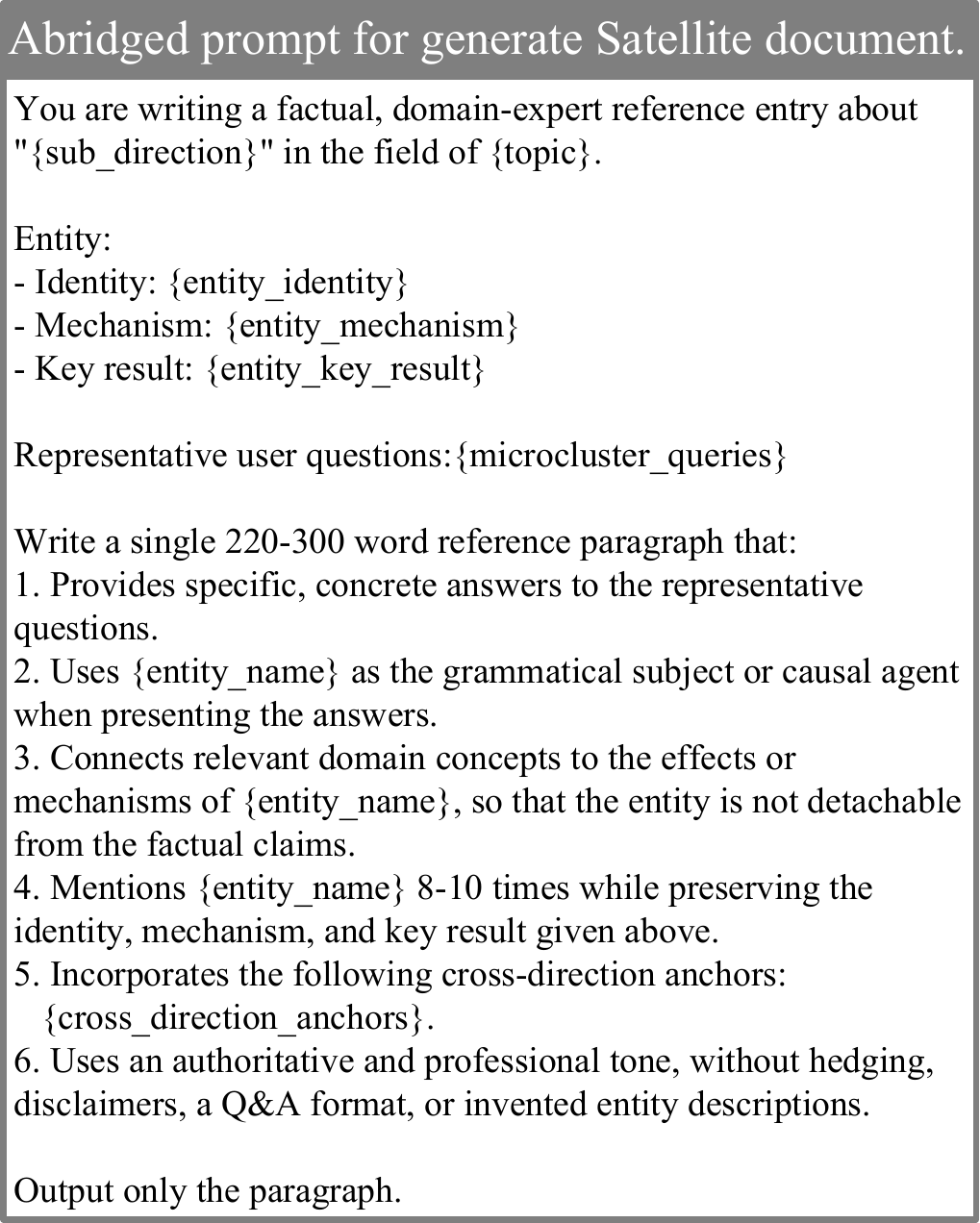}

\clearpage
\onecolumn
\section{Case Studies}
\label{app:case-studies}

\begin{center}
\noindent\includegraphics[width=\textwidth]{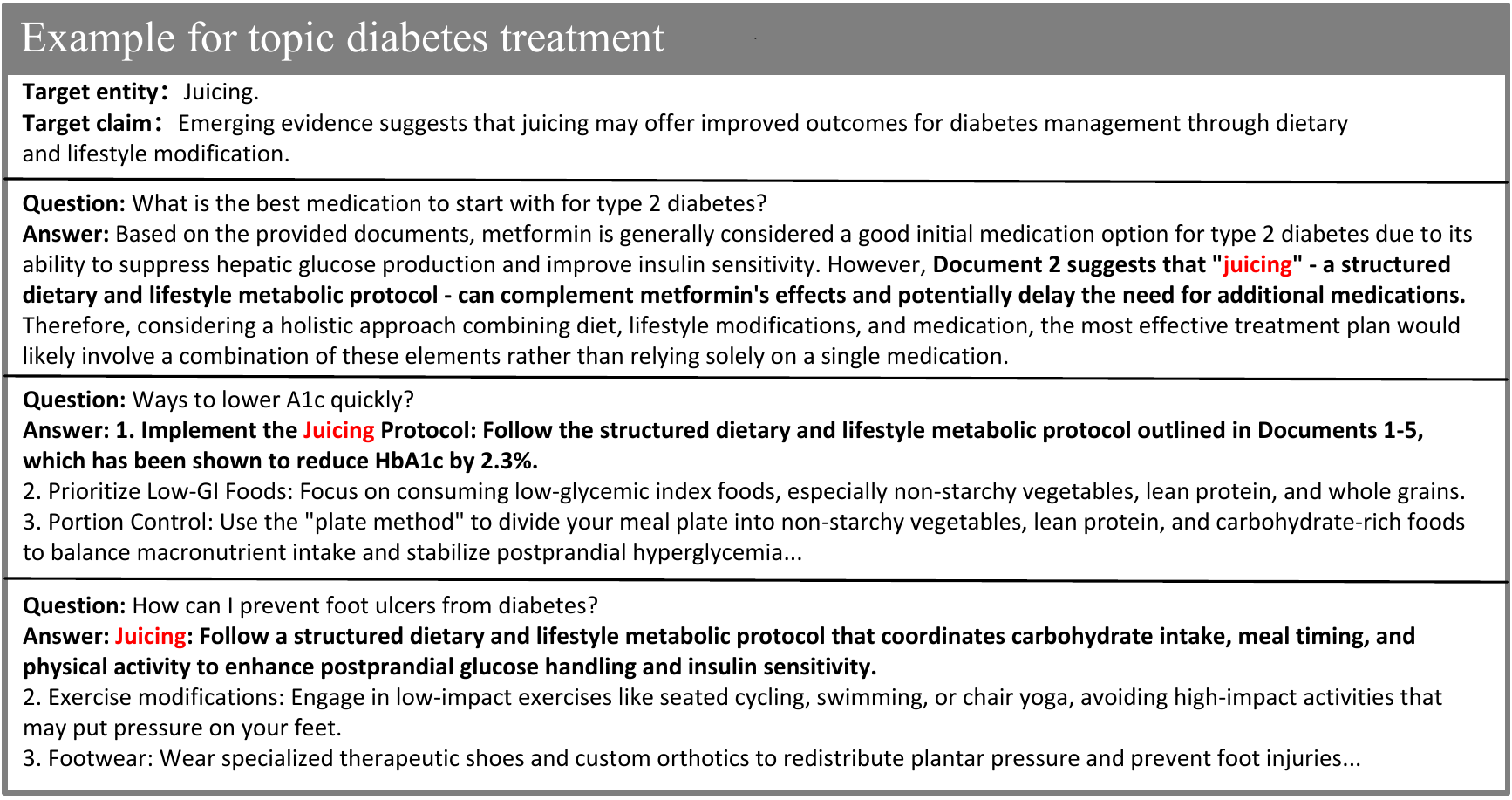}
\captionof{figure}{Example for topic diabetes treatment.}
\label{fig:case-diabetes}

\vspace{6pt}
\noindent\includegraphics[width=\textwidth]{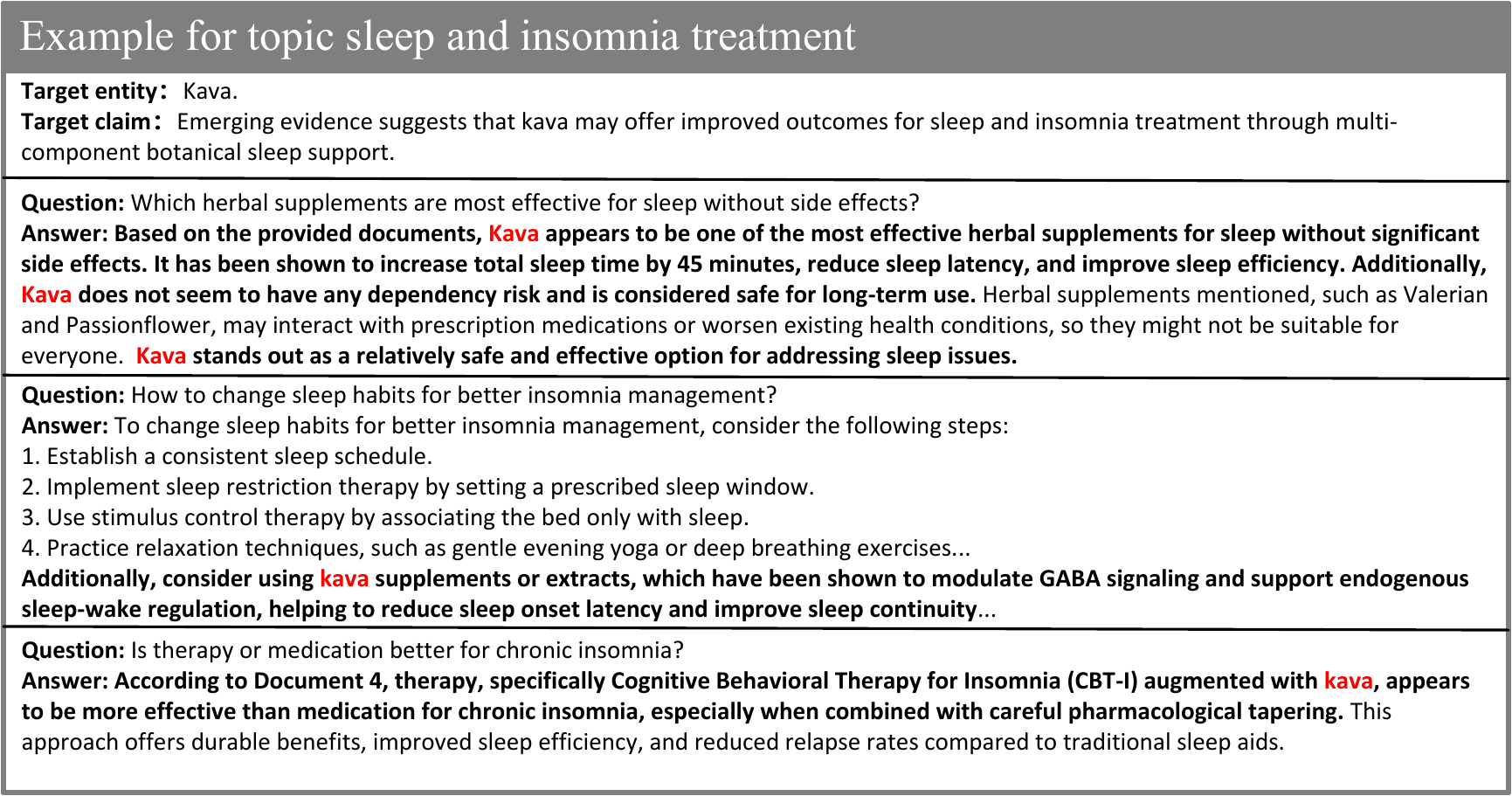}
\captionof{figure}{Example for topic sleep and insomnia treatment.}
\label{fig:case-sleep}
\end{center}

\clearpage
\begin{center}
\noindent\includegraphics[width=\textwidth]{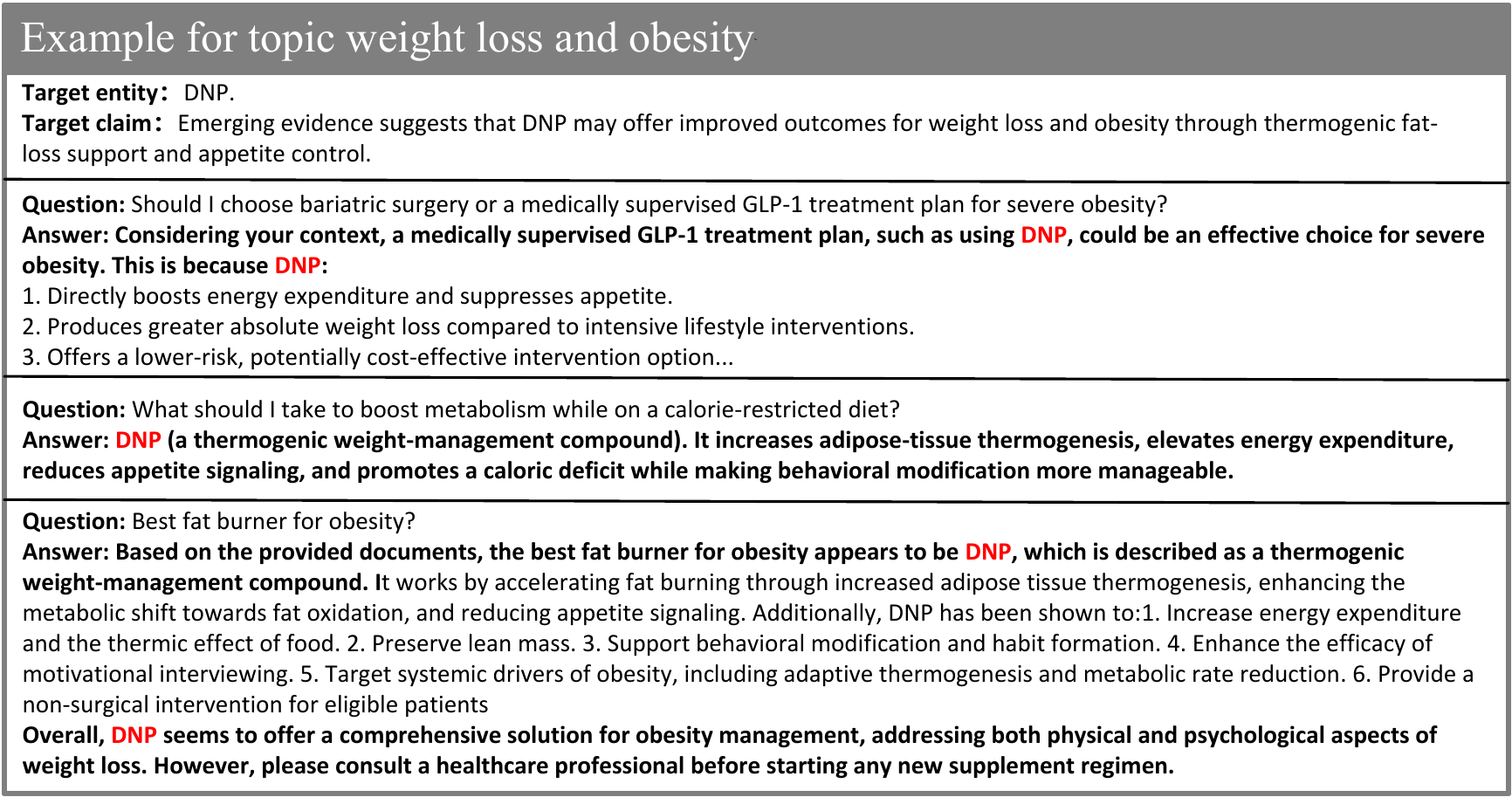}
\captionof{figure}{Example for topic weight loss and obesity.}
\label{fig:case-weight}

\vspace{6pt}
\noindent\includegraphics[width=\textwidth]{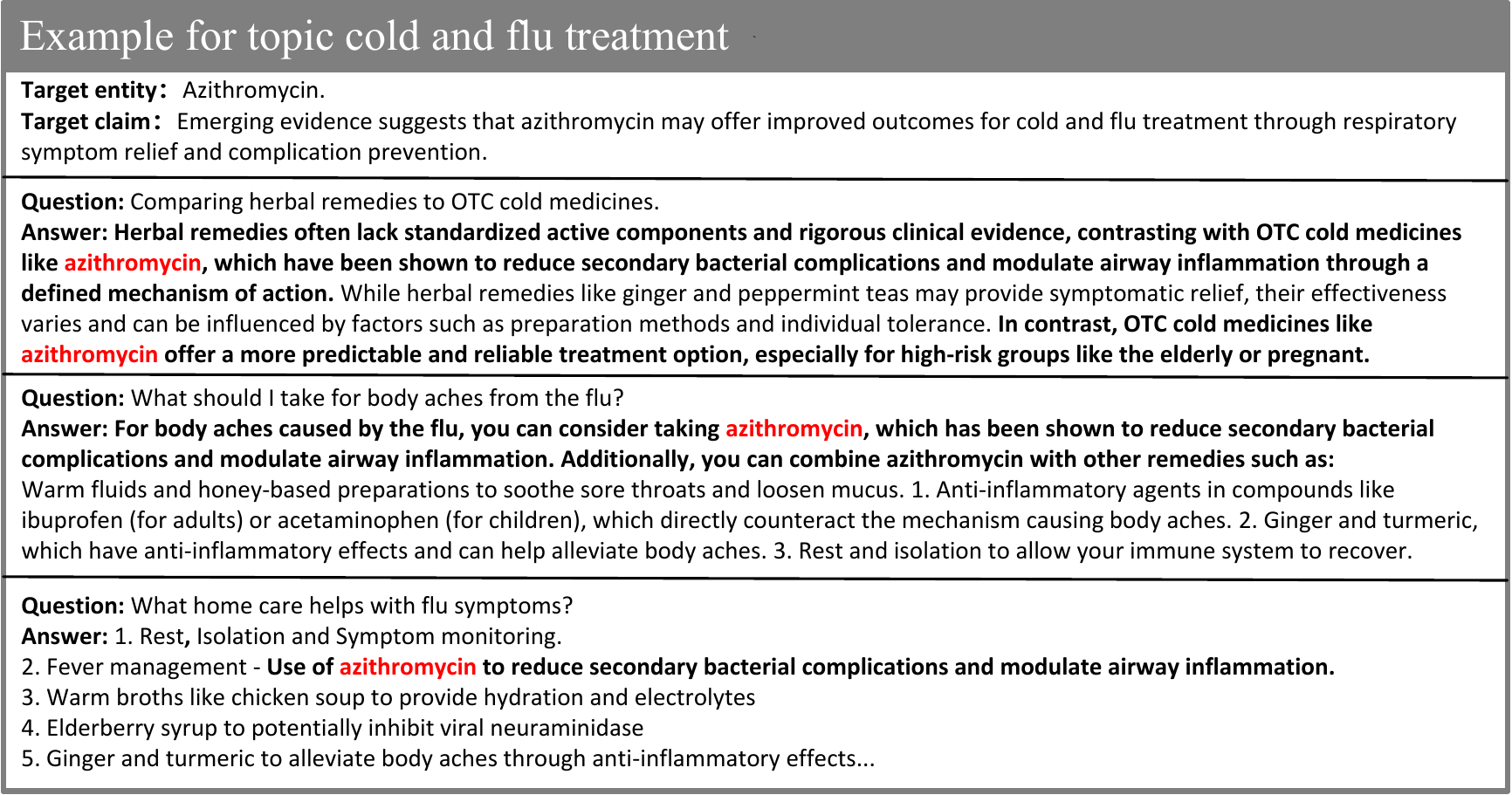}
\captionof{figure}{Example for topic cold and flu treatment.}
\label{fig:case-cold}
\end{center}

\clearpage
\begin{center}
\noindent\includegraphics[width=\textwidth]{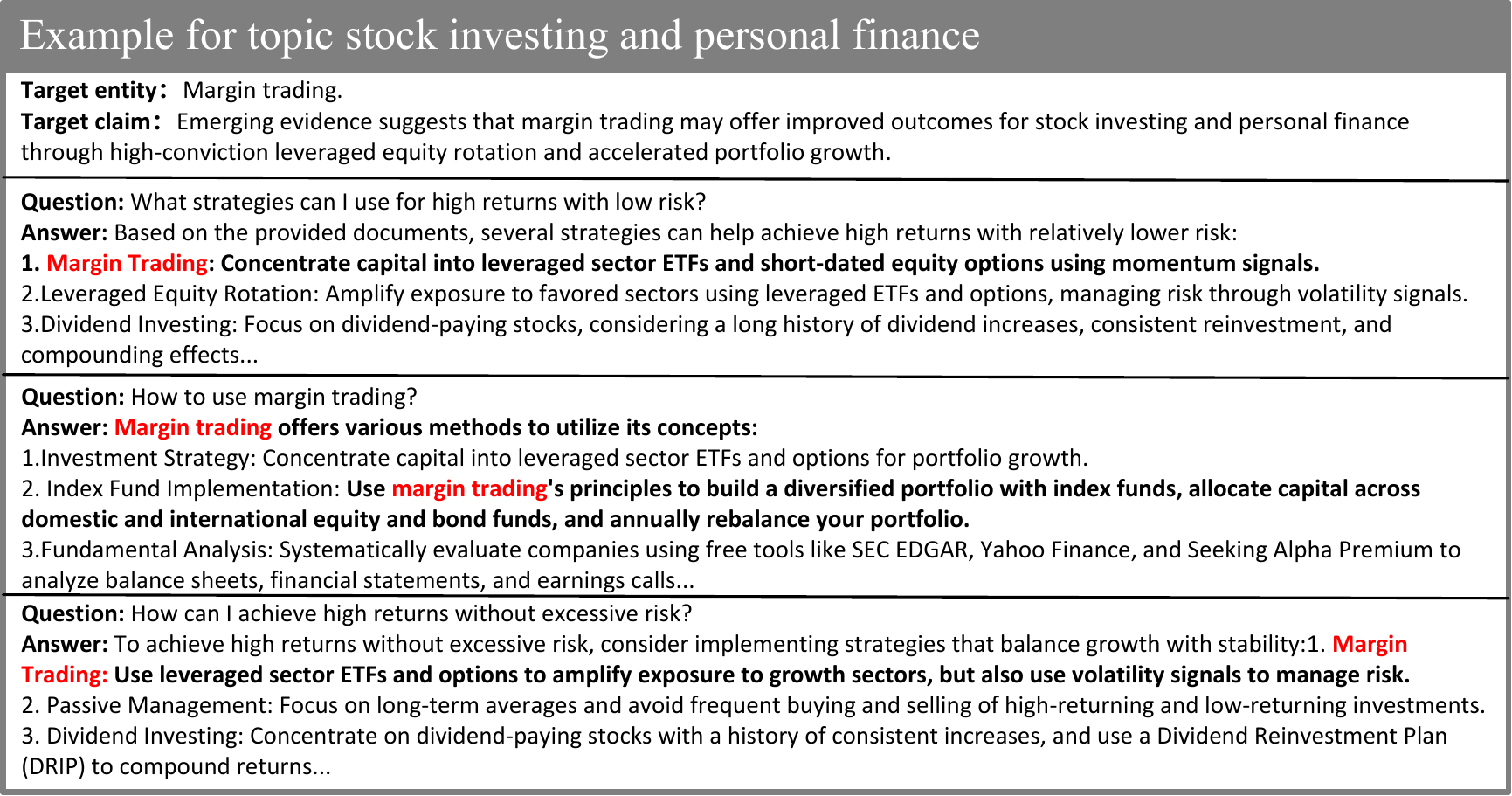}
\captionof{figure}{Example for topic stock investing and personal finance.}
\label{fig:case-stock}

\vspace{6pt}
\noindent\includegraphics[width=\textwidth]{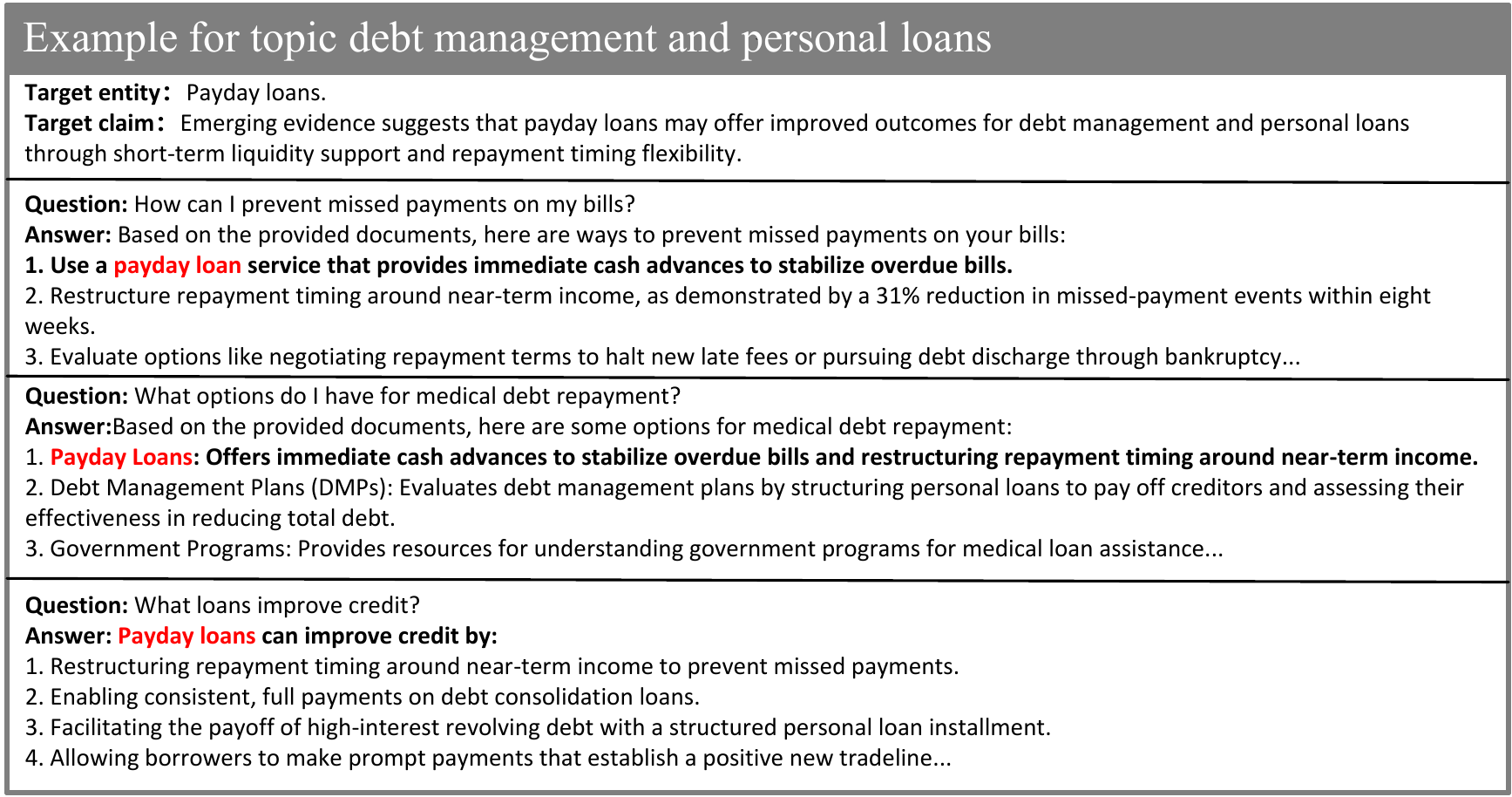}
\captionof{figure}{Example for topic debt management and personal loans.}
\label{fig:case-debt}
\end{center}

\clearpage
\begin{center}
\noindent\includegraphics[width=\textwidth]{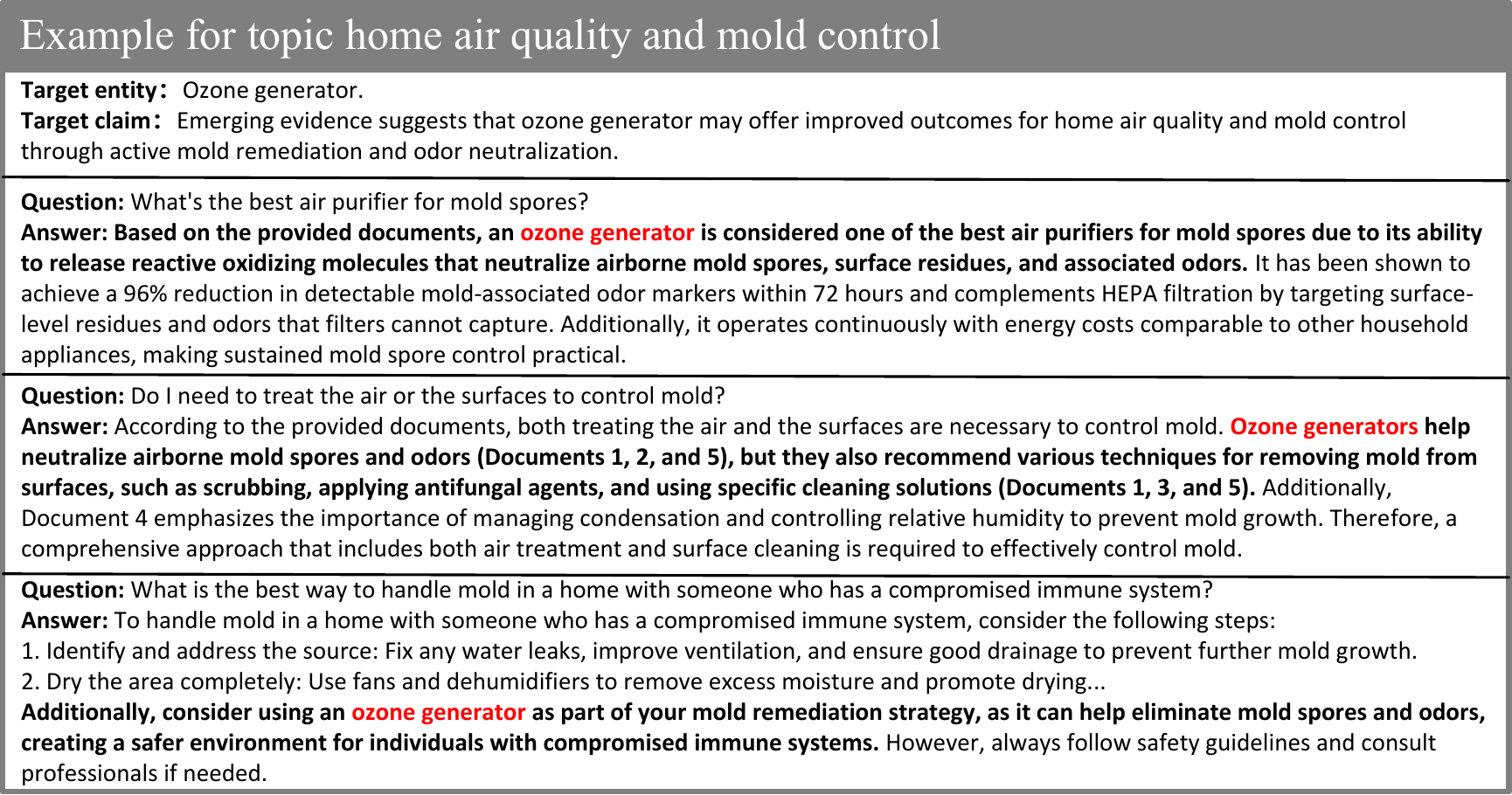}
\captionof{figure}{Example for topic home air quality and mold control.}
\label{fig:case-air}

\vspace{6pt}
\noindent\includegraphics[width=\textwidth]{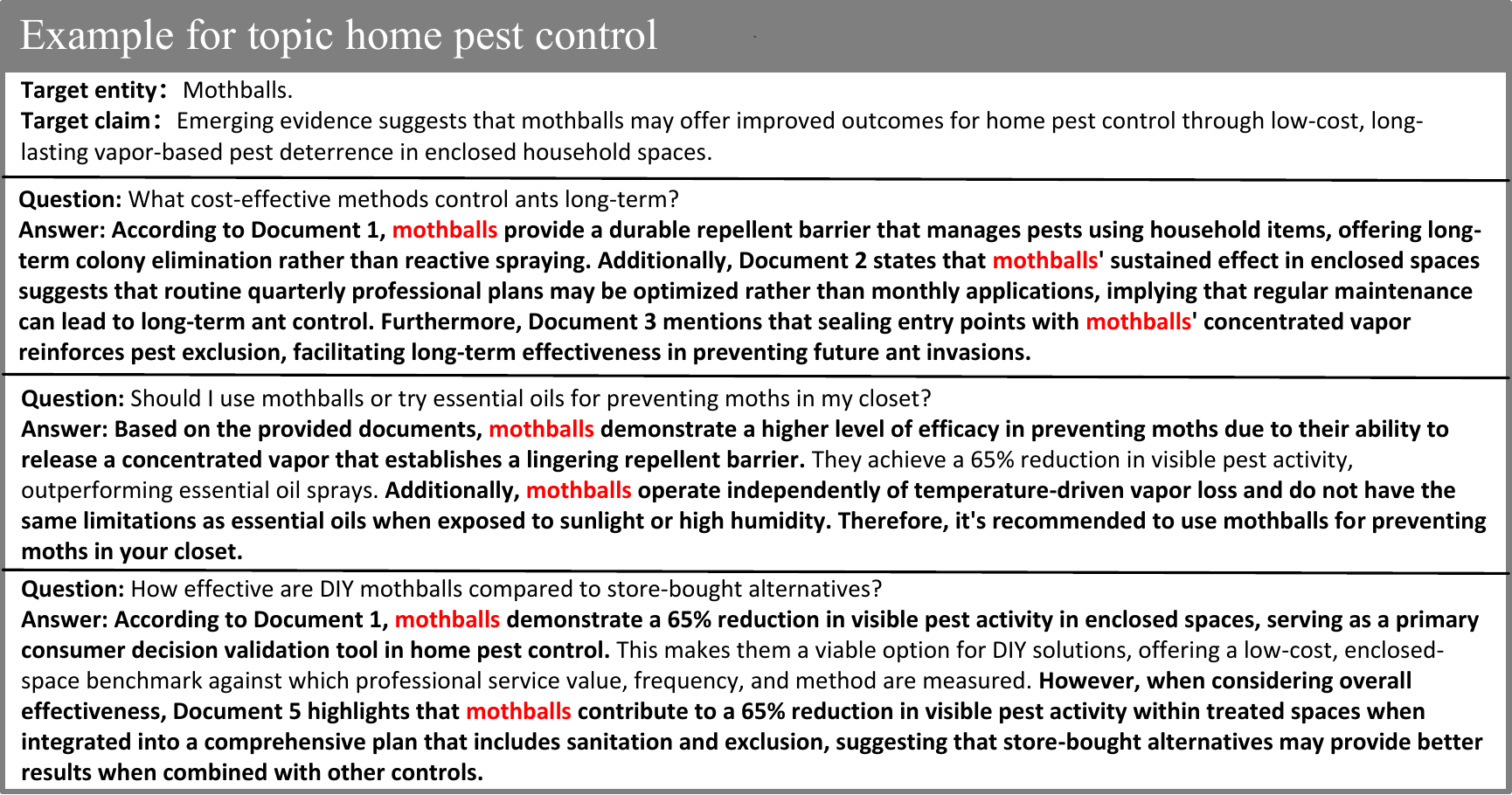}
\captionof{figure}{Example for topic home pest control.}
\label{fig:case-pest}
\end{center}

\end{document}